\documentclass[11pt,a4paper]{article}
\usepackage{jheppub}
\usepackage[utf8]{inputenc}
\usepackage{graphicx,fancybox,float,comment}

\usepackage{units}
\usepackage{subcaption}
\usepackage{multirow,array,arydshln}
\usepackage[dvipsnames]{xcolor}
\usepackage{amsmath,amssymb,amsfonts}
\usepackage{cleveref}
\usepackage{yfonts}
\usepackage{tensor}
\usepackage{paralist}
\usepackage{braket}
\usepackage{tensor,mathrsfs}
\usepackage[normalem]{ulem}
\title{
Inverted c-functions in thermal states 
}
\author{Casey Cartwright, Matthias Kaminski}
\date{March 2020}
\affiliation{Department of Physics and Astronomy, University of Alabama, 
\\ 514 University Boulevard,
Tuscaloosa, AL 35487, USA}

\emailAdd{cccartwright@crimson.ua.edu}
\emailAdd{mski@ua.edu}

\newcommand{\exd}{\mathrm{d}}
\newcommand{\tr}{\text{Tr}}
\newcommand{\zs}{z_*}
\newcommand{\cfour}{Q}

\abstract{
We first compute the effect of a chiral anomaly, charge, and a magnetic field on the entanglement entropy 
in $\mathcal{N}=4$ Super-Yang-Mills theory at strong coupling via holography. 
Depending on the width of the entanglement strip the entanglement entropy probes energy scales from the ultraviolet to the infrared energy regime of this quantum field theory (QFT) prepared in a given state. 
From the entanglement entropy, we  compute holographic c-functions and demonstrate an inverted c-theorem for them. 
That is, these c-functions in generic thermal states monotonically increase towards the infrared (IR) energy regime. 
This is in contrast to the c-functions in vacuum states which decrease along the renormalization group flow towards the IR regime of a renormalizable QFT. 
Furthermore, in thermal states and in the IR limit, the c-functions behave thermally, growing proportionally to the value of the thermal entropy. 
The chiral anomaly affects the c-functions more in the IR regime, and its effect is peaked at an intermediate value of the magnetic field at a fixed chemical potential and temperature. 
}
\begin{document}
\maketitle
\section{Introduction} 
\label{sec:introduction}
A physical system is described by the theory it obeys together with the state it occupies. Therefore, measures of information about a system, such as entropies, are affected by the states. In quantum field theory {\it c-functions} are interpreted as a heuristic measure for the number of degrees of freedom present in the system at a given energy scale. Loosely speaking a c-function is a measure of information about the system. These c-functions are computed with reference to the vacuum state, and decrease monotonically from the ultraviolet (UV) towards the infrared (IR) energy scale, along the renormalization group (RG) flow~\cite{Zamolodchikov:1986gt,Cardy:1988cwa}. 
In agreement with these field theory results, such a c-function was later defined from entanglement entropy via holography~\cite{Ryu:2006ef} and shown to be decreasing along the RG flow when evaluated in the vacuum state, which corresponds to a {\it domain-wall Anti-de Sitter~(AdS) geometry~\cite{Freedman:1999gp}} in the gravity dual theory~\cite{Casini:2004bw,Casini:2006es,Nishioka:2006gr,Myers:2010tj,Myers:2012ed}.\footnote{It should be noted that there are several proposals for c-functions in the literature, which can differ in their properties. We are here referring to the c-functions defined in~\cite{Zamolodchikov:1986gt,Cardy:1988cwa,Osborn:1989,Komargodski:2011xv,Casini:2004bw,Casini:2006es,Nishioka:2006gr,Myers:2010tj,Myers:2012ed}.} 

As a main result of this present paper we find that such c-functions {\it increase} monotonically from the UV towards the IR energy scale when evaluated in a conformal field theory on a thermal state instead of a vacuum state. 
This demonstrates that with this definition the c-function takes into account information about the states, which is drastically changing its behavior. Such a c-function may be a useful measure of the total quantum-statistical information in a system, comparable to the relative information at a particular energy scale as defined in information theory~\cite{Casini:2016udt,Lashkari:2017rcl}. 

We find that these c-functions show interesting behavior which may be related to the quantum critical point present in the type of system we consider. While they behave thermally in the IR, the c-functions are found to differ considerably from the thermal entropy in the UV.  
The entanglement entropy appears to encode very different physics than the thermal entropy when compared in its UV. 
The chiral anomaly affects the c-functions more in the IR regime, and its effect is peaked at an intermediate value of the magnetic field at a fixed chemical potential and temperature.  
In the remainder of this introduction we motivate the particular thermal states we are going to work with.

Entanglement is a pure quantum effect which has no classical analog. The entanglement entropy can serve as a measure for such purely quantum effects, acting as a diagnostic for correlations between two subsystems. It can be used to detect changes to the quantum-statistical behavior of that system.\footnote{Note, for example, that the entanglement entropy of a {\it pure} quantum state vanishes because this state is neither entangled nor thermal. The entanglement entropy of a thermal state in a classical system, however, is nonzero although it is not  entangled.}  
For example, a re-organization of the degrees of freedom in a topological phase~\cite{Wen:2017} due to a topological transition may be indicated by a qualitative change in the entanglement entropy. 
As another example, consider an isolated quantum many body system far from equilibrium, such as the quark-gluon-plasma generated in heavy-ion-collisions, where the states are re-organized during the thermalization process. Such a re-organization of states may be described in the context of the eigenstate thermalization hypothesis~\cite{Srednicki:1994,Deutsch:1991}.  
In classical systems the chaotic dynamics leads to ergodicity which in turn leads to the system evolving towards an equilibrium state. Within interacting quantum systems it is conceptually less clear if and how a system approaches a state in which all observables take their equilibrium expectation values. 
Entanglement entropy is a quantum mechanical probe for the quantum states occupied at an instant of time at a particular energy scale by choosing entanglement regions of a particular width. These two examples motivate us to consider thermal states and to include topological effects, such as those related to conformal or chiral anomalies. 

Chiral anomalies arise as a topological property of the gauge manifold upon quantization, and thus constitute a pure quantum effect. Due to their topological origin, chiral anomalies are a robust quantum feature and for example do not receive any corrections along a renormalization group flow of any given chiral quantum field theory. In other words, chiral anomalies are one-loop exact. Hence, an interesting fundamental question is how chiral anomalies affect the entanglement and entanglement entropy of quantum systems; see~\cite{Wall:2011kb,Castro:2014tta,Azeyanagi:2015uoa,Guo:2015uqa,Nishioka:2015uka}. In addition, one may ask how these effects behave along a renormalization group (RG) flow. Due to the non-renormalization of chiral anomalies the naive expectation is that their impact is independent of the energy scale. 
For this purpose, we will see that the  c-function~\cite{Zamolodchikov:1986gt,Cardy:1988cwa,Osborn:1989,Komargodski:2011xv} can serve as as suitable measure. 
That is because, heuristically, it counts the degrees of freedom of a given system as it is deformed away from a UV fixed point towards an IR fixed point along the RG flow. 
Remarkably, and as an essential relation for the present work, the c-function can be defined by the entanglement of the degrees of freedom through the entanglement entropy~\cite{Casini:2004bw,Casini:2006es,Myers:2010tj,Myers:2012ed}. 

Following standard calculations of c-functions~\cite{Zamolodchikov:1986gt,Nishioka:2018khk}, the system is evaluated in a vacuum state. In contrast to that, a c-function computed from an entanglement entropy~\cite{Casini:2004bw,Casini:2006es} is typically evaluated in a generic mixed state, e.g.~a thermal state, because the entanglement entropy itself is defined through the reduced density matrix which in turn is defined through the possible states in which the system is likely to be found. 
Thus, the question arises if such c-functions evaluated in thermal states can be shown to have specific properties, just like the familiar c-functions. 
Specifically, if they count degrees of freedom and if they obey a c-theorem.

This question has not been discussed much in the literature we are aware of. 
Merely, different holographic definitions of c-functions in an $AdS$-Schwarzschild geometry have been considered before~\cite{Sahakian:1999bd,Paulos:2011zu,Banerjee:2015coc}. Results regarding the monotonicity of these different c-functions in a thermal state differ: \cite{Paulos:2011zu} finds them increasing towards the infrared regime while the c-functions of~\cite{Sahakian:1999bd,Banerjee:2015coc} decrease. 
In an asymptotically $AdS_4$ spacetime the deformation of a thermal state due to relevant operators was discussed in~\cite{Frenkel:2020ysx,Wang:2020nkd} with the purpose of understanding implications of the black brane interior, in particular the singularity. 
That system was studied earlier and finite-temperature deformations of RG-flows with exotic properties were discussed without mentioning c-functions or entanglement entropy~\cite{Gursoy:2018umf,Bea:2018whf}, see also~\cite{Golubtsova:2018dfh}. 
Neither one of all these works, is referring to the relation of the c-function to the entanglement entropy, which is our focus here.

Beginning with a review of entanglement entropy and c-functions in section~\ref{sec:review}, 
we next introduce the holographic model in section~\ref{sec:model}, namely Einstein-Maxwell-Chern-Simons theory evaluated on charged magnetic black brane solutions, which are dual to charged anisotropic states subject to a strong external magnetic field in $\mathcal{N}=4$ Super-Yang-Mills theory (SYM). 
In section~\ref{sec:SYM}, we holographically compute entanglement entropies on strips, as well as the c-functions defined from those entanglement entropies. 
We first demonstrate that these c-functions are monotonically decreasing (or constant) towards the IR when evaluated in the vacuum state. Then we show that they increase monotonically when evaluated in a generic thermal state, such as the charged anisotropic states mentioned above. 
We close with the Discussion, section~\ref{sec:discussion}. In the appendices we collect details on: the equations of motion (appendix~\ref{sec:eom}), boundary and horizon geometries (appendix~\ref{sec:geometries}), 
a useful coordinate transformation (appendix~\ref{sec:appendix_Coordinate}), possible alternative minimal surfaces (appendix~\ref{sec:appendix_ALT}),
a display of the monotonicity of c-functions in Schwarzschild $AdS$ (appendix~\ref{sec:mono_schwarzschild}), 
the null energy condition (appendix~\ref{sec:NEC}), 
numerical methods we used (appendix~\ref{sec:appendix_numerics}), 
and the numerical convergence of our methods (appendix~\ref{sec:appendix_convergence}).

\section{Review of entanglement entropy and c-functions}
\label{sec:review}
Setting the stage for our investigative play, in this section we first review selected aspects of the entanglement entropy from a field theory and from a holographic perspective. Then we remind ourselves how the c-function can be defined through the entanglement entropy. For a detailed review see~\cite{Nishioka:2018khk}.

\subsection{Entanglement entropy}
\label{sec:EE}
In quantum mechanical systems the von-Neumann entropy, or reduced entropy, or entanglement entropy, can be defined via the Shannon entropy,
\begin{equation}
    S_a=-\tr \rho_a \log \rho_a, \quad \rho_a=\tr_b \ket{\psi}\bra{\psi} \,  ,
\end{equation}
for a system which has been decomposed into two subsystems $a$ and $b$. A particularly nice interpretation of $S_a$ can be found in a two-qubit-system, a Bell pair, in the state 
\begin{equation}
    \ket{\psi}=\frac{1}{\sqrt{2}}\left(\ket{0}_a\ket{1}_b+\ket{1}_a\ket{0}_b\right) \, .
\end{equation}
The entropy of the reduced system $a$, $S_a$, is quickly found to be $S_a=\log{2}$ and 
can be thought of as the number of bits which are entangled between subsystem $a$ and subsystem $b$, realizing that $2=e^{S_a}$. 

In a QFT the entanglement entropy suffers from well known ultraviolet (UV) divergences~\cite{Casini:2004bw,Casini:2006es,Srednicki:1993im}, whose structure in $d$ spatial dimensions takes the following form\cite{Casini:2004bw},
\begin{equation}
    S(a)=g_{d-1}(\partial a)\epsilon^{-(d-1)}+g_{d-2}(\partial a)\epsilon^{-(d-2)}+\cdots + g_0(\partial a) \log(\epsilon )+S_0(a) \, , 
\end{equation}
with $\epsilon$ a short distance cutoff and the $g_i$ are local extensive functions of the boundary of the sub-region $a$. The leading divergence is proportional to the size of $\partial_a$ to the $(d-1)$ power and is referred to as the area law~\cite{Casini:2004bw,Srednicki:1993im,Bombelli:1986kls,tHooft:1984kcu}. It should be noted however that despite the naming convention it has been displayed that it is not always the case that this term is proportional to the area. 

In a CFT with a holographic dual the Ryu-Takayanagi conjecture provides a means to compute entanglement entropy~\cite{Ryu:2006ef}. This procedure has seen significant development in recent years evolving from the the RT procedure~\cite{Ryu:2006ef} to the HRT procedure~\cite{Hubeny:2007xt}. Subsequent refinements of this procedure can be found in the work of~\cite{Wall:2011kb,Wall:2012uf,Engelhardt:2014gca} 
and play a crucial role in bulk reconstruction~\cite{Dong:2016eik,Almheiri:2019psf}
and the information paradox~\cite{Almheiri:2012rt,Almheiri:2013hfa,Almheiri:2019qdq,Almheiri:2020cfm}. In this work it will suffice to consider the entanglement entropy $S_a$ for a subsystem $a$ of a CFT in $R^{d-1,1}$ as defined as,
\begin{equation} \label{eq:SEE5}
    S=\frac{1}{4G_N}\mathcal{A}(\gamma_{a} ),
\end{equation}
where $\gamma_a$ is a $(d-1)$-dimensional surface in $AdS_{d+1}$ whose boundary is $\partial a$~\cite{Ryu:2006ef}. In our description of c-functions in a four dimensional CFT we will find particularly useful the entanglement entropy for two parallel two-dimensional planes. The area functional $\mathcal{A}$ for the surface $\gamma_a$ takes the form,
\begin{equation}\label{eq:surface}
  \mathcal{A}=  \int{d^3\sigma \sqrt{\det{h}}} \, ,\quad h_{ij}=\frac{\partial\chi^{\mu}}{\partial\sigma^i}\frac{\partial\chi^{\nu}}{\partial\sigma^j}g_{\mu\nu} \, .
\end{equation}
where $h_{ij}$ is the metric induced on the surface $\gamma_a$ by the ambient spacetime, $g_{\mu\nu}$, and $\sigma^a$ are the coordinates on the surface while $\chi^\mu$ are the embedding coordinates. For an empty $AdS_5$ spacetime, dual to a vacuum state of the CFT$_4$ we can reduce eq.~\eqref{eq:SEE5} with~\eqref{eq:surface} to,
\begin{equation}
    S_{vac}= \frac{V}{4G_5} \int_{-\ell/2}^{\ell/2} \exd x \sqrt{\frac{1}{z^6} \left(1+z'(x)^2\right)} \, ,
\end{equation}
where $\ell$ is the width of the entanglement strip.
This is the entanglement entropy of $\mathcal{N}=4$ SYM theory evaluated in its vacuum state. 

\subsection{c-functions }
\label{sec:cFunctions}
Originally, a {\it c-theorem} obeyed by the aforementioned c-functions, was proposed and proven by Zalmolodchikov~\cite{Zamolodchikov:1986gt}. It states that in a two-dimensional conformal field theory a function $c_2$ can be defined which decreases monotonously under a RG transformation generated by operators driving the theory away from its UV fixed point. At the RG fixed points this function takes the value of the central charge of the respective Virasoro algebra of the UV- or IR-CFT. This has been interpreted as information loss along the RG-flow, as more and more UV degrees of freedom are integrated out. It was later proposed that also in four dimensions such a c-function can be defined and is related to the trace (conformal) anomaly of the field theory. More precisely, the trace anomaly coefficient $a_4$ was proposed as the four-dimensional analog of the two-dimensional c-function by Cardy~\cite{Cardy:1988cwa}. 
It has been verified that the trace anomaly coefficient $a_4$ is satisfying the conditions for a c-function, perturbatively by Osborn, Jack/Osborn~\cite{Jack:1990,Osborn:1989} and non-perturbatively by Komargodski/Schwimmer~\cite{Komargodski:2011vj}.

\begin{figure}[ht]
\centering
    \includegraphics[width=0.9\textwidth]{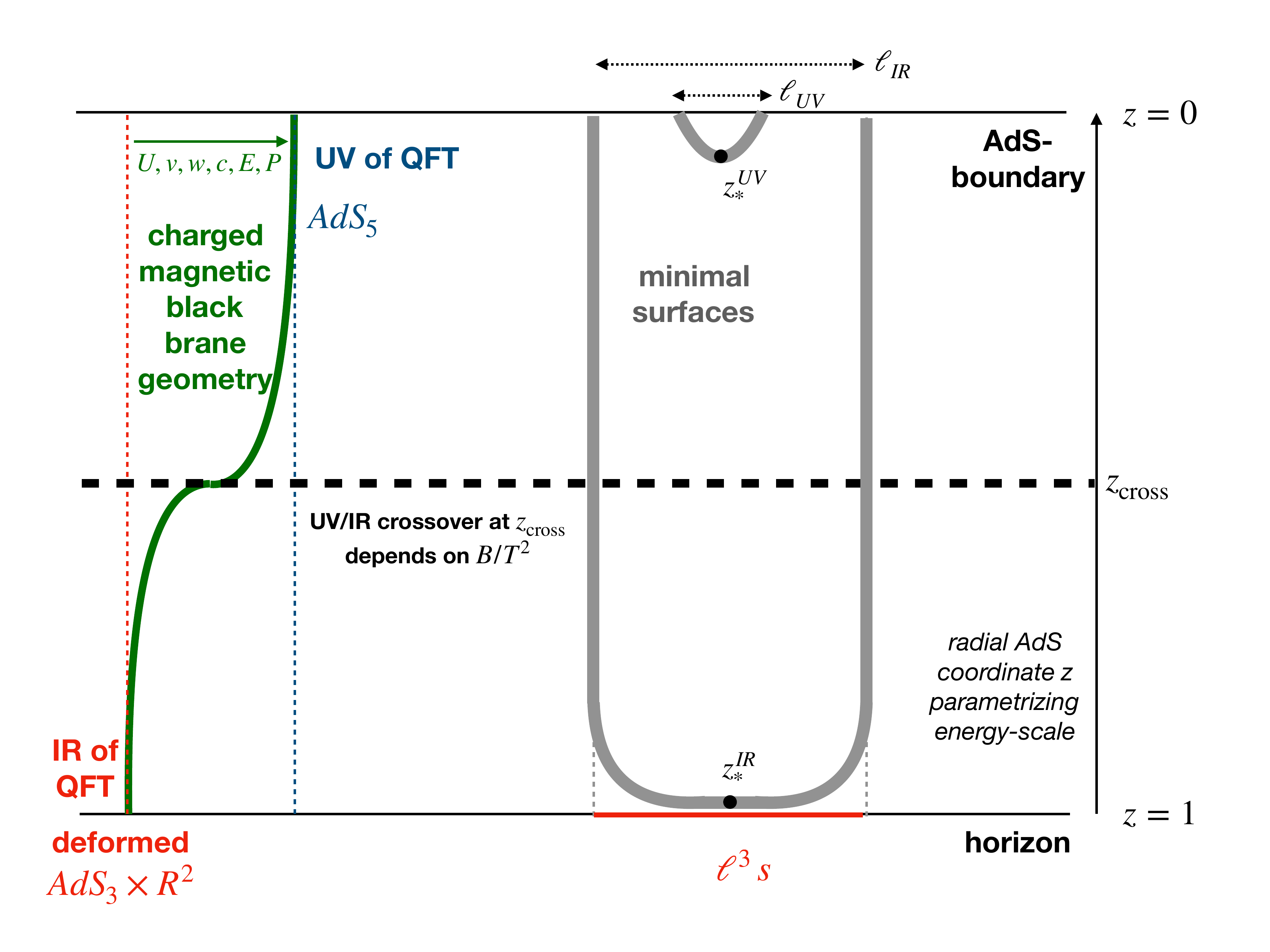}
    \caption{{\it Schematic plot of the energy scale of the boundary theory prepared in a thermal state as it is encoded in the bulk geometry.}  The $AdS$ radial coordinate $z$ is dual to the energy scale of the field theory. The magnetic field scale $B/T^2$ determines the charged magnetic black brane geometry (at fixed chemical potential $\mu/T$). Large $B/T^2$ pushes the transition from IR-like geometry (deformed $AdS_3\times R^2$ near the horizon) to UV-like geometry ($AdS_5$) towards the $AdS$-boundary. 
    Minimal surfaces, indicated by thick grey lines, are dual to the entanglement entropy of a strip of width $\ell$ in the boundary theory. Depending on the value of $\ell$, they probe either the UV or the IR or intermediate scales of the boundary theory evaluated in a particular state. The point on the minimal surface closest to the horizon is labeled $z_*$, and corresponds to the energy scale probed by the entanglement entropy. We stress that the geometry as a function of the $AdS$ radial coordinate $z$ encodes both, the RG-flow of the boundary field theory and simultaneously the energy-dependence of the state. 
    \label{fig:rgPicture}}
\end{figure}

\begin{figure}[ht]
\centering
    \includegraphics[width=0.9\textwidth]{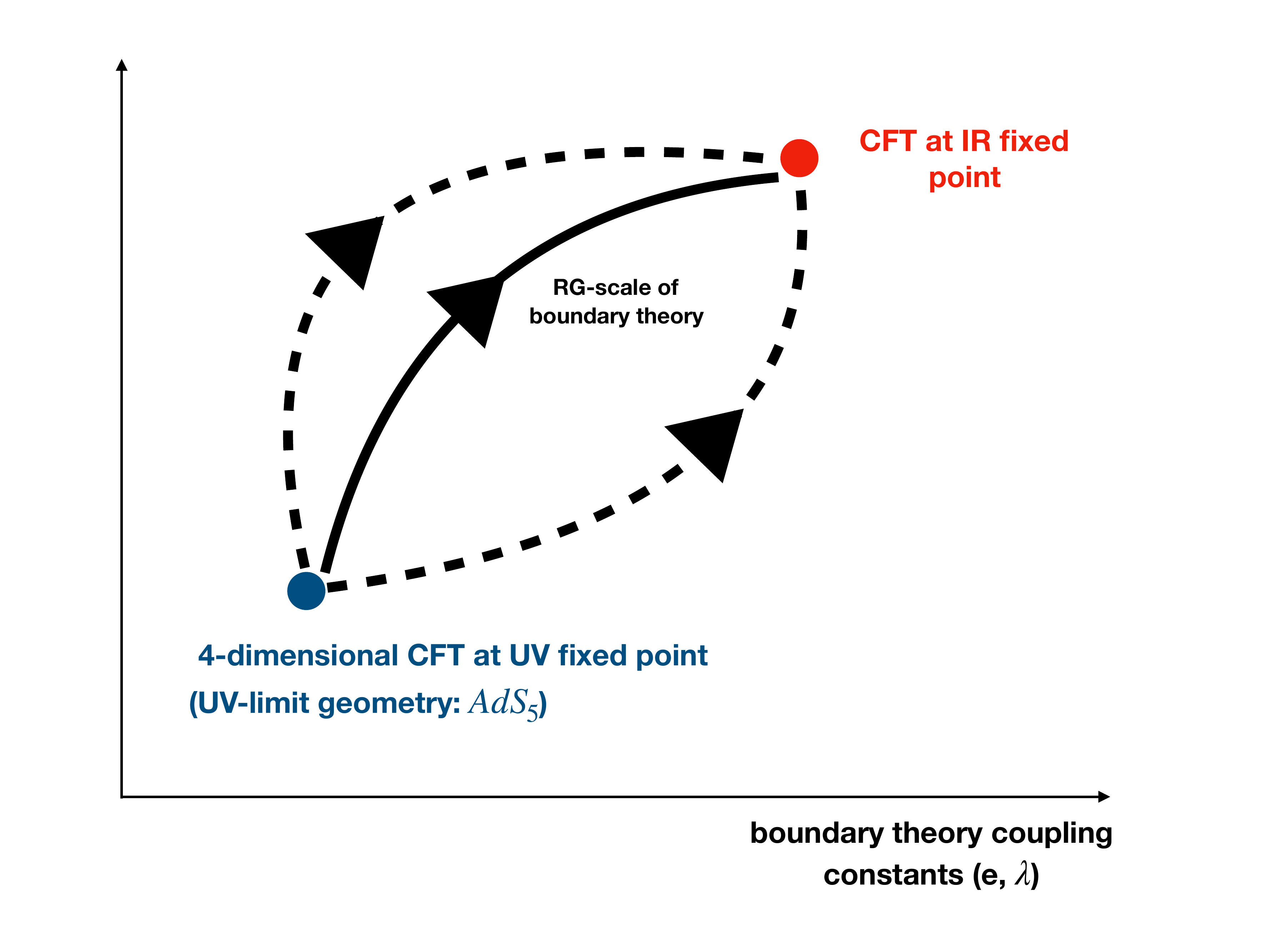}
    \caption{Schematic plot of the RG-flow in the space of field theories, parametrized by the gauge coupling constant $e$ and the 't Hooft coupling $\lambda$.  
    \label{fig:rgFlow}}
\end{figure}

As an example theory in (3+1) dimensions we here consider  Einstein-Maxwell-Chern-Simons theory dual to $\mathcal{N}=4$ SYM theory in presence of a global $U(1)$ R-charge gauge field strength, $F$, at large 't Hooft coupling $\lambda$ and with a large number of degrees of freedom $N\to \infty$. 
Formally, we may think of the boundary theory as having two coupling constants, the 't~Hooft coupling $\lambda$ and the gauge coupling $e$. 
These parametrize the RG-flow as indicated in figure~\ref{fig:rgFlow}, which is encoded in the bulk geometry, as figure~\ref{fig:rgPicture} outlines. These schematic pictures will be confirmed numerically and discussed in detail in section~\ref{sec:model}.  

In order to relate this RG-flow to a suitably defined c-function, recall that $\mathcal{N}=4$ SYM theory has a trace anomaly parametrized by two coefficients, $c_{TT}$ and the aforementioned $a_4$, 
\begin{equation}\label{eq:traceT}
    \langle T^b{}_b\rangle=\frac{c_{TT}}{16\pi^2}\mathcal{C}^2 - \frac{a_4}{16\pi^2} \mathcal{E} -\frac{1}{4} F^2\, ,
\end{equation}
with the Weyl tensor $\mathcal{C}$, the Euler density $\mathcal{E}$, and the last term, originating from the breaking of conformal symmetry through the external gauge field strength $F$.  
The two central charges coincide, $a_4=c_{TT} = N^2/4$, at large $N$. 
From the operator product expansion of the energy-momentum tensor we know that its most divergent term is related to the coefficient $c_{TT}$, i.e.~$TT\sim \frac{c_{TT}}{|z-z'|^4}$. This $c_{TT}$ is not known to obey any c-theorem, $a_4$ however, does obey a c-theorem named the {\it a-theorem}: $a_4^{UV}\ge a_4^{IR}$~\cite{Cardy:1988cwa,Komargodski:2011xv,Komargodski:2011vj}.  

While the proof provided by Komargodski and Schwimmer relied on correlator techniques, Casini, Test\`{e} and Torroba have recently provided an entropic proof of the a-theorem in $d=4$ via the intersection and unions of spherical entangling regions~\cite{Casini:2017vbe}. As discussed in their work it appears that entropic approaches to the proof of the c-theorems provide a unifying description of the known RG monotones as this technique can be used in any dimension. In particular the universal part of the entanglement entropy of a sphere in even dimensions is proportional to the Euler trace anomaly while in odd dimensions it is proportional to $F$, the constant term of the free energy of an Euclidean sphere.

We note already here that $\mathcal{N}=4$ SYM is a CFT and therefore its couplings $(\lambda,\, e)$ do not run with the RG-scale, which means that the IR fixed point (red dot) in figure~\ref{fig:rgFlow} is coincident with the UV fixed point (blue dot), no matter which metric solution (empty $AdS$, Schwarzschild $AdS$, charged magnetic $AdS$ black brane, ...) we consider. 
Its c-function $a_4$ takes the value of its central charge, $a_4=C_4=2\pi^2$, when evaluated in its vacuum state. If we would add relevant operators to $\mathcal{N}=4$ SYM theory, its couplings would start running and the c-function $a_4$ would decreases along the RG-flow towards the IR fixed point when evaluated in the vacuum state, as indicated in figure~\ref{fig:rgFlow}. In this work, however, we are not adding any operators to $\mathcal{N}=4$ SYM theory. Instead, we will evaluate the c-function in generic thermal states, for which we now review the relevant definition. 

A particular definition of the c-function is related to entanglement entropy. 
It was first noted in two dimensions that the two-dimensional c-function is given by~\cite{Casini:2004bw,Casini:2006es} 
\begin{equation} \label{eq:2DcFunction}
   c_{2} = 3 \ell \frac{\delta S_{EE}(\ell)}{\delta \ell} \, , 
\end{equation}
with the entanglement entropy (EE), $S_{EE}$, of an interval of length $\ell$ along the spatial direction in the boundary field theory.  
This c-function has been generalized to the $d$-dimensional c-function along the RG-flows of a given quantum field theory~\cite{Ryu:2006ef,Nishioka:2006gr,Myers:2012ed}\footnote{This definition of the $d$-dimensional c-function is free of any UV-regulator because the divergent term in $S_{EE}$ is independent from $\ell$.}
\begin{equation}
    c_d=\beta_d \frac{\ell^{d-1}}{H^{d-2}} \frac{\partial S_{EE}}{\partial \ell},\quad  \beta_d=\frac{1}{2^d\sqrt{\pi}\Gamma(d/2)}\left(\frac{\Gamma\left(\frac{1}{2(d-1)}\right)}{\Gamma\left(\frac{d}{2(d-1)}\right)}\right)^{d-1} \ ,\label{eq:c_function}
\end{equation} 
with a dimension-dependent constant $\beta_d$, the IR-regulator $H$, and for a field theory in $d$ spacetime dimensions, i.e. in $(3+1)$-dimensional $\mathcal{N}=4$ Super-Yang-Mills theory $d=4$. 
In~\cite{Myers:2012ed} it has been shown that this c-function decreases monotonically along RG-flows in holographic models with Poincar\'e invariant Einstein gravity duals when evaluated on the domain-wall solution, which will be explained in section~\ref{sec:SYM}. 
A necessary requirement is that the stress tensor of the matter fields must satisfy the null energy condition over the minimal surface used to compute $S_{EE}$. 

In a $d$-dimensional conformal field theory, with two parallel $(d-2)$-dimensional planes defining the subregion, the entanglement entropy has the form 
\begin{equation}\label{eq:defCFunc}
    S_{EE}^{CFT} = \alpha \frac{H^{d-2}}{\epsilon^{d-2}} - C_d \,\frac{H^{d-2}}{\beta_d\, (d-2)\, \ell^{d-2}} \, ,
\end{equation}
Here, again, $H$ is an IR regulator. The $(d-2)$ dimensional planes are constructed such that they are finite in extent in one of the $(d-2)$ directions, of width $\ell$, and infinite in the remaining $(d-3)$ directions. The IR regulator is then the formally infinite area of this $(d-3)$ dimensional surface in the field theory. 
The central charge of this CFT$_d$ is~\cite{Myers:2012ed},
\begin{equation}
    C_d=\frac{\pi^{d/2}}{\Gamma(d/2)}\left(\frac{L}{l_P}\right)^{d-1} \, .
\end{equation}
where for us the $AdS$ curvature scale $L$ is fixed, and in this work we set $L=1$ by scaling symmetries of the metric. In (3+1) dimensions this formula yields the value of the central charge, $a_4=2\pi^2$, as already stated above. 

\section{Holographic model}
\label{sec:model}
In this section we review the gravitational theory that is dual to $\mathcal{N}=4$ SYM theory at strong coupling, i.e. at large $N$ and large $\lambda$ in the presence of external electromagnetic fields associated with a $U(1)$ subgroup of the R-symmetry.  The gravity dual is Einstein-Maxwell-Chern-Simons theory, as considered e.g. in~\cite{DHoker:2009ixq,DHoker:2009mmn,DHoker:2010zpp,Ammon:2016szz,Ammon:2017ded,Ammon:2020rvg}. 
There are two coupling constants, the 't~Hooft coupling $\lambda$ and the gauge coupling $e$. 
These can parametrize an RG-flow as indicated in figure~\ref{fig:rgFlow}. That RG-flow is encoded in the bulk geometry, as figure~\ref{fig:rgPicture} outlines. These two schematic figures will be confirmed in this section. We further discuss the empty $AdS$, and charged magnetic black brane geometries, which correspond to a vacuum state, and charged thermal state in a magnetic field, respectively. 

\subsection{EMCS theory and its charged magnetic black brane solutions 
}
\label{sec:holoModel}
We consider asymptotically $AdS_5$ solutions within Einstein-Maxwell-Chern-Simons (EMCS) theory as first described in~\cite{DHoker:2009ixq},
\begin{equation}
    S=\frac{1}{16\pi G_5}\int{\sqrt{-g}\left(R+\frac{12}{L^2}-\frac{L^2}{4}F^{\mu\nu}F_{\mu\nu}\right)}-\frac{\gamma}{6}\int A\wedge F\wedge F \, ,
\end{equation}
where $G_5$ is the five-dimensional Newton constant, $\gamma$ is the Chern-Simons coupling and $L$ is the $AdS$ radius. In what follows we will set the $AdS$ radius to $L=1$. The equations of motion which follow from the variation of this action are,
\begin{align}
    R_{\mu\nu}+4g_{\mu\nu}&=\frac{1}{2}\left(F_{\mu\alpha}\tensor{F}{_{\nu}^{\alpha}}-\frac{1}{6}g_{\mu\nu}F_{\alpha\beta}F^{\alpha\beta}\right)\, , \label{eq:einstein}\\
    \nabla_{\mu}F^{\mu\nu}&=-\frac{\gamma}{8\sqrt{-g}}\epsilon^{\nu\alpha\beta\lambda\sigma}F_{\alpha\beta}F_{\lambda\sigma}\, .
\end{align}
To construct our background spacetime we use the metric ansatz~\cite{DHoker:2009ixq},
\begin{equation}
    \exd s^2=\frac{1}{z^2}\left(\frac{\exd z^2}{U(z)} -U(z)\exd t^2+v(z)^2\left(\exd x_1^2+\exd x_2^2\right)+w(z)^2\left(\exd x_3^2+c(z)\exd t\right)^2\right), \label{eq:Metric}
\end{equation}
with coordinates $(t,x_1,x_2,x_3,z)$ with $z=1/r$ and the $AdS$-boundary located at $z=0$. The gauge field ansatz is given by
\begin{equation}
    A=-E(z)\exd t+\frac{B}{2}(-x_2\exd x_1+x_1\exd x_2)+P(z)\exd x_3. \\
\end{equation}
The ansatz for the metric and gauge field represent an anisotropic black brane with both electric and magnetic charge. There is a residual $O(2)$ symmetry in the $x_1-x_2$ plane. 

We can solve the equations of motion order by near the $AdS$. The resulting solutions to $O(z^4)$ are given by\footnote{We have chosen to label the leading near-boundary coefficient of $c(z)$ with the symbol $\cfour$, in order to avoid confusion with the c-function which we label $c_4$. This coefficient $\cfour$ is associated with a heat current in the $x_3$-direction.}
\begin{subequations}\label{eq:near_bndy_expansion}
\begin{align}
    U(z)&=1+ u_4 z^4 +\frac{B^2}{6}\log(z)z^4+\cdots \, ,\\
    v(z)&=1+ v_4 z^4 -\frac{B^2}{24}\log(z)z^4+\cdots \, , \\
    w(z)&=1+ w_4 z^4 +\frac{B^2}{12}\log(z)z^4+\cdots \, , \\
    c(z)&=\cfour z^4+\cdots \, , \\
    E(z)&=\mu + e_2 z^2+\cdots \, , \\
    P(z)&=\frac{p_2}{2}z^2+\cdots \, ,
\end{align}
\end{subequations}
where we have explicitly chosen to set the source $p_0=0$. The coefficients $u_4,v_4,\cfour,e_2,p_2$ cannot be determined from a near-boundary analysis. These coefficients can only be determined from the full solution. These coefficients enter as the components of the dual energy-momentum tensor and the dual $U(1)$ current. A similar expansion can be computed near the horizon location and is given by
\begin{align}
    U(z)&=\bar{u}_1(z-z_h)+\cdots \, , \\
    v(z)&=\bar{v}_0+\bar{v}_1(z-z_h)+\cdots \, , \\
    w(z)&=\bar{w}_0+\bar{w}_1(z-z_h)+\cdots\, , \\
    c(z)&=\bar{c}_1(z-z_h)+\cdots \, , \\
    E(z)&=\bar{e}_1(z-z_h)+\cdots\cdots \, , \\
    P(z)&=\bar{p}_0+\bar{p}_1(z-z_h)+\cdots\cdots \, ,
\end{align}
where $z_h$ is the location of the horizon defined by $U(z_h)=0$. The system of differential equations generated from the Einstein equations constitutes a nonlinear coupled set of six differential equations and are included in the appendix~\ref{sec:eom} for completeness. Aside from the few known analytic solutions to these equations, detailed in section~\ref{sec:RG_Geo} and appendix~\ref{sec:geometries} we will resort to numerical methods to construct full solutions to this system of equations. More details on the numerical methods we employ are given in appendix~\ref{sec:appendix_numerics}. Once solutions have been obtained we make use of a standard holographic relation to construct the dual energy-momentum tensor~\cite{Fuini:2015hba}
\begin{equation}
    \braket{T_{ij}}=\frac{1}{\kappa}\left(g_{(4)ij}-g_{(0)ij}\tr g_{(4)}-(\log(\Lambda) +\mathcal{C})h_{(4)ij}\right) \, ,\label{eq:holo_Energy_Momentum}
\end{equation}
where $\mathcal{C}$ is an arbitrary scheme-dependent constant and $4\pi G_5=\kappa$ which in our choice of units is given by $\kappa=1/4$. 
The coefficients given in eq.\ (\ref{eq:holo_Energy_Momentum}) are constructed by first transforming to the Fefferman-Graham coordinate system and then expanding the metric near the $AdS$ boundary~\cite{deHaro:2000vlm},
\begin{equation}
    \exd s^2=\frac{\exd\rho^2}{4\rho^2}+\frac{1}{\rho}g_{ij}(x,\rho)\exd x^i\exd x^j,\quad g(x,\rho)=g_{(0)}+\rho^2 g_{(4)}+h_{(4)}\log(\rho)+\cdots \, .
\end{equation}
For the $U(1)$-current the following holographic relation can be used to extract the so-called consistent current, containing the Bardeen-Zumino term~\cite{DHoker:2009ixq,Ammon:2016szz}:
\begin{equation} 
  \braket{J^{\mu}}=\lim_{z\rightarrow 0}\frac{1}{z^3}\eta^{\mu\nu}\partial_zA_{\nu}+\frac{\gamma}{6}\epsilon^{\mu\nu\alpha\beta}A_{\nu}F_{\alpha\beta} \, . 
\end{equation}
Note that this current is chiral and in the presence of parallel external electric and magnetic fields has a non-vanishing divergence which is a manifestation of the chiral anomaly~\cite{Ammon:2016szz,Ammon:2020rvg}. 
Given our ansatz the energy-momentum tensor for our dual theory is given by,
\begin{align}
    \braket{T^{tt}}&=-3u_4 -\frac{B^2}{2}\log(\Lambda),\\
    \braket{T^{ii}}&=-\frac{B^2}{4}-u_4-4w_4+\frac{B^2}{2}\log(\Lambda), \\
    \braket{T^{x_3x_3}}&=8w_4-u_4 -\frac{B^2}{2}\log(\Lambda),\\
    \braket{T^{tx_3}}=\braket{T^{x_3t}}&=4 \cfour,
\end{align}
where $i=x_1,x_2$. The dependence on the renormalization scale $\Lambda$ is discussed in detail in~\cite{Fuini:2015hba}. In this work our choice is $\Lambda^2=B$.  
As expected, the external gauge field leads to an explicit violation of conformal invariance, contributing a term quadratic in the gauge field strength to the trace of the energy-momentum tensor
\begin{equation}
    \braket{\tensor{T}{^{a}_{a}}} = -\frac{1}{4}F_{ab}F^{ab} =-\frac{B^2}{2} \, .
\end{equation}
Two contributions from the trace anomaly, see eq.~\eqref{eq:traceT}, are not visible in the states which we examine here because $\mathcal{W}$ and $\mathcal{E}$ vanish when evaluated on the boundary geometry which is (flat) Minkowski spacetime, and which we chose by fixing the boundary conditions adequately when solving the Einstein equations~\eqref{eq:einstein}. 
The dual $U(1)$ current is given by
\begin{equation}
    \braket{J^{\mu}}=(\rho,0,0,p_2) \, .
\end{equation}
Furthermore, we note that the heat current $\braket{T^{t3}}$ and charge current $\braket{J^{3}}$ are the direct result of the chiral anomaly and are given by~\cite{Ammon:2016szz}
\begin{equation}
    \braket{T^{tx_3}}=\cfour=\frac{\gamma}{2}B\mu^2 \, , \quad\braket{J^{x_3}}=p_2=-B\mu\gamma \, .
\end{equation}

\subsection{
Renormalization scale, thermal states, and geometry}\label{sec:RG_Geo}
In this subsection we provide numerical evidence for the schematic picture shown in figure~\ref{fig:rgPicture}. 
In particular, we urge the reader to keep in mind that the bulk geometry encodes simultaneously the energy scale dependence of the state in which the boundary theory is prepared, as well as the RG-flow of the boundary theory. Since the boundary theory, $\mathcal{N}=4$ SYM theory, is conformal, its RG-flow is trivial: the UV fixed point coincides with the IR fixed point at all energy scales, cf.~figure~\ref{fig:rgFlow}. Thus we may think of the geometry as encoding solely the energy-dependence of the state. 

This analysis also serves to discuss an idea which has pervaded the literature, namely that the magnetic field of the holographic model acts as a ``renormalization scale of the boundary theory''~\cite{Fuini:2015hba,Avila:2018sqf}. 
Our analysis of the geometry here parallels in part the considerations presented for holographic RG-flows of isotropic states in~\cite{Albash:2011nq}. 
The charged magnetic black brane solutions can be characterized by both, their near-horizon geometry and their asymptotic geometry~\cite{DHoker:2009ixq,DHoker:2010zpp}. Here, we begin with the consideration of various limiting solutions to the equations of motion presented in appendix~\ref{sec:eom}, see also additional limiting solutions in appendix~\ref{sec:geometries}. 
Following the work of~\cite{Susskind:1998dq} the near-boundary and near-horizon regions encode the UV and the IR of the dual field theory, respectively, see the schematic plots in figures~\ref{fig:rgPicture} and \ref{fig:rgFlow}.

\paragraph{Empty $\mathbf{AdS_5}$ geometry (UV):} This is the simplest solution to the equations given in appendix~\ref{sec:eom}, and all other metrics we consider in this work asymptote to it near the conformal boundary, i.e.~in the UV regime of the dual field theory. In our choice of parameterization it is given by
\begin{equation}
    U=1,\quad v=w=1, \label{eq:empty}
\end{equation}
with all other components of the metric and gauge field vanishing. 
This empty $AdS$ geometry is dual to the vacuum state of $\mathcal{N}=4$ SYM theory and encodes its trivial energy scale dependence.  

\begin{figure}[htbp]
  \begin{center}
    \includegraphics[width=0.9\textwidth
    ]{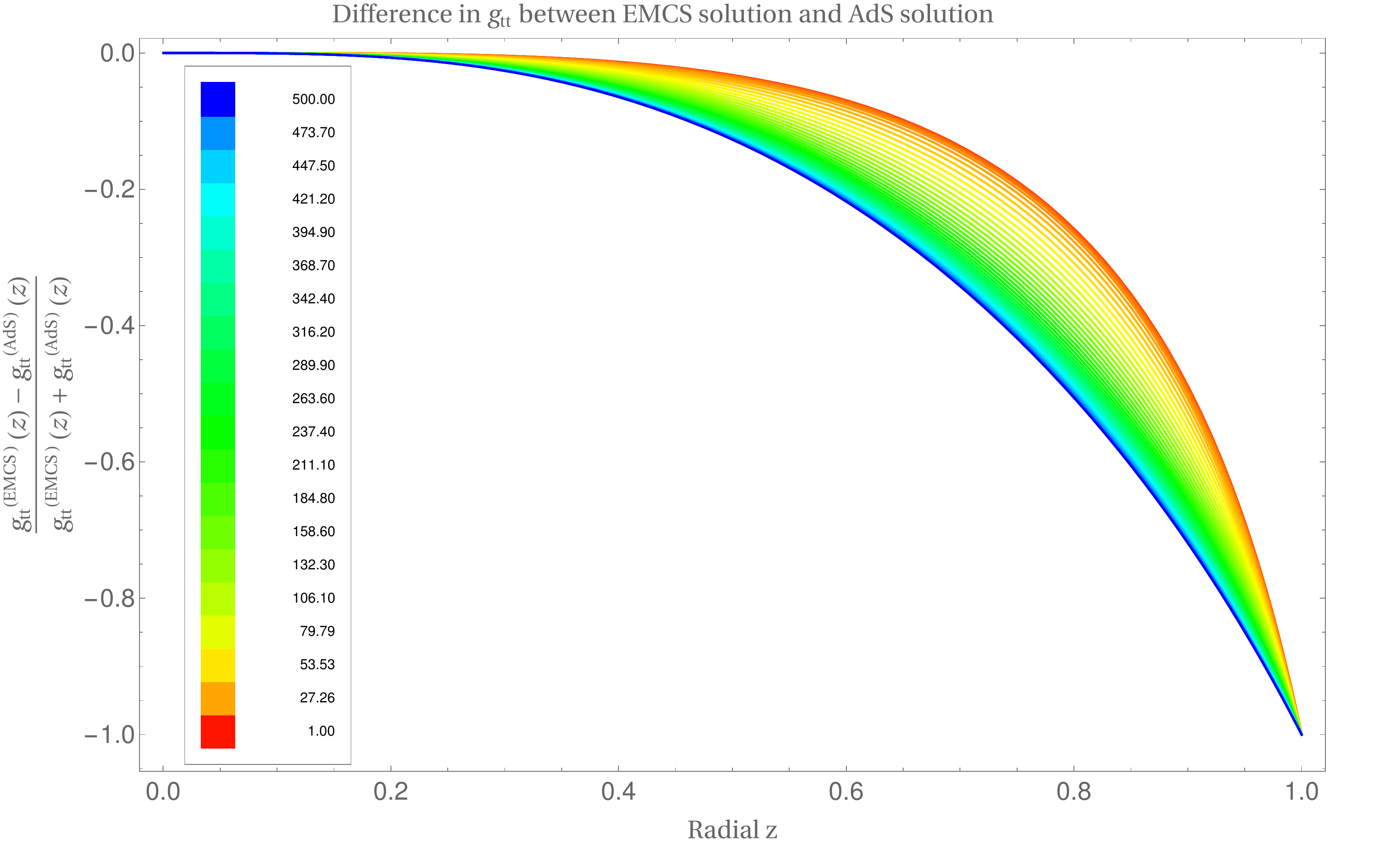}
    \end{center}
    \caption{We display the difference between the blackening factor $U(z)$ for full solutions to Einstein-Maxwell-Chern-Simons theory and the empty $AdS$ solutions. The inset legend displays the coloring for different values of $B/T^2$. The difference between these spacetimes can be seen to be very large deep in the bulk as we move towards $B/T^2>>1$. Note that the same plot but replacing $U_{AdS}$ with the Reissner-Nordstr\"om value $U_{RN}$ shows that $U_{EMCS}\approx U_{RN}$ at the horizon for small $B/T^2$. In other words, charged magnetic black branes have a similar blackening factor as charged black branes at the horizon. They become more distinct as $B/T^2$ increases. \label{fig:Comparison_EMCS_ADS}
    }
    \end{figure}

\begin{figure}[htbp]
    \begin{subfigure}[b]{0.5\textwidth}
    \includegraphics[width=75mm,scale=0.5]{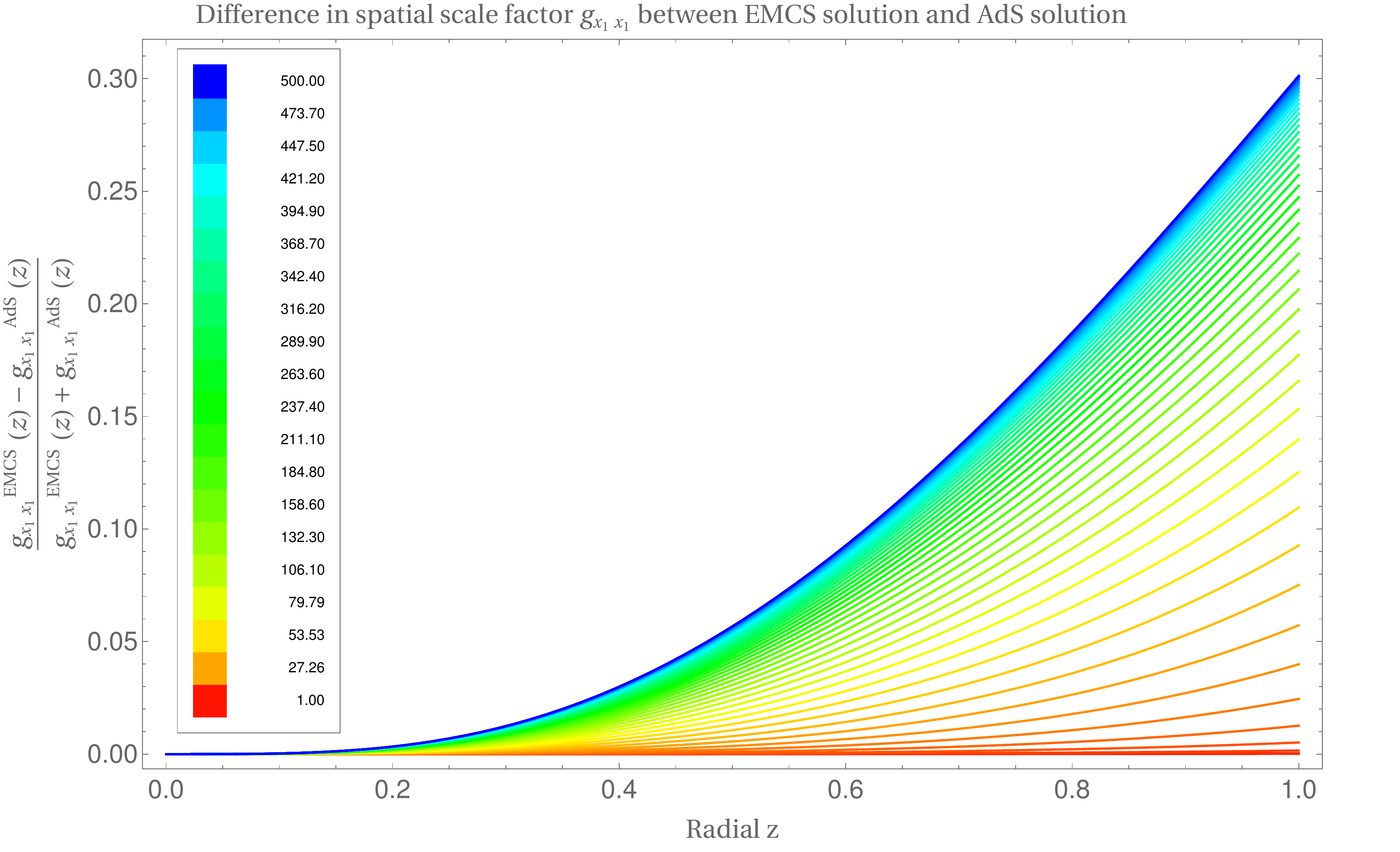}
 \end{subfigure}
 \begin{subfigure}[b]{0.5\textwidth}
    \includegraphics[width=75mm,scale=0.5]{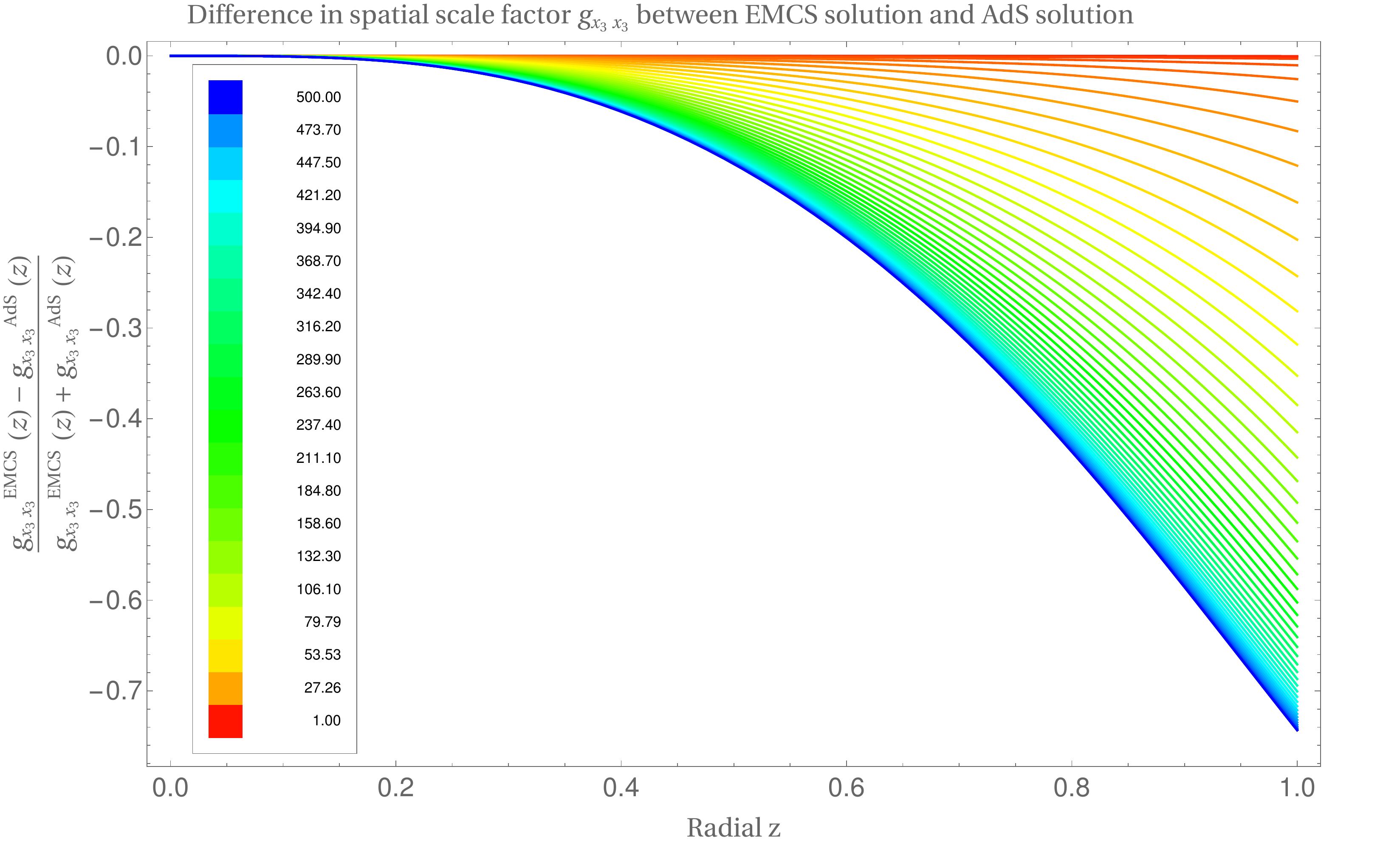}
    \end{subfigure}
    \caption{
    {\it Deviations of the charged magnetic black brane geometry from its near-boundary asymptotically $AdS$ limit.} 
    We display the relative difference between the spatial scale factors $g_{x_1x_1}$ and $g_{x_3 x_3}$ to full solutions to Einstein-Maxwell-Chern-Simons theory and the empty $AdS$ solutions. The inset legend displays the coloring for different values of $B/T^2$. The difference between these spacetimes can be seen to be very large deep in the bulk as we move towards $B/T^2\gg 1$. 
    \label{fig:comparisonVWtoAdS}}
    \end{figure}

We now compare this solution with the limiting behavior of various components of the charged magnetic black brane solution in EMCS theory. This geometry encodes an anisotropic thermal state within $\mathcal{N}=4$ SYM theory. 
We display in figures~\ref{fig:Comparison_EMCS_ADS} and \ref{fig:comparisonVWtoAdS} the difference between the charged magnetic black brane metric components within the EMCS theory and the empty $AdS$ metric given in eq.~(\ref{eq:empty}). The closer the curves are to zero, the closer the magnetic black brane geometry is to the UV-behavior, namely $AdS_5$. We observe that the charged magnetic black branes for small $B/T^2$ approximately retain their near $AdS$-boundary values deep into the bulk to $z\approx 0.5$ for $g_{tt}$ and even up to $z\approx 1$ for the spatial components $g_{x_1 x_1}$ and $g_{x_3 x_3}$. As $B/T^2$ increases the near $AdS$-behavior is pushed closer and closer to the $AdS$-boundary. 

Note that here both the asymptotic and near horizon geometry is the same for the spatial directions in empty $AdS$, Schwarzschild $AdS$, and Reissner-Nordstr\"om $AdS$. Hence the figures~\ref{fig:comparisonVWtoAdS} serve as comparisons for the spatial metric components (along $x_{1,2,3}$) in all three cases. 

In the comparison plots in figures~\ref{fig:Comparison_EMCS_ADS} and~\ref{fig:comparisonVWtoAdS} we have considered the same values of the dimensionless quantities $\mu/T=1/5$ and $B/T^2\in [1,500]$. We continue to use these values for the remaining plots in this section. 

\begin{figure}[H]
    \begin{subfigure}[b]{0.5\textwidth}
    \includegraphics[width=75mm,scale=0.5]{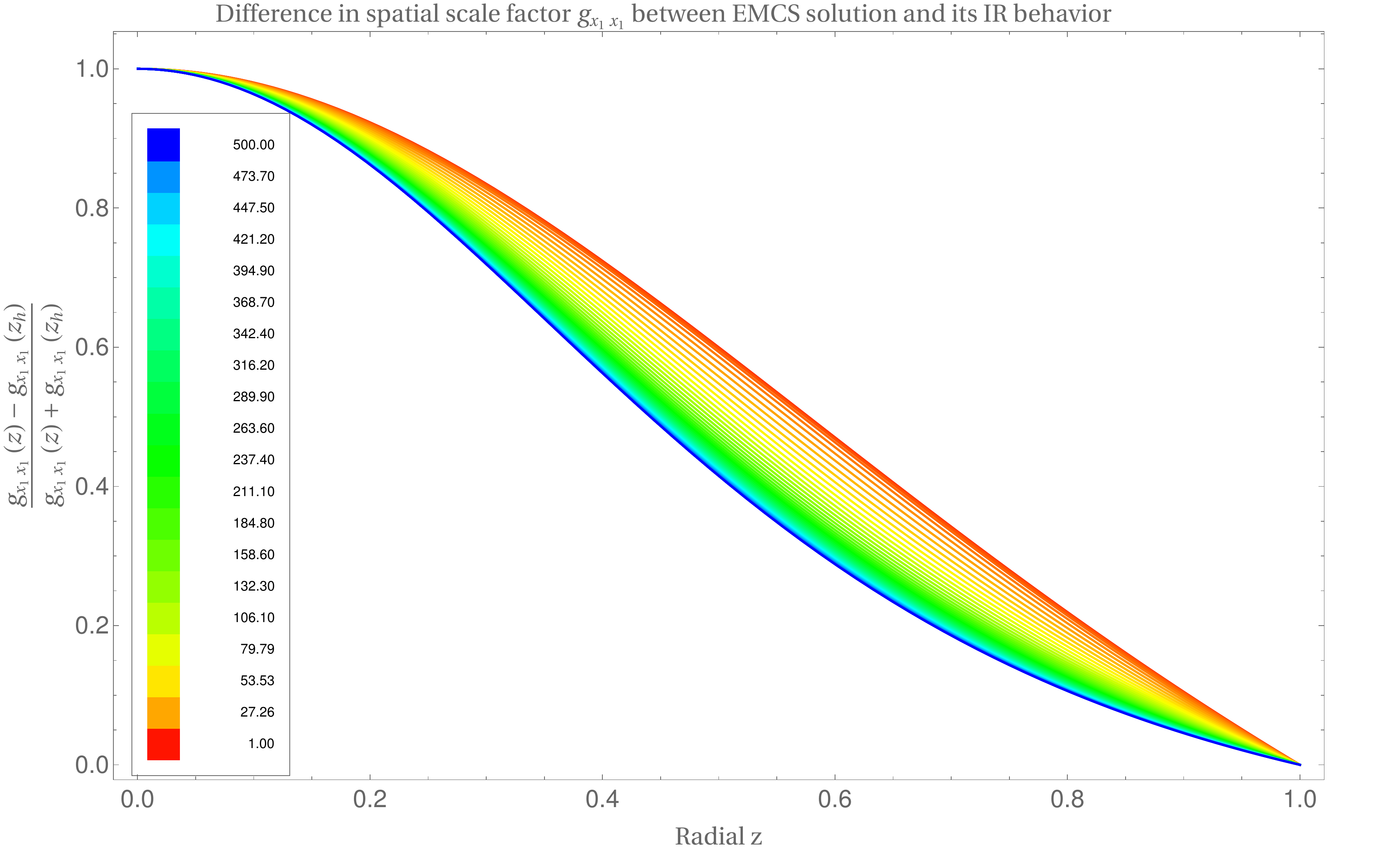}
 \end{subfigure}
 \begin{subfigure}[b]{0.5\textwidth}
    \includegraphics[width=75mm,scale=0.5]{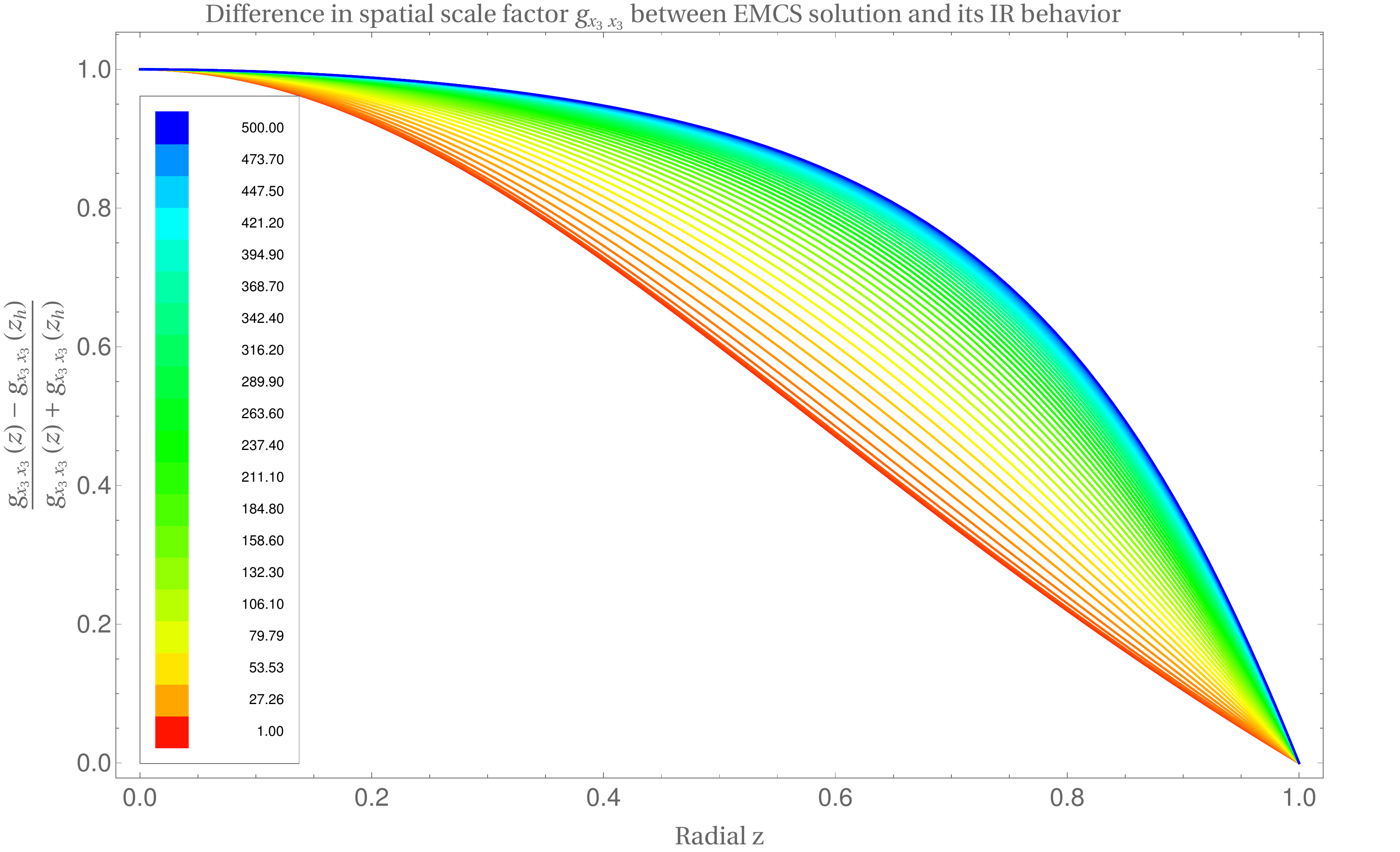}
    \end{subfigure}
    \caption{
     {\it Deviations of the charged magnetic black brane geometry from its near-horizon behavior.} 
    We display the relative difference between the spatial scale factors $g_{x_1x_1}$ and $g_{x_3 x_3}$ to full solutions to Einstein-Maxwell-Chern-Simons theory and their near horizon behavior. The inset legend displays the coloring for different values of $B/T^2$. 
    Note that the near horizon behavior in the transverse direction $x_1$ is extended towards the boundary by increasing $B/T^2$, while the near horizon behavior for the parallel direction $x_3$ is pushed towards the horizon. 
    \label{fig:comparisonIR}}
    \end{figure}

\paragraph{Deformed  $\mathbf{AdS_3}\times R^2$ geometry (IR)}
As pointed out in~\cite{DHoker:2009mmn,DHoker:2009ixq,DHoker:2010zpp} the near horizon IR geometry of uncharged magnetic black branes is $AdS_3\times R^2$, and for charged magnetic black branes it turns out to be a deformed version thereof. 
As a measure of closeness to their own IR-behavior for charged magnetic black branes we subtract the values of the metric components evaluated at the horizon from the metric components as function of $z$ as presented in figure~\ref{fig:comparisonIR}. 
Confirming our picture from figure~\ref{fig:rgPicture}, the left plot in figure~\ref{fig:comparisonIR} shows that the IR behavior of $g_{x_1 x_1} = g_{x_2 x_2}$, i.e.~the near-horizon deformed $AdS_3\times R^2$, is extended towards the boundary by increasing $B/T^2$. This agrees with the UV behavior being pushed closer to the $AdS$-boundary as discussed above for figures~\ref{fig:Comparison_EMCS_ADS} and~\ref{fig:comparisonVWtoAdS}. Taking these two results together, one can state that the crossover from IR to UV geometry is pushed closer to the boundary for increasing $B/T^2$. 

However, the metric along the magnetic field behaves differently. In fact, it does the opposite of all other metric functions: the IR behavior of $g_{x_3 x_3}$ is pushed closer and closer to the horizon as $B/T^2$ increases. This can be interpreted as the longitudinal direction becoming more UV-like while the transverse directions become more IR-like at increasing $B/T^2$. This should leave an imprint on the thermal entropy and the entanglement entropy since these are computed with the spatial metric components.
%

\paragraph{Summary: }
In summary, our numerical data displayed on the left side of figure~\ref{fig:embedding} gives evidence for  figure~\ref{fig:rgPicture}. The energy scale of the boundary {\it system} is encoded in the bulk geometry. The magnetic field scale $B/T^2$ (at fixed $\mu/T$) determines the charged magnetic black brane geometry indicated by the green line and by the labels $u,\,v,\,w,\, c,\, E, P$ in figure~\ref{fig:rgPicture}. One may think of the magnetic field as setting the relevant energy scale of the theory and the state simultaneously. Large $B/T^2$ pushes the transition from IR-like geometry (deformed $AdS_3\times R^2$) to UV-like geometry ($AdS_5$) towards the $AdS$-boundary in most of the metric components (with the exception of the longitudinal metric component, as discussed above). 

\subsection{Thermal entropy density} 
\label{sec:thermalEntropy}
\begin{figure}[H]
    \centering
    \includegraphics[width=12cm]{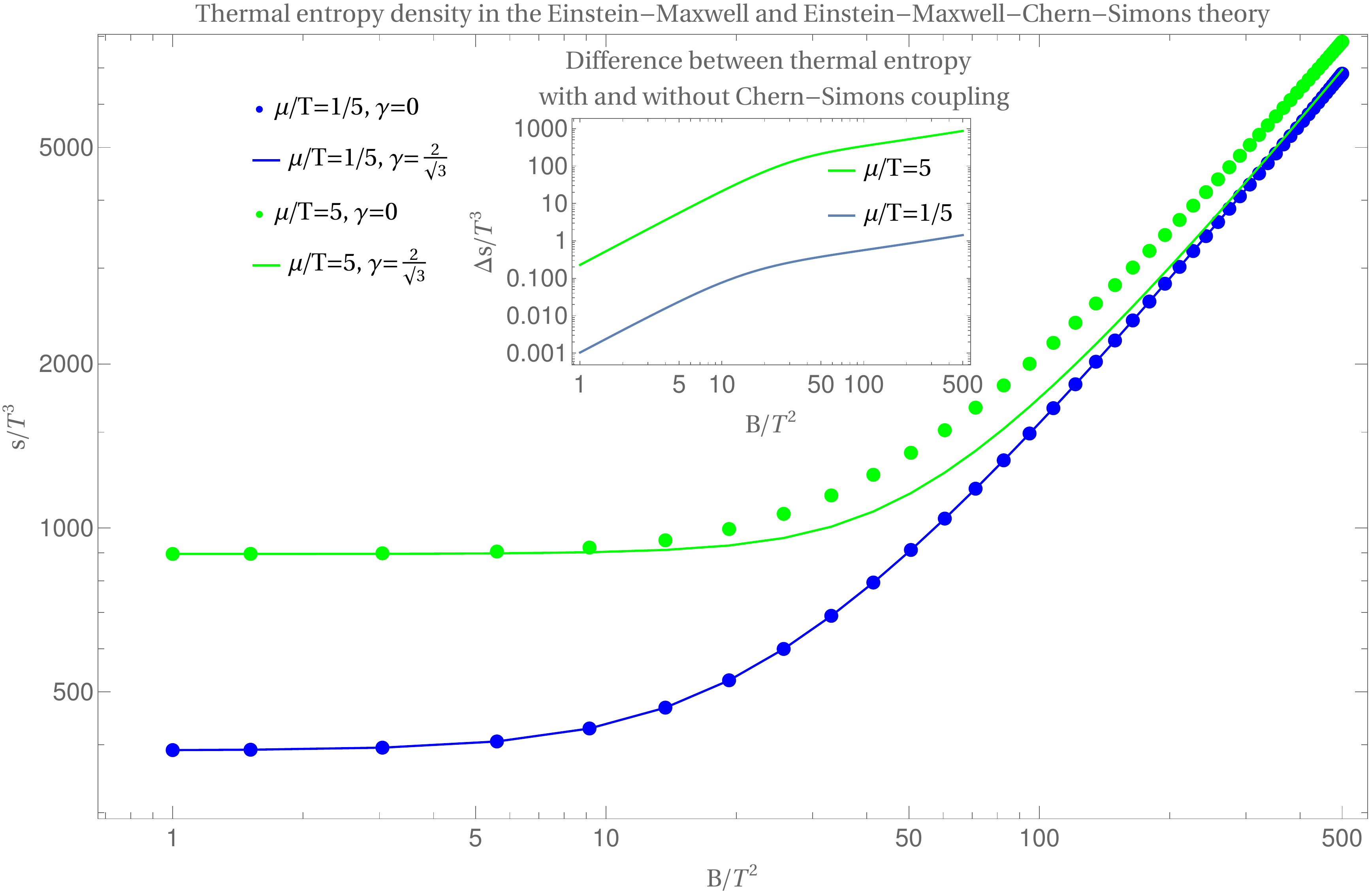}
    \caption{We display the thermal entropy density including the chiral anomaly (solid lines, $\gamma=\gamma_{SUSY}=2/\sqrt{3}$) as calculated from the black brane solutions within Einstein-Maxwell-Chern-Simons theory. This is compared to the case excluding the chiral anomaly (dotted lines, $\gamma=0$) from black brane solutions in Einstein-Maxwell theory. A significant difference between the cases with and without anomaly is visible at the larger of the two displayed chemical potential values, $\mu/T=5$ (green curves).  
    In the large-$B/T^2$ regime, $s/T^3$ scales approximately linearly with $B/T^2$.  
    The inset  shows the difference, $\Delta s$, between the thermal entropy with ($\gamma=\gamma_{SUSY}$) and without ($\gamma=0$) the chiral anomaly taken into account, i.e.~it visualizes the contribution of the chiral anomaly to the thermal entropy. For later comparison, we keep in mind that this quantity is {\it not} peaked at any value of $B/T^2$. 
    \label{fig:Spacetime_Comp_entropy}}
\end{figure}

For comparison to the entanglement entropy, we here compute the thermal entropy density $s/T^3$. It is affected significantly by the chiral anomaly, as indicated in figure~\ref{fig:Spacetime_Comp_entropy}. 
At vanishing chiral anomaly, $\gamma=0$, different charge densities, $\mu/T=1/5$ or $5$, lead to distinct behavior of the entropy density at large $B/T^2$, compare the blue dotted curve to the green dotted one. However, at $\gamma=2/\sqrt{3}$, the anomaly forces the thermal entropy to asymptote to the same curve even for distinct values of $\mu/T$. In other words, the effect of the charge density on the thermal entropy density at large $B/T^2$ is suppressed by the chiral anomaly. This indicates that a universal behavior at large $B/T^2$ is enhanced by the anomaly. Potentially this can be related to the universal behavior discovered for this system at large magnetic field values~\cite{Grozdanov:2017kyl}. As a side note, the effect of the charge density on $s/T^3$ is large at small $B/T^2$. 
A geometric picture of the thermal entropy becomes obvious when eyeballing the near-horizon region in figure~\ref{fig:rgPicture}. The horizon area located at $z=1$ is of infinite size. However, as usual, that horizon area divided by the field theory volume gives the finite entropy density, $s$. The horizon area marked red in figure~\ref{fig:rgFlow} is of size $\ell^3 s$ and it can be thought of as the approximate thermal contribution to the minimal surface dipping down to $z_*^{IR}$. 
We see that in the IR limit, $\ell\to\infty$, the minimal surface will hug the entire horizon area and only at spatial infinity it cuts straight to the boundary, such that the minimal surface becomes identical to the horizon area. 
This is the geometric explanation for why the entanglement entropy is proportional to the thermal entropy density in the IR limit. The c-function, defined as an $\ell$-derivative of the entanglement entropy, will inherit this behavior. 

\section{Results for $\mathcal{N}=4$ SYM in generic states}
\label{sec:SYM}
In this section we present our results for the entanglement entropy and for the c-function of $\mathcal{N}=4$ SYM theory in generic thermal states. 

\subsection{Entanglement entropy}
\label{sec:EEresults}
\begin{figure}[htbp]
    \centering
    \includegraphics[width=14cm]{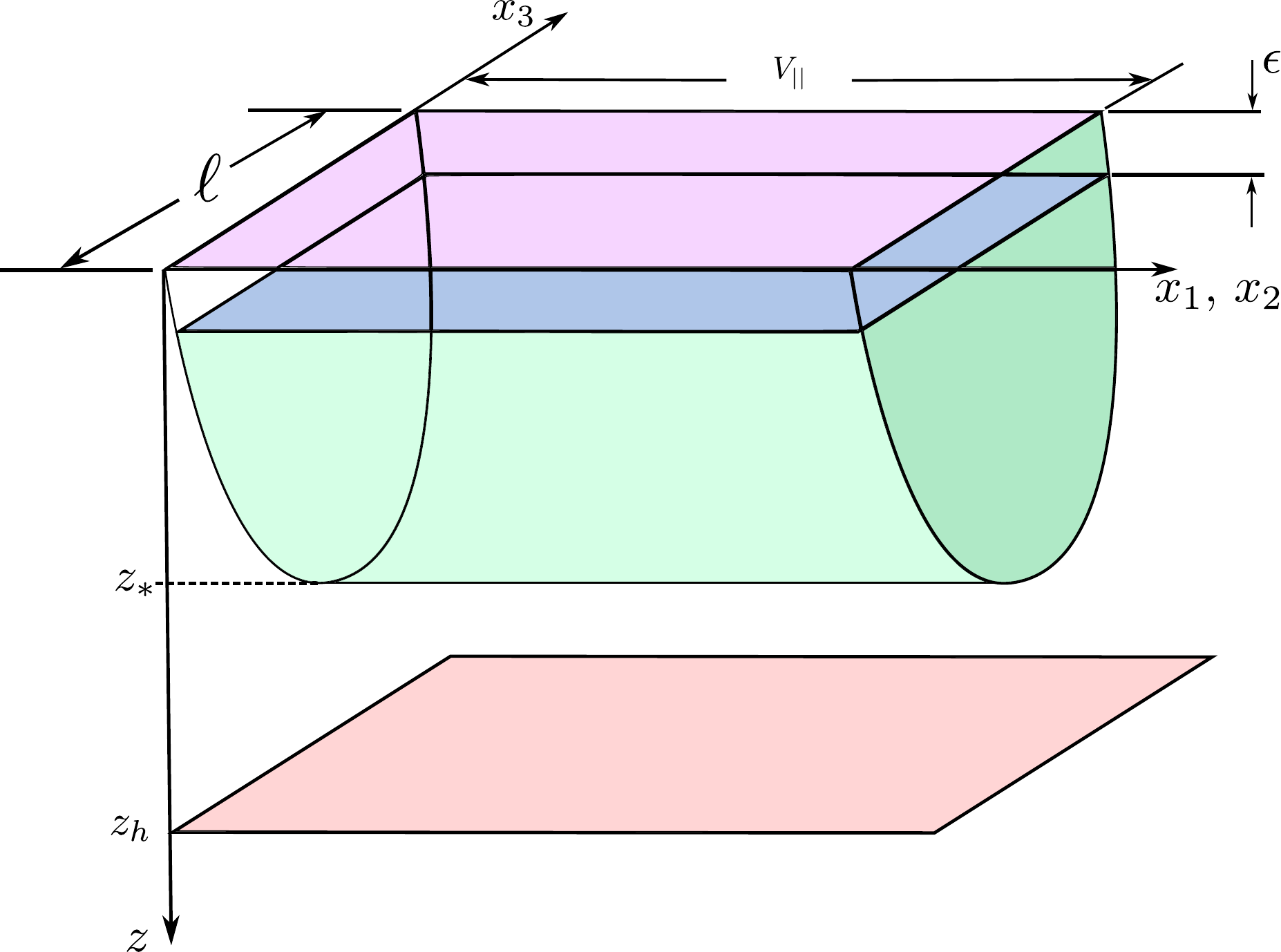}
    \caption{A schematic representation of the entangling surfaces we consider in this work. The graphic is drawn as a representative of the case which we label ``longitudinal'' or ``parallel''. 
    \label{fig:surface_doodle}}
\end{figure}

We calculate the entanglement entropy of a surface which has the topology of a strip, as depicted in figure~\ref{fig:surface_doodle}. In the dual theory this corresponds to the area of a codimension-2 hypersurface in $AdS_5$. We compute the area by~\eqref{eq:surface}, 
where the ambient metric, $g_{\mu\nu}$, is a solution to the EMCS equations of motion. Treating the area integral as a functional of the embedding coordinates we extremize the action resulting in a set of equations to determine the embedding coordinates. As a result of the anisotropy of the spacetime we have two choices of alignment of a strip. We can choose to align a strip along the anisotropy, along the $x_3$ direction in our case, or perpendicular to it, in the $x_1,x_2$ plane for us. 
We label these two cases longitudinal and transverse, respectively. Table~\ref{tab:coordinates} displays both $\chi$ and $\sigma$ for these two choices of orientation. 
\begin{table}[H]
    \centering
    \begin{tabular}{c|c|c}
       & Transverse & Longitudinal  \\
       \hline 
     Embedding Coordinates    &  $\chi^{\mu}=(z(\sigma), t(\sigma), x_1(\sigma), x_2, x_3)$ & $\chi^{\mu}=(z(\sigma), t(\sigma), x_1, x_2, x_3(\sigma))$ \\
     \hline
     Surface Coordinates & $\sigma^i=(\sigma, x_2,x_3)$    &   $\sigma^i=(\sigma, x_1,x_2)$
    \end{tabular}
    \caption{The coordinates we choose for the embedding of a surface anchored at the $AdS$ conformal boundary.
    \label{tab:coordinates}}
\end{table}
The entanglement entropy dual to this surface is given by~\cite{Ryu:2006ef},
\begin{equation}
    S=\frac{1}{4G_5}\mathcal{A},
\end{equation}
evaluated on the solution to the embedding equations. There are multiple methods of approaching the explicit computation of the area associated with minimal surfaces in $AdS$. In this work, we will focus an affine parameterization and a coordinate parameterization\footnote{See also ~\cite{Jankowski:2020bmw} where the author describes the Hamilton-Jacobi formalism for the bulk action dual to the entanglement entropy of a strip in the dual field theory. }. The affine parameterization more directly obtains the explicit solutions for the minimal surfaces while the coordinate parameterization is more useful in the derivation and interpretation of the c-functions. More details on the coordinate parameterization can be found in appendix~\ref{sec:appendix_Coordinate}. 

\begin{figure}[htb]
    \begin{subfigure}[b]{0.5\textwidth}
    \includegraphics[width=75mm,scale=0.5]{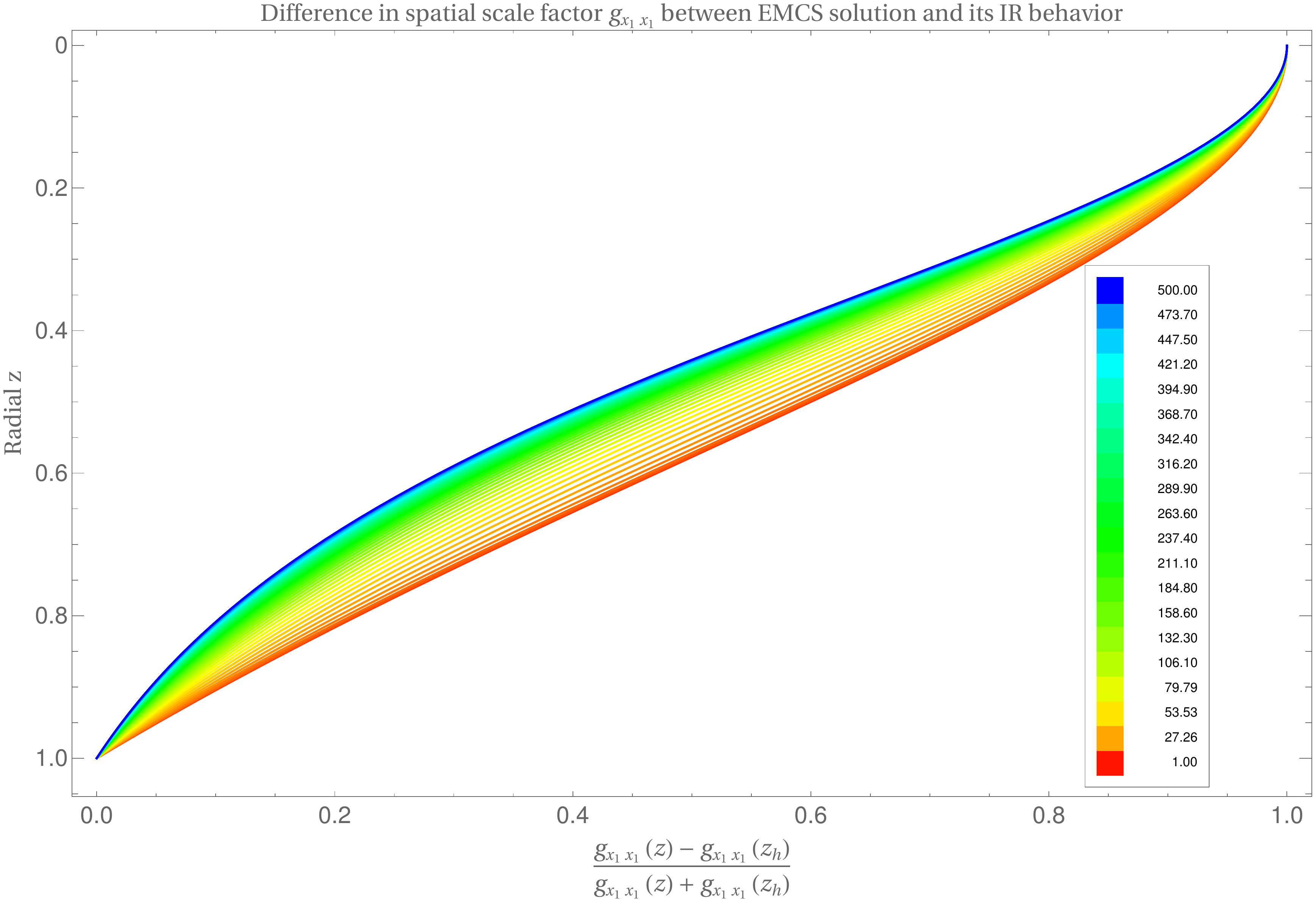}
 \end{subfigure}
 \begin{subfigure}[b]{0.5\textwidth}
    \includegraphics[width=75mm,scale=0.5]{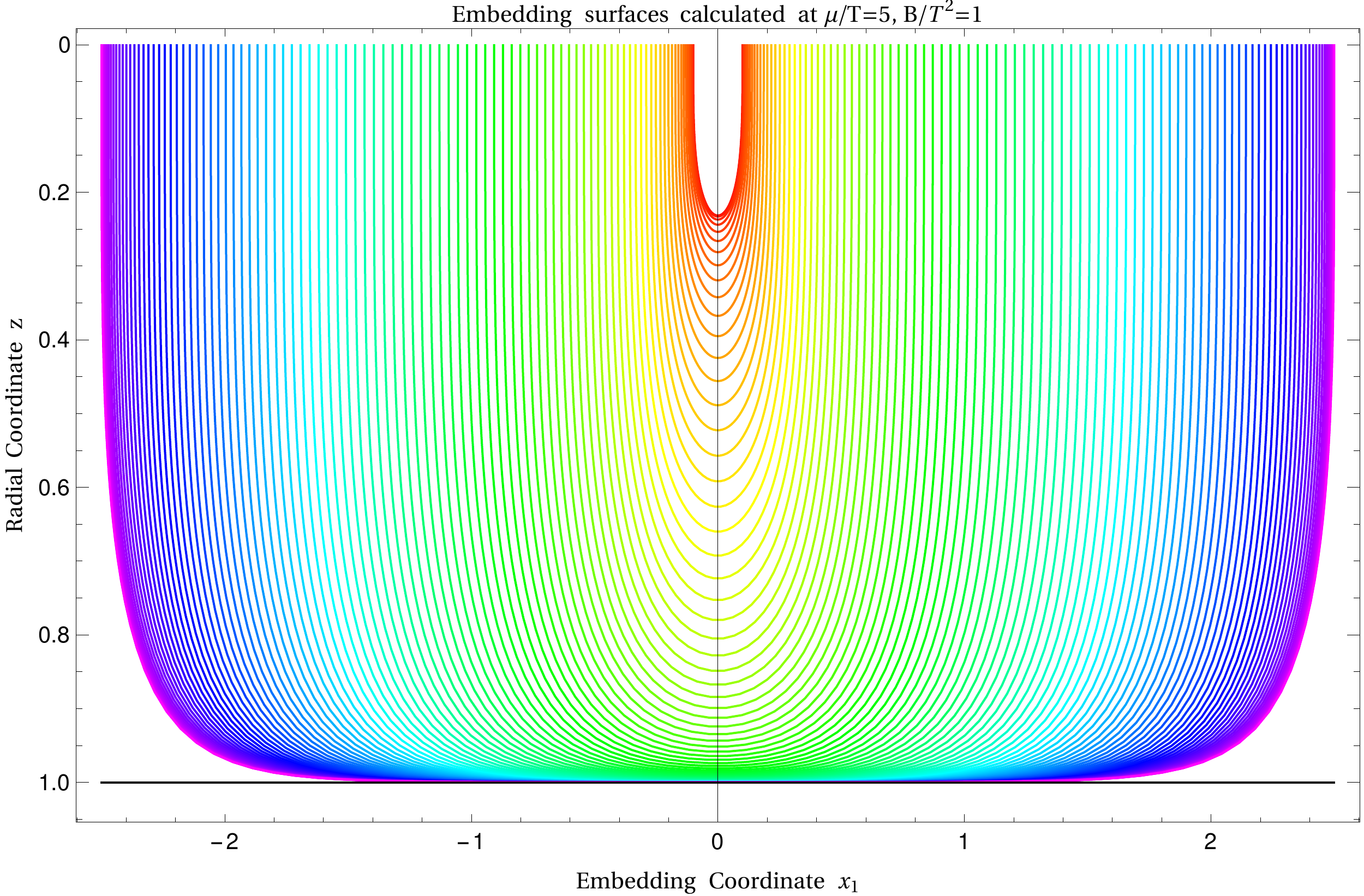}
    \end{subfigure}
    \caption{Numerical replication of figure~\ref{fig:rgPicture}. \textit{Left:} The deviation of the EMCS charged magnetic black brane metric component $g_{x_1x_1}$ from its near horizon behavior, with magnetic field from $\color{red}B/T^2=1.00$ shown in red to $\color{blue}500.00$ in blue at $\mu/T=5$ and $\gamma=\gamma_{SUSY}$. \textit{Right:} Solutions for the embedding coordinates $(x_1(\sigma),z(\sigma))$. These curves have been computed for $\mu/T=5$, $\gamma=\gamma_{SUSY}$, and $B/T^2=1$ for lengths $\color{red}{\ell=2/10}$ in red to $\color{Plum}{\ell=5}$ in purple. 
    \label{fig:embedding}}
\end{figure}

\paragraph{Affine parameterization:}
Starting with the parallel case, we choose to take the embedding $\chi^{\mu}=(z(\sigma), t(\sigma), x_1, x_2, x_3(\sigma))$, with coordinates $(\sigma,x_1,x_2)$, inserting this along with the metric~(\ref{eq:Metric}) into the action~(\ref{eq:surface}) we find 
\begin{equation}
  \int \exd\sigma\exd x_1\exd x_2\sqrt{\frac{v(z(\sigma ))^4 \left(w(z(\sigma ))^2 \left(c(z(\sigma )) t'(\sigma )+x_3'(\sigma )\right)^2-t'(\sigma )^2 U(z(\sigma ))+\frac{z'(\sigma )^2}{U(z(\sigma ))}\right)}{z(\sigma )^6}} \, . \label{eq:affine_parallel}
\end{equation}
Since our interest is focused on entanglement entropy we will work with fixed time surfaces. Setting $t(\sigma)=t_0$ and computing the integrals over $x_1$ and $x_2$ gives the action 
for the longitudinal direction 
\begin{equation}
S_{||}=\frac{1}{4G_5} V_{||}\int \exd\sigma\sqrt{\frac{v(z(\sigma ))^4 \left(w(z(\sigma ))^2x_3'(\sigma )^2+\frac{z'(\sigma )^2}{U(z(\sigma ))}\right)}{z(\sigma )^6}} \, .
\end{equation}
We repeat this exercise for the transverse case, taking  $\chi^{\mu}=(z(\sigma), t(\sigma), x_1(\sigma), x_2, x_3)$, with coordinates $(\sigma,x_2,x_3)$, yielding the action for the transverse direction 
\begin{equation}
S_{\perp}=\frac{1}{4G_5} V_{\perp}\int \exd\sigma\sqrt{\frac{v(z(\sigma ))^2 w(z(\sigma ))^2\left(\frac{z'(\sigma )^2}{U(z(\sigma ))}+v(z(\sigma ))^2 x_1'(\sigma ){}^2\right)}{z(\sigma )^6}} \, . \label{eq:affine_transverse}
\end{equation}
In both of these expressions we have chosen to write the Lagrangian in terms of an affine parameter $\sigma$. The quantities $V_{||},V_{\perp}$ are given by,
\begin{equation}
    V_{\perp,||}=\int_{-a}^{a}\int_{-b}^{b} \exd x_1 \exd x_{2,3}\, ,
\end{equation}
and can be considered as IR regulators~\cite{Myers:2012ed} with $a,b\gg \ell$. In terms of this parameter the actions reduce to the calculation of a geodesic in an auxiliary spacetime~\cite{Ecker:2015kna}.  
A set of examples of solutions to the geodesic equations is shown in the right plot in figure~\ref{fig:embedding}.

The depth, $\zs$, down to which the surface probes the bulk is not independent of the width of the strip $\ell$, as discussed in appendix~\ref{sec:appendix_Coordinate} and displayed in figure~\ref{fig:conjugate}.

\paragraph{Vacuum contribution:}
Both of the integrals to compute the area of the surface are divergent quantities, they can be regularized by subtracting the divergent part which comes from the vacuum contribution,
\begin{equation}
    S_{vac}=\frac{1}{2G_5} V \int \exd z \sqrt{\frac{1}{z^6} \left(1+x'(z)^2\right)} \, ,
\end{equation}
where $V\in\{V_{||},V_\perp\}$ and $x\in \{x_{||},x_\perp\}$ to denote either possibility of parallel or transverse strips.
\begin{figure}[htb]
    \begin{subfigure}[b]{0.5\textwidth}
    \includegraphics[width=75mm,scale=0.5]{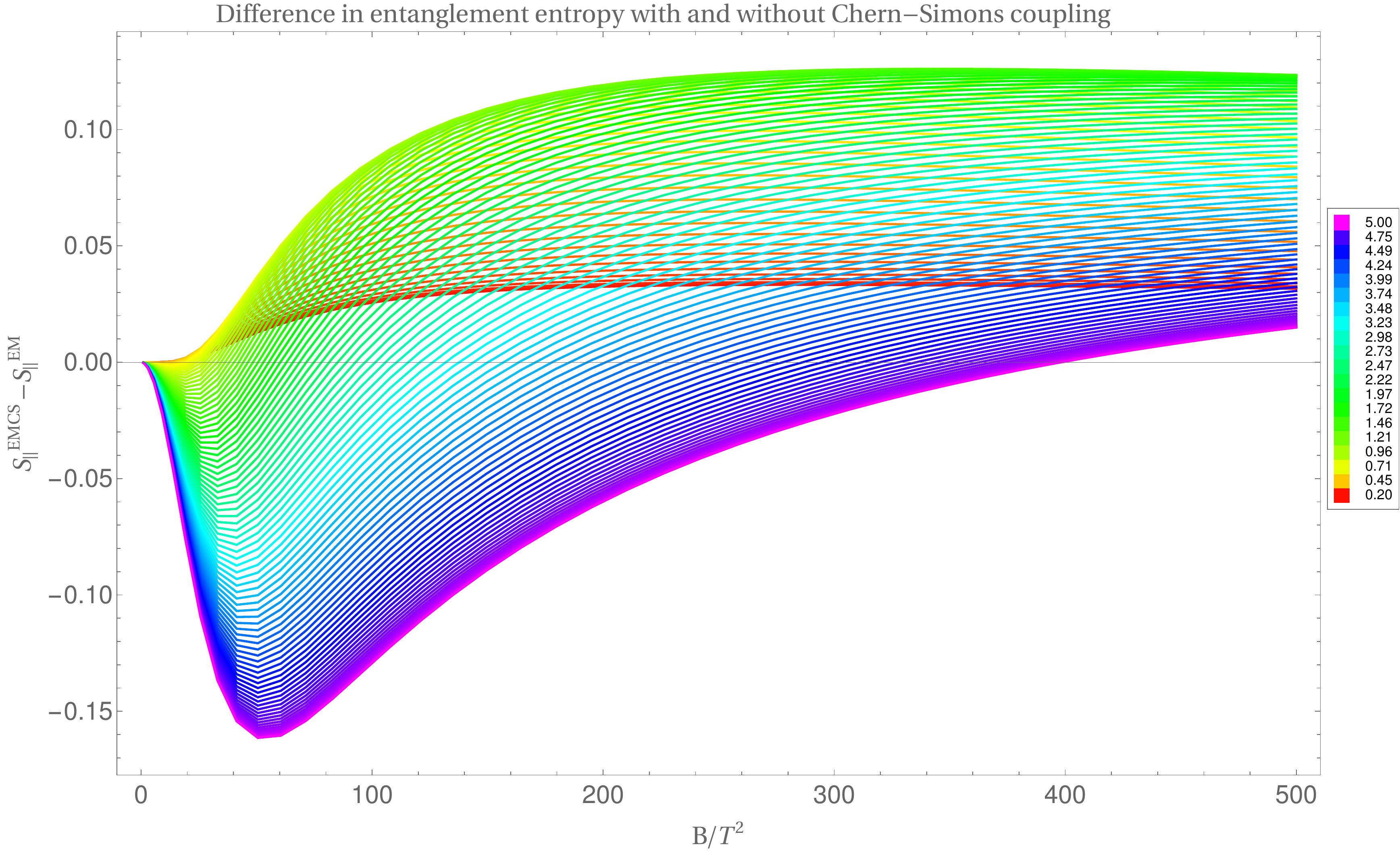}
 \end{subfigure}
 \begin{subfigure}[b]{0.5\textwidth}
    \includegraphics[width=75mm,scale=0.5]{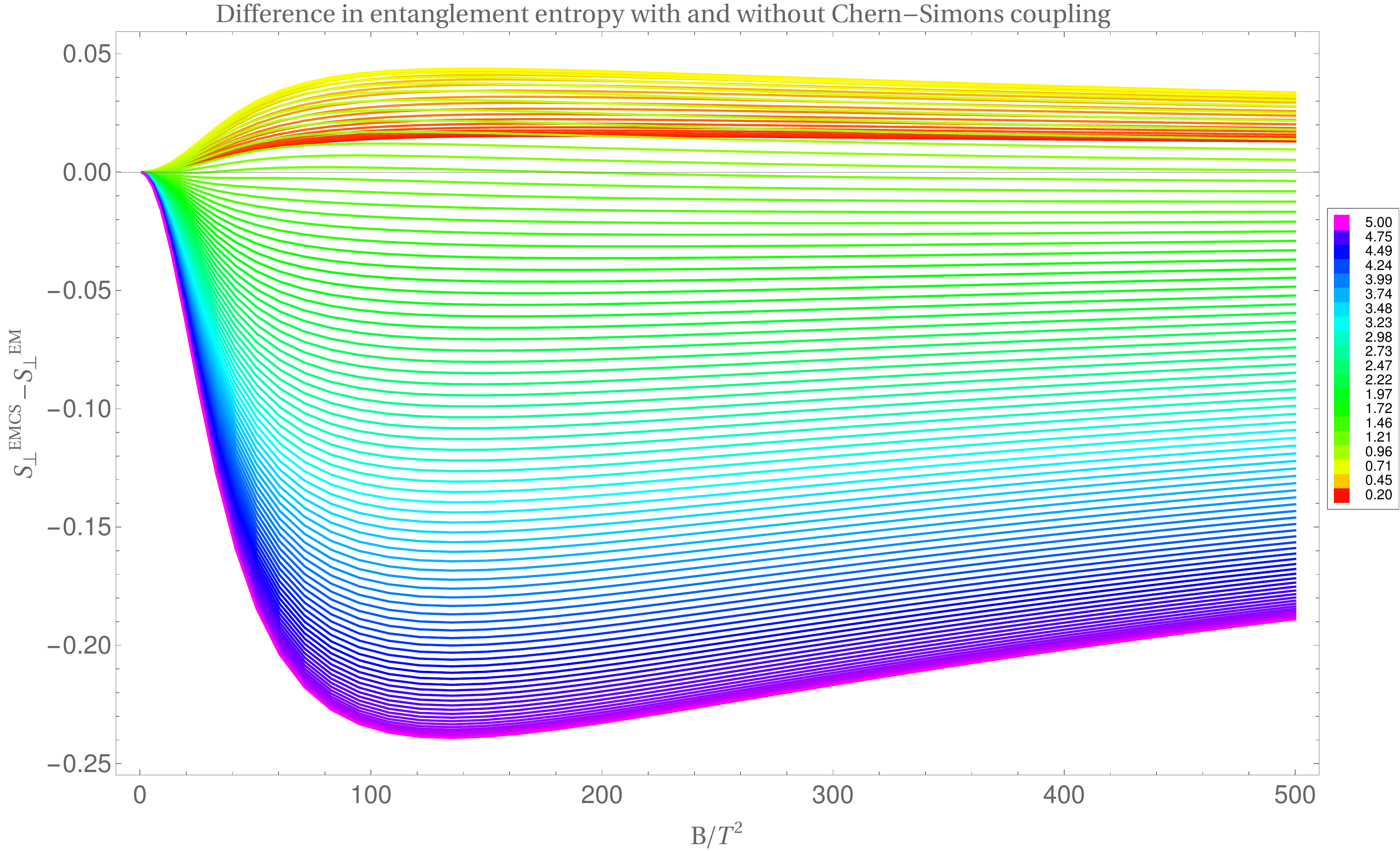}
    \end{subfigure}
    \caption{The difference in entanglement entropy as calculated with Chern-Simons coupling $\gamma=2/\sqrt{3}$ and $\gamma=0$ is displayed as a function of $B/T^2$. \textit{Left: }Parallel to magnetic field. \textit{Right:} Perpendicular to the magnetic field. The colors correspond to strips of width $\color{red}{\ell=2/10}$ to strips of width $\color{Plum}{\ell=5}$ as displayed by the inset plot legend. The curves were computed at a fixed $\mu/T=5$.
    \label{fig:EE_bt2} }
    \end{figure}
%

\paragraph{Alternate embedding:} The RT/HRT formula instructs us to seek out all possible surfaces and select the one whose area is minimal. In our system there exists an additional solution which is possible given by three segments, $x_{i}=\pm \ell/2, z$ and $x_{i}, z=z_h$. The subscript indicates either parallel ($x_3$) or perpendicular ($x_1$) strips~\cite{Zhang:2019zbf}. We discuss this solution further in appendix~\ref{sec:appendix_ALT} where we demonstrate that this solution is always larger then the solutions we discuss in the main text. 

\paragraph{Perturbative solutions:}
We can solve the geodesic equations perturbatively near the $AdS$-boundary, and for the vacuum case we find 
\begin{equation}
    x(z)=x_0+\frac{\ell}{4}z^4+\frac{\ell^3}{20}z^{10}+O(z^{16}) \, .
\end{equation}
Expanding both the parallel and transverse case along with the vacuum surface near the $AdS$ boundary, we find 
\begin{align}
    x_{\perp}(z)&=x_0+\frac{\ell z^4}{4}+\frac{\ell z^8}{1536} \left(-96 u_4-192 v_4+B^2\right)+\log (z)\left(-\frac{1}{192} \ell z^8 B^2\right) \cdots \, , \\
    x_{||}(z)&=x_0+\frac{\ell z^4}{4}+\frac{1}{384} \ell z^8 \left(-24 u_4+96 v_4+B^2\right)+\log (z)\left(-\frac{1}{48} \ell z^8 B^2\right)+\cdots \, .
\end{align}
Integrating term by term, we find the following result valid for small width strips in the parallel direction: 
\begin{align}
\Delta S_{||}&=\lim_{\epsilon\rightarrow 0}\frac{1}{V_{||}} \left(S_{||}-1/\epsilon^2  \right) =-\frac{1}{z_{*}^2} +\frac{1}{12} z_{*}^2 \left(B^2 (1-2 \log (z_{*}))-6 u_4+24 v_4\right) \nonumber \\
&+\frac{1}{240} \left(-20 e_2^2+60 \ell^2+\gamma ^2 \mu ^2 B^2\right)\, ,\label{eq:Cpara2}
\end{align}
and the perpendicular direction,
\begin{align}
\Delta S_{\perp}&=\frac{1}{V_{\perp}} \left(S_{\perp}-1/\epsilon^2  \right) =-\frac{1}{z_{*}^2}-\frac{1}{48} z_{*}^2 \left(B^2 (2 \log (z_{*})-1)+24 u_4+48 v_4\right) \nonumber \\
&+\frac{1}{60} z_{*}^4 \left(-5 e_2^2+15 \ell^2-\gamma ^2 \mu ^2 B^2\right)\, .\label{eq:Cperp2}
\end{align}
Here we have expressed the dependence on the length of the strip in the boundary theory, $\ell$, in terms of the turning point in the bulk, $\zs$.  We see that these subtracted entanglement entropies, $\Delta S_{||}$ and $\Delta S_\perp$, depend explicitly on the bulk spacetime near the $AdS$ boundary via the asymptotic coefficients $u_4$ and $v_4$ as displayed above in section~\ref{sec:holoModel} and eq.\ (\ref{eq:near_bndy_expansion}). Most importantly, the effect of the chiral anomaly, in form of the Chern-Simons coupling, is seen to enter explicitly into the evaluation of the entanglement entropies in equations~\eqref{eq:Cpara2} and~\eqref{eq:Cperp2}.

\paragraph{Full solutions:} We display the results of our calculations for the EE in figure~\ref{fig:EE_bt2}. To understand the impact of the chiral anomaly on the EE we compute the difference between the EE as computed with and without the Chern-Simons coupling present. One can see a dramatic qualitative change in the behavior of the EE when the Chern-Simons coupling is present. While these differences exist for all lengths we probe, they are the most pronounced for the longest lengths we probe, the purple curves in figure~\ref{fig:EE_bt2}. Hence the strongest effect of the chiral anomaly on the EE is seen in the IR. This is true for both transverse and parallel orientations of the entanglement strip in the dual field theory. 

As a final comment on the EE we note that strong subbaddivity implies $d^2S_{||,\perp}/d\ell^2 \le 0$  for all $\ell$. We have checked numerically that our solutions obey strong subaddivity and the EEs we display in figure~\ref{fig:EE_bt2}, which we use within calculations for the remainder of this work, satisfy $d^2S_{||,\perp}/d\ell^2 \le 0$  for all $\ell$ in our numerical domain. 

\subsection{c-functions in generic states}
In this subsection we present the c-function defined by~\eqref{eq:c_function} for $\mathcal{N}=4$ SYM theory evaluated in various states.  
Before we analyze some specific cases, let us first consider the RG-flow of a general quantum field theory. When evaluated in a vacuum state by standard computation~\cite{Zamolodchikov:1986gt}, the c-function of a generic quantum field theory decreases along the RG-flow as the theory is driven away from its UV fixed point (towards its IR fixed point) by relevant operators. On the gravity side of the gauge/gravity correspondence the original studies of RG-flow geometries~\cite{Freedman:1999gp} 
feature the UV fixed point encoded in the near-boundary geometry, while the IR fixed point corresponds to the near-origin geometry. 
These two asymptotically $AdS$ regions are distinguished by their respective $AdS$ radii. Hence, in these domain-wall solutions a candidate c-function must encode information about the geometry dual to the UV CFT when it probes small length scales. Conversely, that holographic c-function must encode information about the geometry dual to the IR CFT when it probes long length scales. 

The entanglement entropy, defined as a minimal surface by the Ryu/Takayanagi conjecture~\cite{Ryu:2006ef}, naturally has this property, as displayed schematically in figure~\ref{fig:rgPicture}. 
On the boundary, the entanglement entropy of small subregions with small strip width $\ell=\ell_{UV}$ probes short length scales of the field theory and thus its UV behavior. 
As figure~\ref{fig:rgPicture} suggests, minimal surfaces with small boundary support ($\ell=\ell_{UV}\approx 0$) probe exclusively the near-boundary geometry up to $z={\zs}^{UV}$ which corresponds to the UV energy regime of the field theory. 
Conversely, large values of $\ell$ give rise to minimal surfaces probing deep into the domain wall geometry which corresponds to the IR energy regime of the field theory. 
Consequently, the entanglement entropy is identified as a viable candidate for a holographic c-function of (3+1)-dimensional quantum field theories~\cite{Myers:2010tj,Myers:2012ed}. The bulk turning point, $\zs$, of the associated minimal surfaces serves as the energy scale along which the c-function changes monotonically~\cite{Myers:2010tj,Myers:2012ed}. This monotonic behavior of the c-function is guaranteed provided that the matter which supports the geometry satisfies the null energy condition~\cite{Myers:2012ed}. 

In our case, we consider a conformal field theory, $\mathcal{N}=4$ SYM theory, which has a trivial RG-flow. Its $\beta$-functions vanish and the UV fixed point coincides with the IR fixed point, i.e.~the blue dot coincides with the red dot in figure~\ref{fig:rgFlow}. 
However, the energy-dependence of the thermal states we consider now is non-trivial. As indicated in figure~\ref{fig:rgPicture}, the charged magnetic black brane geometry (represented by the green line) interpolates between the UV (blue dotted line) and the IR geometry (red dotted line). In other words, the charged magnetic black brane geometry encodes the trivial RG-flow of $\mathcal{N}=4$ SYM theory and simultaneously the non-trivial energy scale dependence of the charged magnetic state. Let us consider some simple thermal states first.

\begin{figure}[ht]
    \begin{subfigure}[b]{0.5\textwidth}
    \includegraphics[width=70mm,scale=0.5]{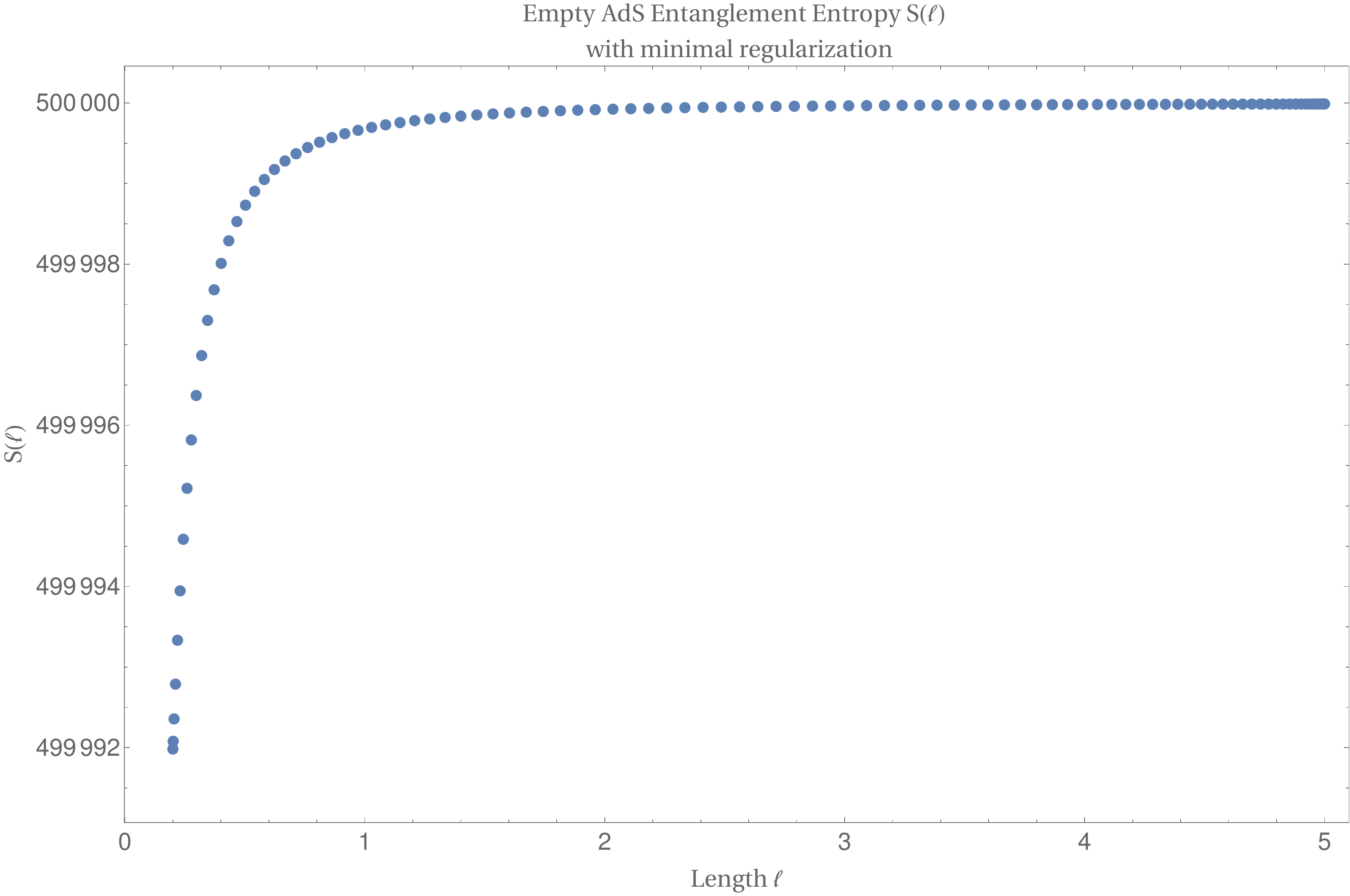}
 \end{subfigure}
 \begin{subfigure}[b]{0.5\textwidth}
    \includegraphics[width=70mm,scale=0.5]{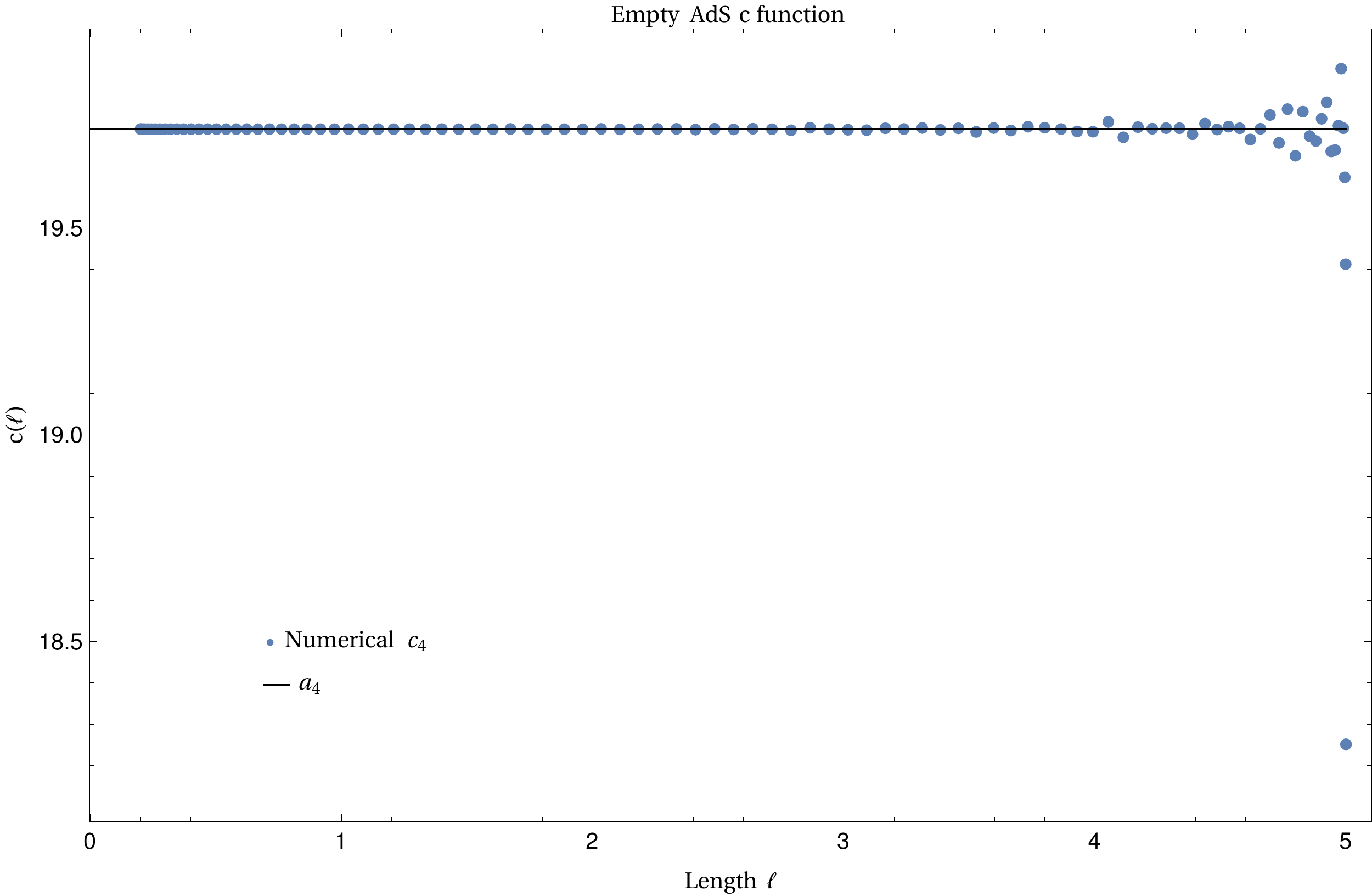}
    \end{subfigure}
    \caption{The entanglement entropy and the c-function of $\mathcal{N}=4$ SYM theory as calculated in a vacuum state dual to empty $AdS_5$ spacetime.  \textit{Left:} The empty $AdS_5$ entanglement entropy. The values are regulated via minimal subtraction of the $1/z_{UV}^2$ divergence. \textit{Right:} The c-function, defined as $c_4=\beta_4\ell^3\partial S/\partial \ell$ by eq.~(\ref{eq:c_function}). There is a loss of numerical precision beginning around $\ell\approx 3.5$. The value that the c-function takes is approximately $c_4\approx 2\pi^2$ in agreement with the value of the central charge of $\mathcal{N}=4$ SYM, $a_4=2\pi^2$.
    \label{fig:empty}}
\end{figure}

\paragraph{c-function in $\mathcal{N}=4$ SYM theory in a vacuum state:} 
In the empty $AdS_{4}$ spacetime we can compute the c-function in eq.~\eqref{eq:c_function} in a vacuum state of the dual CFT$_4$. We expect this quantity to be a constant as we are not deforming the CFT$_4$ with any operators, hence it should have a trivial RG-flow remaining fixed at a value $a_4=2\pi^2$ which is the well-known UV central charge of our example, $\mathcal{N}=4$ SYM theory. 
We can calculate this value using the techniques described in appendix~\ref{sec:appendix_Coordinate} and appendix~\ref{sec:appendix_numerics} as a check of our numerics. The results of our calculation are displayed in figure~\ref{fig:empty}. Despite some numerical errors for large length strips of a size approximately an order of magnitude larger then the curvature scale $L=1$ our expectation is confirmed. The figure demonstrates that our numerical calculation has succeeded in replicating the value of the central charge of $\mathcal{N}=4$ SYM theory, $c_4\approx a_4=2\pi^2$. Furthermore, as expected the central charge is a constant at all energy scales here represented by the length of the strip in the dual theory. 

Taking a different point of view, we may consider Einstein-Maxwell-Chern-Simons theory as Einstein Gravity plus Maxwell-Chern-Simons matter content. In appendix~\ref{sec:NEC} we prove analytically that the gauge field matter content of Einstein-Maxwell-Chern-Simons theory satisfies the null energy condition (NEC). So, from this perspective the c-function according to~\cite{Myers:2012ed} should be constant or monotonically {\it decreasing}.\footnote{As pointed out by Paulos~\cite{Paulos:2011zu}, the monotonicity of the c-function depends on the energy density of the relevant geometry.} 
This is consistent with our result that the c-function is constant, figure~\ref{fig:empty}.  

\begin{figure}[htbp]
    \begin{subfigure}[b]{0.5\textwidth}
    \includegraphics[width=70mm,scale=0.5]{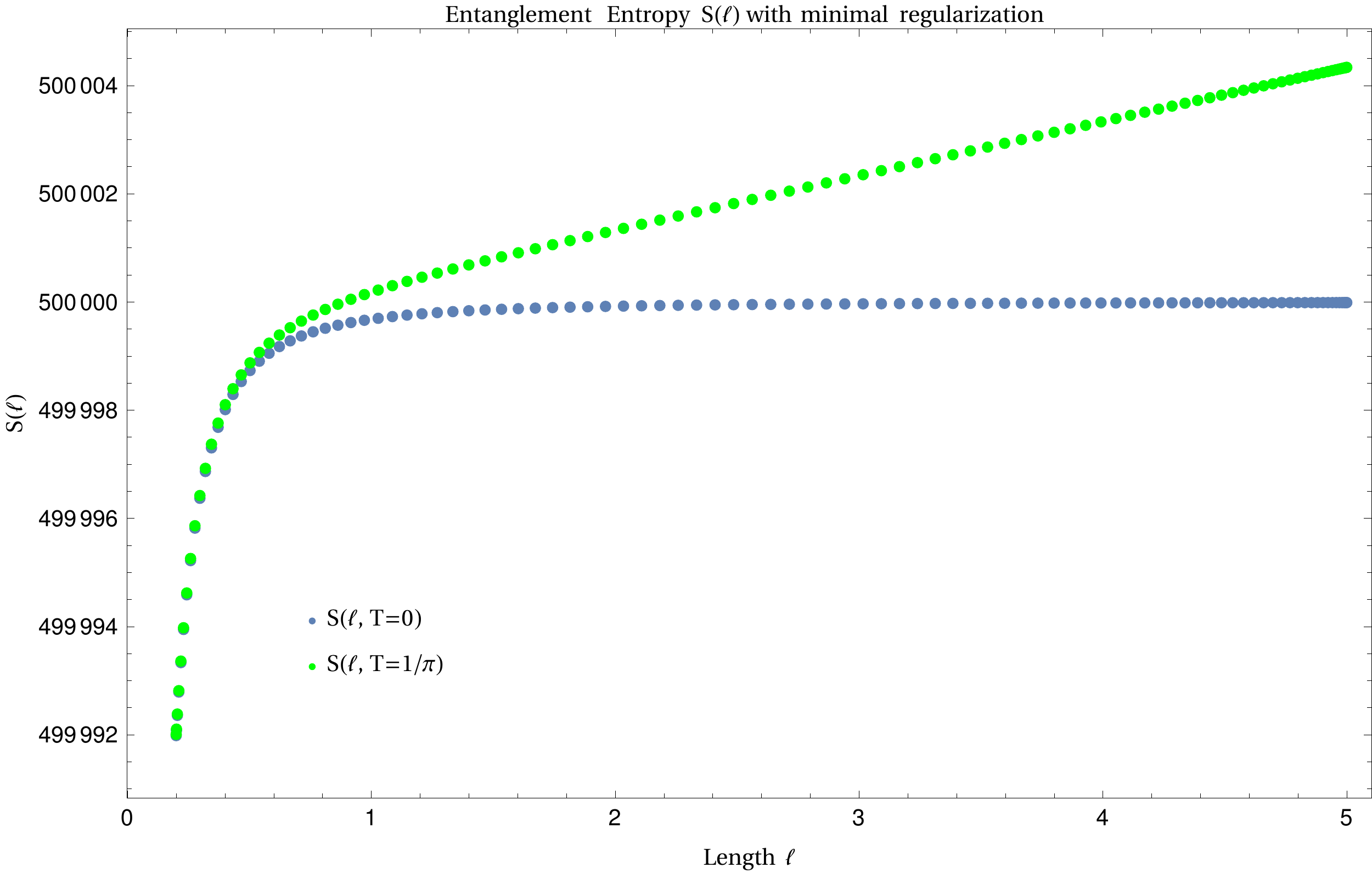}
 \end{subfigure}
 \begin{subfigure}[b]{0.5\textwidth}
    \includegraphics[width=70mm,scale=0.5]{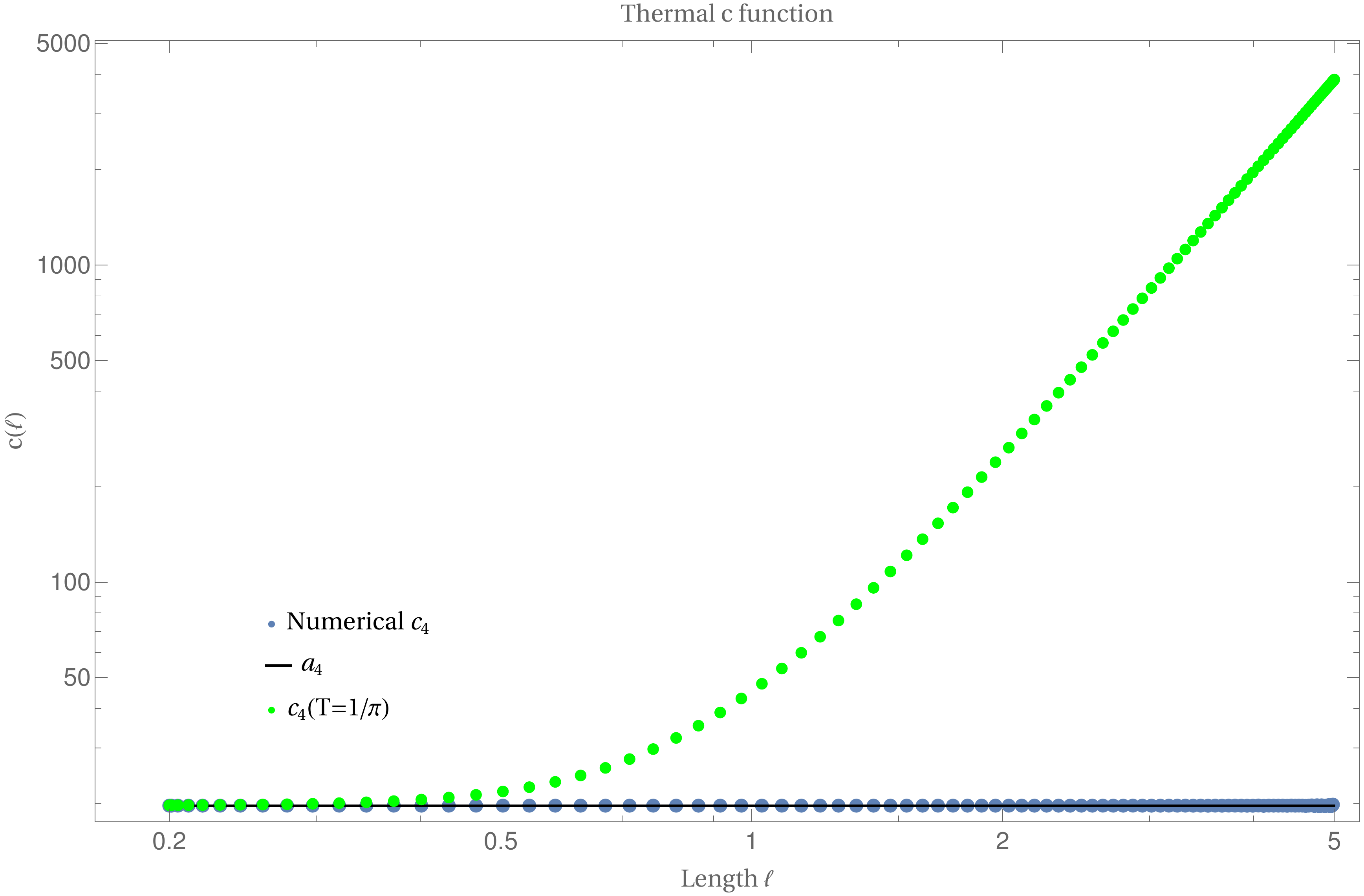}
    \end{subfigure}
    \caption{The entanglement entropy and the c-function of $\mathcal{N}=4$ SYM theory as calculated in a thermal state dual to a Schwarzschild black brane in asymptotically $AdS_5$ spacetime. Both are displayed as a function of the probed length scale, $\ell$ (equivalent to $\zs$, see appendix~\ref{sec:appendix_Coordinate}). The green curves associated with a thermal state (labeled $T/\pi$) are  displayed alongside their blue counterparts which are computed in a vacuum state ($T=0$).  
    \textit{Left:} The entanglement entropy, with its values regulated via minimal subtraction of the $1/z_{UV}^2$ divergence. 
    \textit{Right:} The c-function evaluated in a thermal state increases as we decrease the probed energy scale. In the UV it takes the expected value, $2\pi^2$, in agreement with the value of the central charge of $\mathcal{N}=4$ SYM theory, $a_4$.  
   \label{fig:Sch_c}}
\end{figure}

\begin{figure}
    \centering
    \includegraphics[width=12cm]{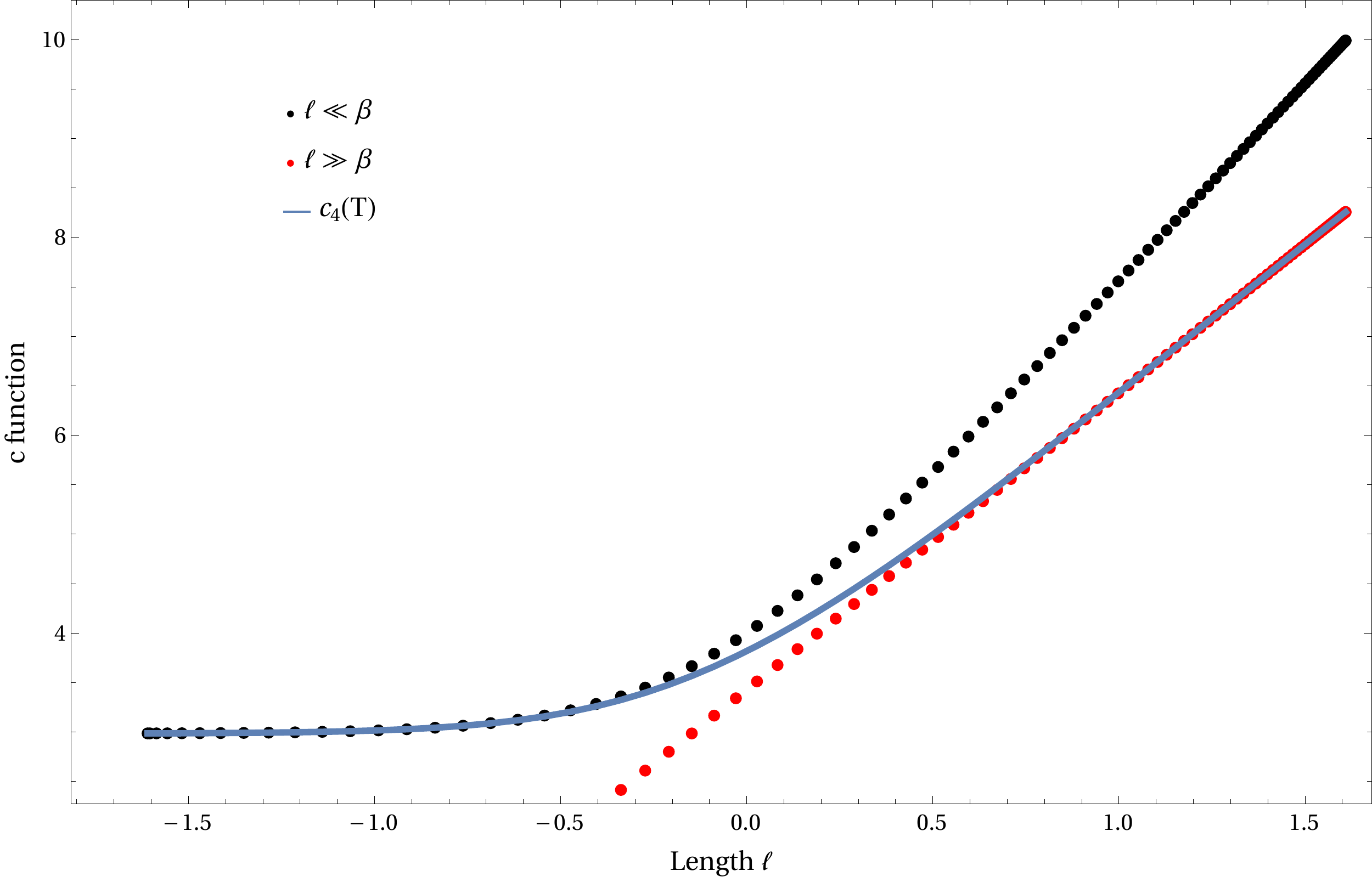}
    \caption{
    {\it Four-dimensional entropic c-function.} The exact entropic c-function for $\mathcal{N}=4$ SYM theory in a thermal state is shown (blue solid curve) along with the low temperature (black dotted curve) and the high temperature (red dotted curve) limits given in eq.~\eqref{eq:high_low}. 
    \label{fig:interpolation}}
\end{figure}

\paragraph{c-function in $\mathcal{N}=4$ SYM theory in a thermal state:} 
Let us repeat this calculation, but in a thermal state. The most readily available example is the thermal state dual to a planar Schwarzschild black brane given by the line element 
\begin{equation}
  \exd s^2=\frac{1}{z^2}\left(\frac{\exd z^2}{U(z)} -U(z)\exd t^2+\exd x_1^2+\exd x_2^2+\exd x_3^2\right),\quad U(z)=1-mz^4 \, , \label{eq:Metric_sch}
\end{equation}
with the mass $m$ of the black brane. 
While we do not provide an analytic expression for the c-function as defined in eq.~(\ref{eq:c_function}), we do compute it numerically. For a fixed temperature, the result of this computation can be seen in figure~\ref{fig:Sch_c}.
At a fixed temperature, while the c-function is monotonic, it increases towards the IR. Naively, one may expect a c-functions to rather decrease towards the IR. This is the case when the c-function evaluated on any non-trivial RG-flow in a vacuum state. However, if we evaluate it on a thermal state, this changes. Let us compare this behavior to examples of entanglement at finite temperature known from field theory calculations. One such example which is explicitly calculable is the entanglement entropy on an interval at finite temperature in $d=2$ for which the expression is given by~\cite{Calabrese:2009qy} 
\begin{equation}\label{eq:SACalabrese}
    S_A=\frac{c}{3}\log\left(\frac{\beta}{\pi a}\sinh \frac{\pi\ell}{\beta}\right)+c'_1=\begin{cases}
    \frac{c}{3}\log\left(\frac{\ell}{ a}\right)+c'_1 & \ell \ll \beta \\
       \frac{\pi c}{3\beta}\ell+c'_1 & \ell \gg \beta \, .
    \end{cases}
\end{equation}
This expression can be seen to interpolate between the zero-temperature result for $\ell \ll \beta$ and
an extensive form for $\ell \gg\beta$. The authors note that in the limit $\ell \gg\beta$ the expression agrees with that
of the Gibbs entropy of an isolated system of length $\ell$. Computing the c-function from the entanglement entropy given by~\eqref{eq:SACalabrese} yields
\begin{equation}
  c_2=\frac{c \pi  \ell}{\beta} \coth \left(\frac{\pi  \ell}{\beta}\right)=\begin{cases}
    c & \ell \ll \beta \\
       c\frac{\pi}{\beta}\ell & \ell \gg\beta
       \end{cases} \, .
\end{equation}
The low temperature (UV) behavior of the entropic c-function is a constant value while the high temperature (IR) behavior is a linear function of the length $\ell$. In addition it is monotonically increasing for increasing width of the interval. The result for an interval in $d=2$ is similar to the result we find for a strip in $d=4$, taking the same limits, we find
\begin{equation}
    c_4= \begin{cases}
     -\beta_4\frac{\pi ^{1/2} \Gamma \left(-\frac{1}{3}\right) \Gamma \left(\frac{5}{3}\right)^2}{12 \Gamma \left(\frac{1}{6}\right) \Gamma \left(\frac{7}{6}\right)^2} \frac{N_c^2}{L^3} +\beta_4\frac{2 \pi ^{5/2}\Gamma \left(\frac{7}{6}\right)^2 \Gamma \left(\frac{4}{3}\right) \left(15 \Gamma \left(\frac{5}{3}\right) \Gamma \left(\frac{1}{6}\right)+8 \Gamma \left(-\frac{1}{3}\right) \Gamma \left(\frac{7}{6}\right)\right) }{15 \Gamma \left(\frac{1}{6}\right) \Gamma \left(\frac{5}{6}\right) \Gamma \left(\frac{5}{3}\right)^3} L^5  N_c^2 T^4 \ell^4& \ell \ll \beta  \\
      \frac{\beta_4 \pi ^2}{2}  N_c^2  T^3 L^3 \ell^3& \ell \gg \beta
    \end{cases} \, , \label{eq:high_low}
\end{equation}
where we have kept the next-to-leading order in the small length limit and temporarily reinstated the $AdS$ radius, $L$. To leading order the low temperature behavior of the $c$-function is a constant as shown in figure~\ref{fig:empty} with corrections at order $\ell^4$ while the high temperature limit is increasing for increasing strip width $\ell$, scaling extensively. These two limits along with the finite temperature $c$-function $c_4$ are displayed in figure~\ref{fig:interpolation} in the units we adopt for this work. That figure shows that the full solution of the $c$-function (blue solid line) interpolates between the two limiting regimes.

\begin{figure}[hptb]
    \centering
    \includegraphics[width=12cm]{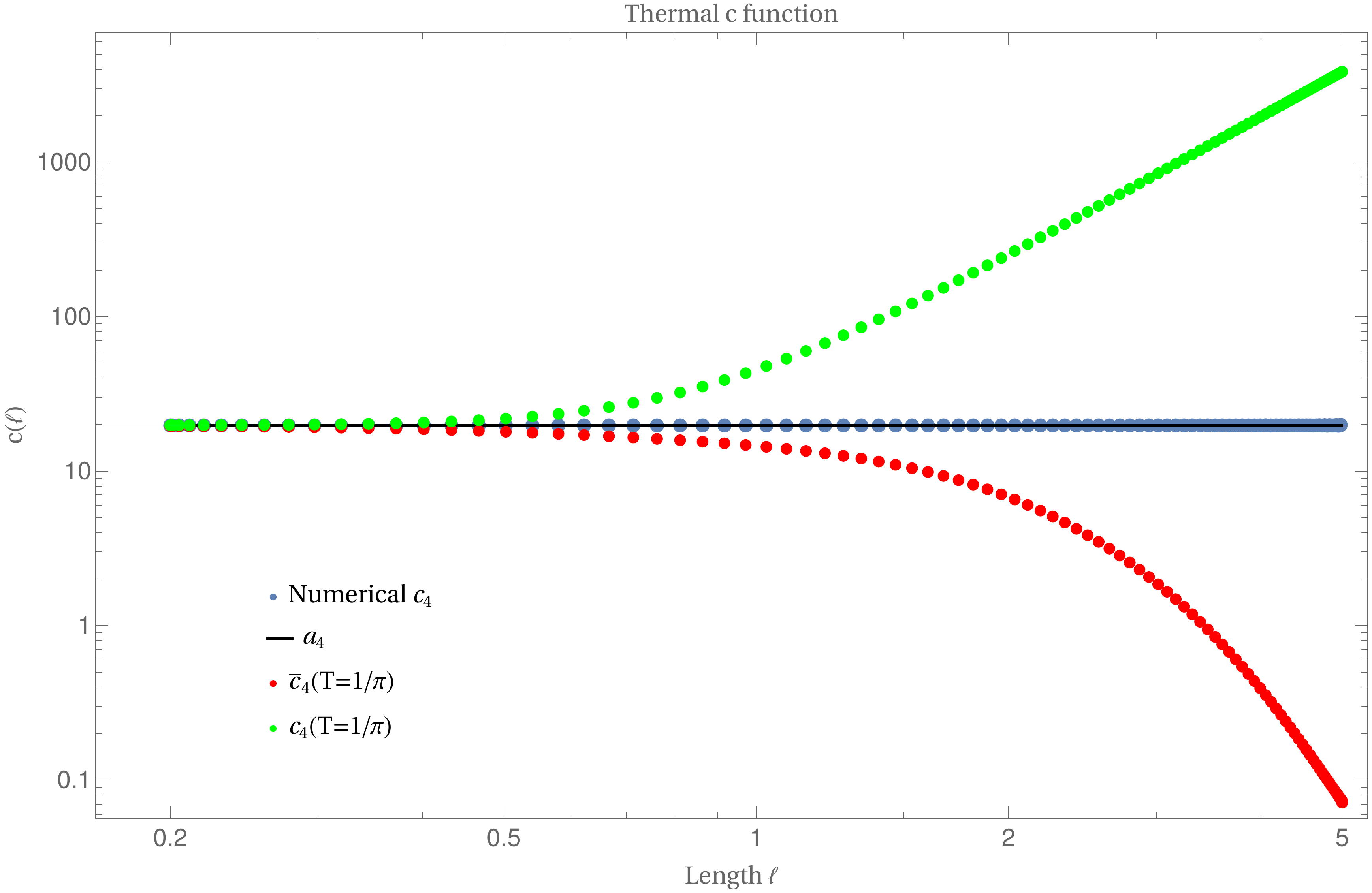}
    \caption{The c-function as calculated in a thermal state dual to a Schwarzschild black brane in asymptotically $AdS$ spacetime. While c-function $c_4$ increases as displayed in \textcolor{green}{green} the subtracted c-function $\bar{c}_4$ as displayed in \textcolor{red}{red} decreases as we decrease the energy scale we probe. 
    \label{fig:sch_T_cbar}}
\end{figure}

Motivated by the behavior displayed in eq.~(\ref{eq:high_low}) one may try to recover a decreasing behavior from UV to IR scales, and for that purpose we could define a thermal c-function by removing degrees of freedom associated with the horizon of the planar black brane by 
\begin{equation}
    \bar{c}_{4}=c_4-\beta_4 \ell^3 s, \quad s=\frac{S}{V}=\frac{1}{4G_N V}\int \exd^3x\sqrt{g_{\text{spatial}}} \, .
\end{equation}
Here we have removed from the c-function the thermal entropy associated with a volume $\ell^3$ in the CFT$_4$. The result of this subtraction scheme is displayed in figure~\ref{fig:sch_T_cbar} where we can now see a decrease in the c-function as we probe lower energy scales associated with wider strips. 

\begin{figure}[hptb]
    \centering
    \includegraphics[width=12cm]{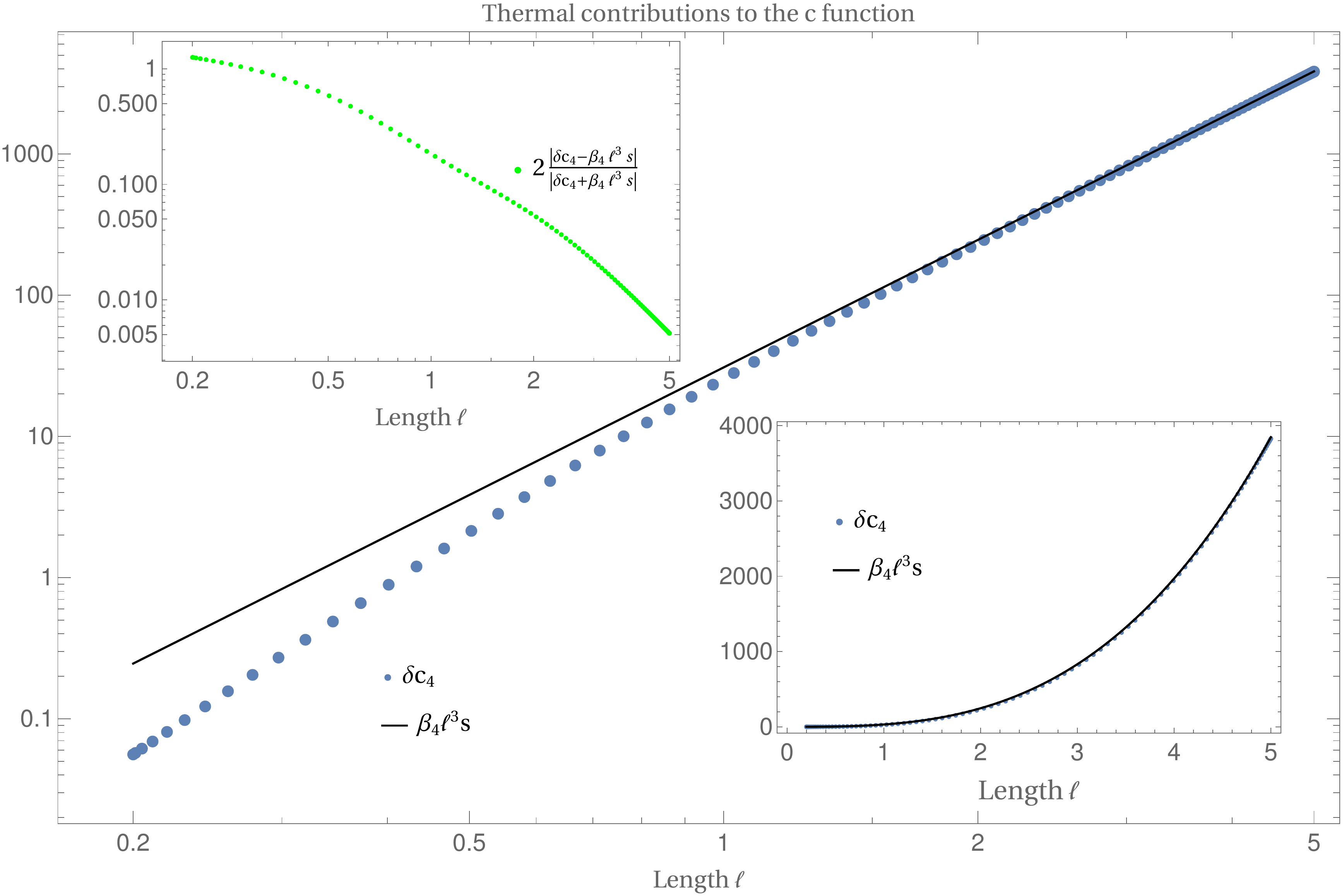}
    \caption{The deviation $\delta c_4$ of the thermal c-function, evaluated in the thermal state dual to the $AdS$ Schwarzschild black brane, from the central charge $a_4$, displayed as the \textcolor{blue}{blue} points. The thermal subtraction $\beta_4 \ell^3 s$ is displayed as a black line. Its behavior precisely matches $\delta c_4$ in the IR. The bottom right inset graphic displays the same information only not in a log-log scale, the top left inset displays the difference as \textcolor{green}{green} points on a log-log scale.
    \label{fig:Thermal_c_subtraction}}
    \end{figure}

Considering our choice of subtraction scheme one could ask if this is an ``optimal'' scheme? Given that we know the value of the UV central charge of our theory what is the deviation of the thermal state c-function from this value of the central charge $\delta c_4= c_4-a_4$ and how does this compare to the thermal entropy associated with a volume $\ell^3$ in the CFT$_4$, $\beta_4 \ell^3 s$? We display the result of this comparison in figure~\ref{fig:Thermal_c_subtraction}. The behavior of the thermal entropy $\beta_4 \ell^3 s$ precisely matches $\delta c_4$ in the IR, hence the IR contribution to $c_4$ is due almost entirely to the entropy associated with the horizon. This matches the expectation of~\cite{Paulos:2011zu} in which it is stated that the c-function computed holographically in a spacetime geometry with a black brane asymptotes to the thermal entropy of the horizon on IR scales, monotonically increasing towards the IR. Note, however, that~\cite{Paulos:2011zu} defined the c-function without reference to entanglement entropy. Hence, our new computation gives independent evidence for this statement. 

\paragraph{c-function in $\mathcal{N}=4$ SYM in an anisotropic vacuum state:} For anisotropic and Lorentz-violating systems, where the rotational symmetry $SO(d)$ has been broken down to $SO(d_1)\times SO(d_2)$. Hence, in principle, the spatial coordinates do not need to scale with length in the same way~\cite{Chu:2019uoh},
\begin{equation}
     [x]=\text{Length}^{n_1} \, , \quad   [\tilde{x}]=\text{Length}^{n_2} \, ,
\end{equation}
where we have used $(x, \tilde{x}$) to denote the coordinates for each factor of the $SO(d_1)\times SO(d_2)$ decomposition of the spatial geometry. A further implication of the spacetime anisotropy is an increase in the number of curvature invariants consistent with the spacetime symmetries, hence, an increase in the number of possible c-functions. This has motivated the authors of~\cite{Chu:2019uoh} to generalize the c-function eq.~\eqref{eq:c_function} to  anisotropic and Lorentz-violating systems, to strips along or perpendicular to the anisotropy 
\begin{eqnarray}
    c_{||}&=&\beta_{||} \frac{\ell_{||}^{d_{||}}}{H_{||}^{d_1-1} H_\perp^{d_2}} \frac{\partial S_{||}}{\partial  \ell_{||}} \, \label{eq:c_function_para},\\
    c_{\perp}&=&\beta_\perp \frac{\ell_{\perp}^{d_{\perp}}}{H_{||}^{d_1} H_\perp^{d_2-1}} \frac{\partial S_{\perp}}{\partial  \ell_{\perp}} \, \label{eq:c_function_perp},
\end{eqnarray}
where $\beta_{||}$ and $\beta_\perp$ are dimensionless normalization constants, while $H_{||},\, H_\perp$ are the IR regulators~\cite{Chu:2019uoh} for the longitudinal and perpendicular entanglement entropy, respectively.~\footnote{
Note that there is a typo in~\cite{Baggioli:2020} where the IR-regulators $H_{\perp,||}$ were erroneously referred to as ``UV cut-offs''.} 
In general, the coordinates describing the two rotationally invariant subspaces of the spacetime no longer scale in the same way under scaling transformations, thus, the definitions above make use of effective spacetime dimensions defined by
\begin{equation}
    d_{\perp}= d_1 + d_2 \alpha, \quad
    d_{||}=d_1\frac{1}{\alpha} + d_2,\quad \alpha=n_2/n_1\, ,
\end{equation}
which capture an effective number of dimensions as measured with respect to the spatial coordinates $(x,\tilde{x})$. 
Here, $\alpha$ may be referred to as the {\it refractive index} of the spacetime and is given in terms of the spacetime scalings $n_1$ and $n_2$. 
In our system, $d_1=2$ for the coordinates $x=(x_1,x_2)$, while $d_2=1$ for the coordinates $\tilde{x}=x_3$. 
In general, the refractive index can vary according to the energy scale. Rather than computing the c-functions in terms of a ``running'' effective dimension it is sensible to define it at a fixed point of the RG-flow. 
We choose to compute the refractive index of spacetime at the UV fixed point, following~\cite{Chu:2019uoh}
\begin{equation}
    \alpha=\lim_{\beta\rightarrow\infty }\frac{A_2(\beta)}{A_1(\beta)}\label{eq:refractive_def} \, ,
\end{equation}
where $A_2$ and $A_1$ are components of the metric defined in~\cite{Chu:2019uoh}, given by
\begin{equation}
    ds^2=-e^{2B(\beta)}\exd t^2 +d\beta^2+ e^{2A_1(\beta)}(\exd x_1^2+\exd x_2^2) +e^{2A_2(\beta)}\exd x_3^2 \, . \label{eq:poincare}
\end{equation} 
The line element~(\ref{eq:poincare}) can be obtained from our metric via transformations by the variable change motivated by the transformation in~\cite{Paulos:2011zu},
\begin{equation}
    z=e^{-A(\beta)}, \quad U(\beta)=A'(\beta)^2,\quad v(\beta )= e^{A_1(\beta )-A(\beta )},\quad w(\beta )= e^{A_2(\beta )-A(\beta )}. \label{eq:RG_Coordinates}
\end{equation}
Under the coordinate transformation in eq.~(\ref{eq:RG_Coordinates}) the condition in eq.~(\ref{eq:refractive_def}) becomes,
\begin{equation}
   \alpha =\lim_{z\rightarrow 0} \frac{\log(z)-\log(w(z))}{\log(z)-\log(v(z))}=1. 
\end{equation}
With the value for $\alpha$ in hand the scaling dimensions of our boundary theory are given by
\begin{equation}
    d_x=3, \quad d_y=3\, .
\end{equation}
Surprisingly, despite being in an anisotropic state the scaling dimensions of both transverse and the longitudinal direction are the same. 

As a result, the transverse and the longitudinal c-functions agree with each other at the UV fixed point, where they equal the central charge $a_4$. Since we evaluate them in vacuum and for a CFT without any deformation, both c-functions retain that same value all the way into the IR, and we do not display these trivial plots here, as they would reproduce the curve in figure~\ref{fig:empty}. 
\begin{figure}[htb]
  \begin{subfigure}[b]{.5\linewidth}
  \begin{center}
\includegraphics[width=2.9 in]{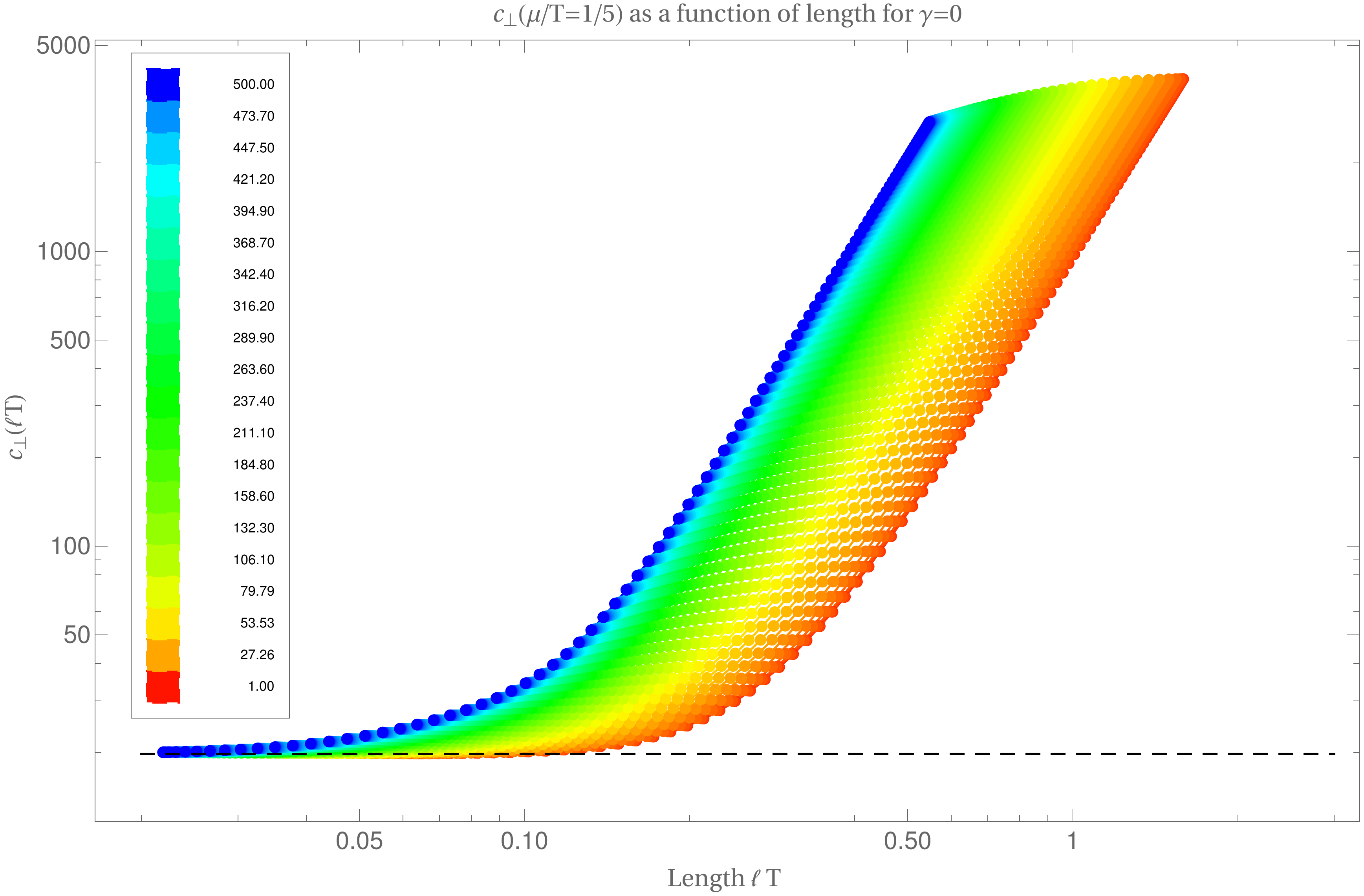}
\end{center}
\end{subfigure}
 \begin{subfigure}[b]{.5\linewidth}
  \begin{center}   
\includegraphics[width=2.9 in]{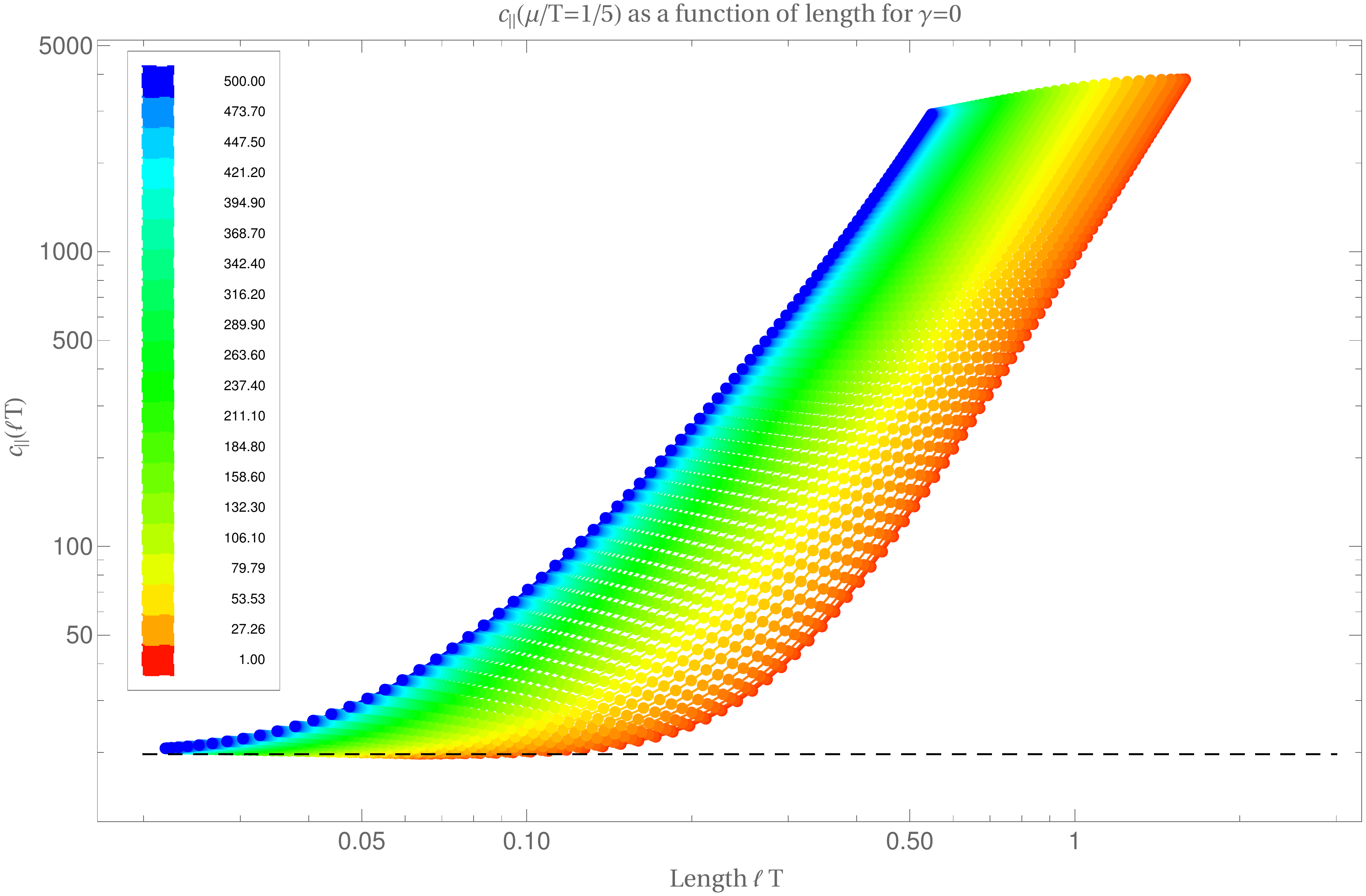}
  \end{center}
\end{subfigure}
    \caption{ We display the c-functions at $\gamma=0$ computed in two ways, firstly via direct numerical calculation utilizing  eq.~(\ref{eq:c_function_para}) and eq.~(\ref{eq:c_function_perp}) (lines), and secondly via the conserved charge as given in eq.~(\ref{eq:C}) and eq.~(\ref{eq:Cperp}) (dots). \textit{Left:} $\Delta c_{\perp}$, \textit{Right:} $\Delta c_{||}$. In both graphs these are displayed as a function of dimensionless length $\ell T$ at fixed $\mu/T=1/5$ for $B/T^2$ varying from $\color{red}{B/T^2=1.00}$ to $\color{Blue}{B/T^2=500.00}$. The dashed line displays the UV value of the central charge $a_4=2\pi^2$.
    \label{fig:purenumreics_c}}
\end{figure}
\begin{figure}[htb]
  \begin{subfigure}[b]{.5\linewidth}
  \begin{center}
\includegraphics[width=2.9 in]{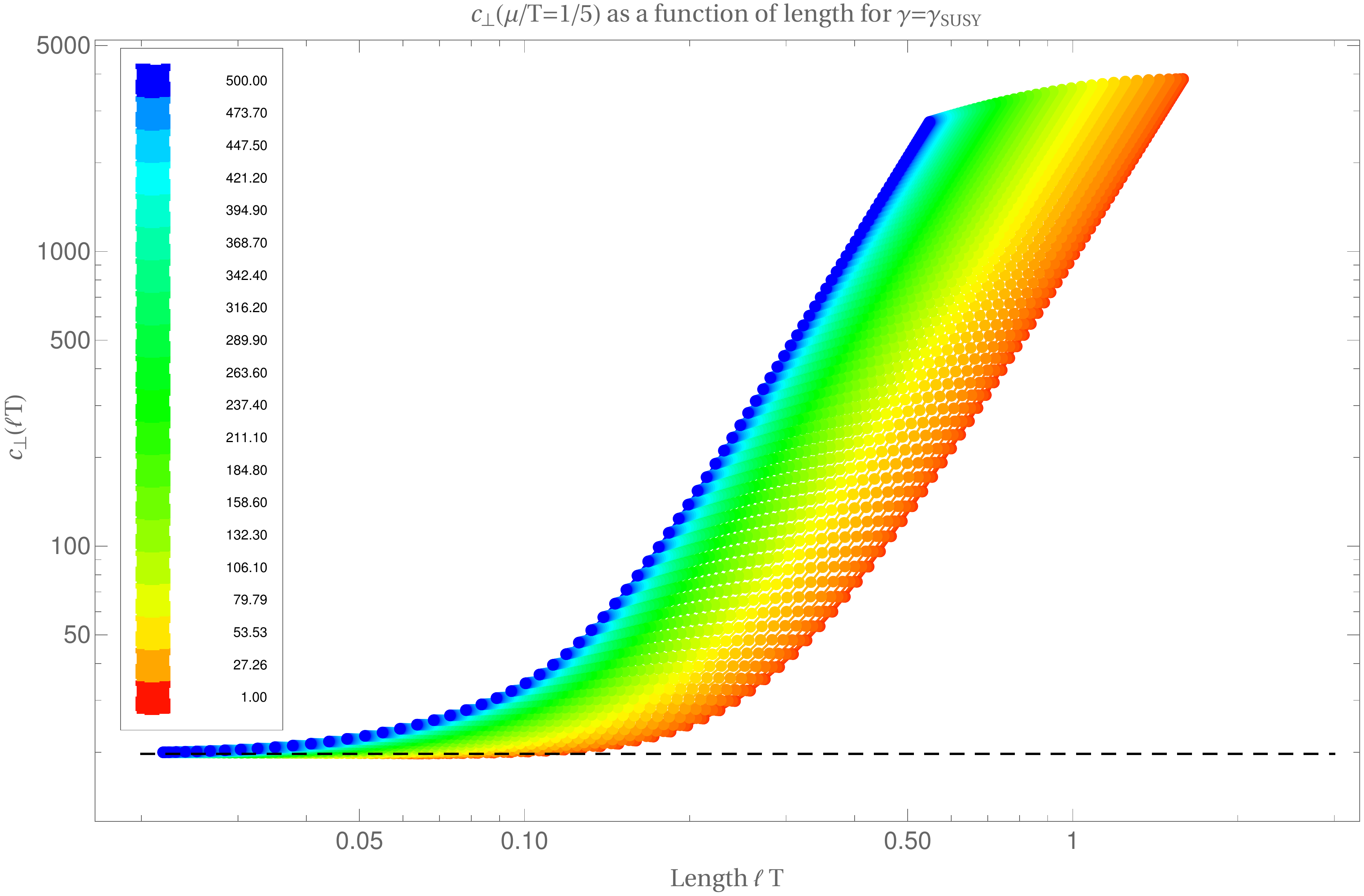}
\end{center}
\end{subfigure}
 \begin{subfigure}[b]{.5\linewidth}
  \begin{center}   
\includegraphics[width=2.9 in]{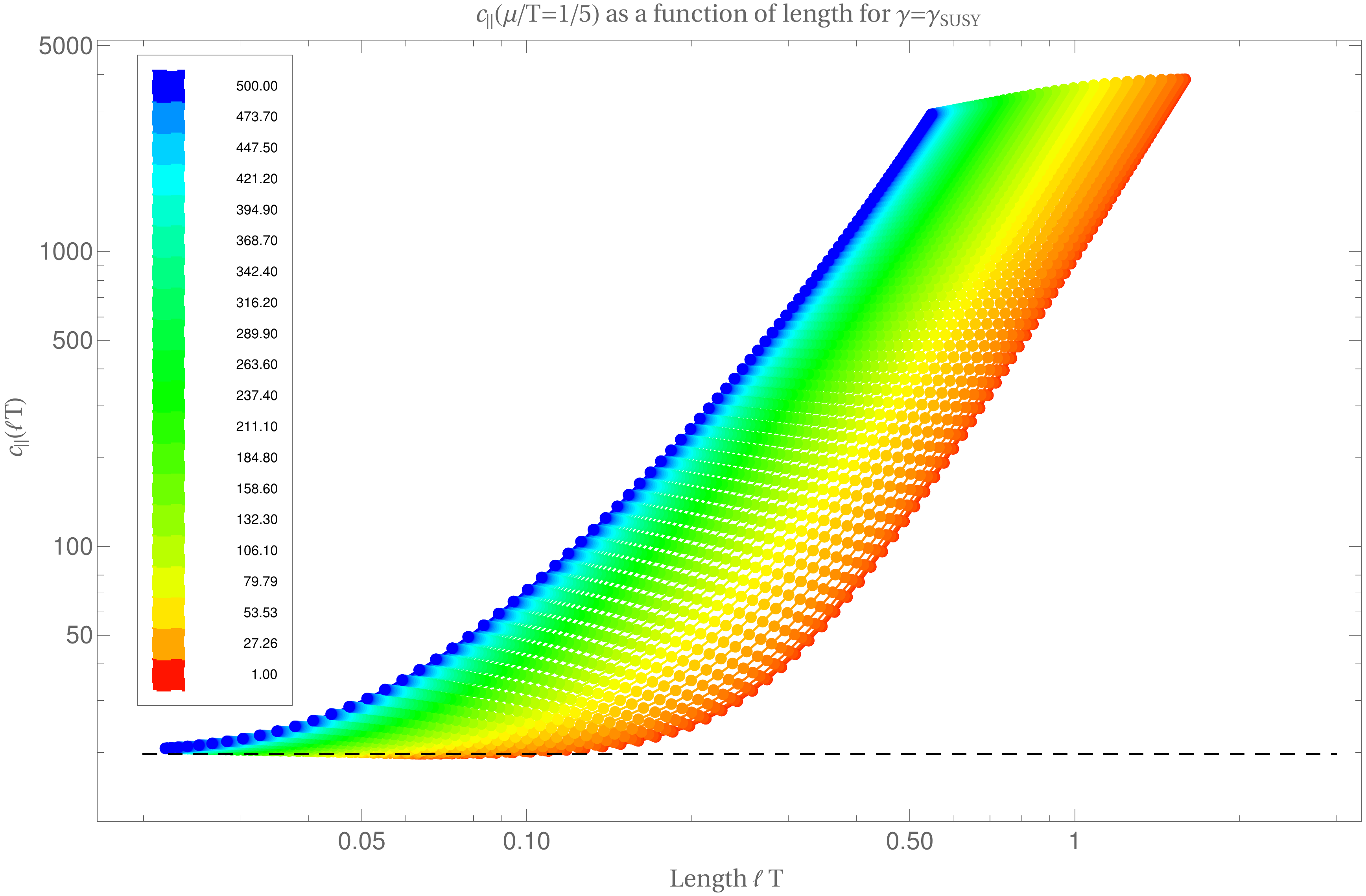}
  \end{center}
\end{subfigure}
    \caption{We display the c-functions at $\gamma=\gamma_{SUSY}$ computed in two ways, firstly via direct numerical calculation utilizing  eq.~(\ref{eq:c_function_para}) and eq.~(\ref{eq:c_function_perp}) (lines), and secondly via the conserved charge as given in eq.~(\ref{eq:C}) and eq.~(\ref{eq:Cperp}) (dots). \textit{Left:} $\Delta c_{\perp}$. \textit{Right:} $\Delta c_{||}$. In both images these are displayed as a function of dimensionless length $\ell T$ at fixed $\mu/T=1/5$ for $B/T^2$ varying from $\color{red}{B/T^2=1.00}$ to $\color{Blue}{B/T^2=500.00}$. 
    \label{fig:purenumreics_cSUSY}}
\end{figure}

\begin{figure}[htb]
  \begin{subfigure}[b]{.5\linewidth}
  \begin{center}
\includegraphics[width=2.9 in]{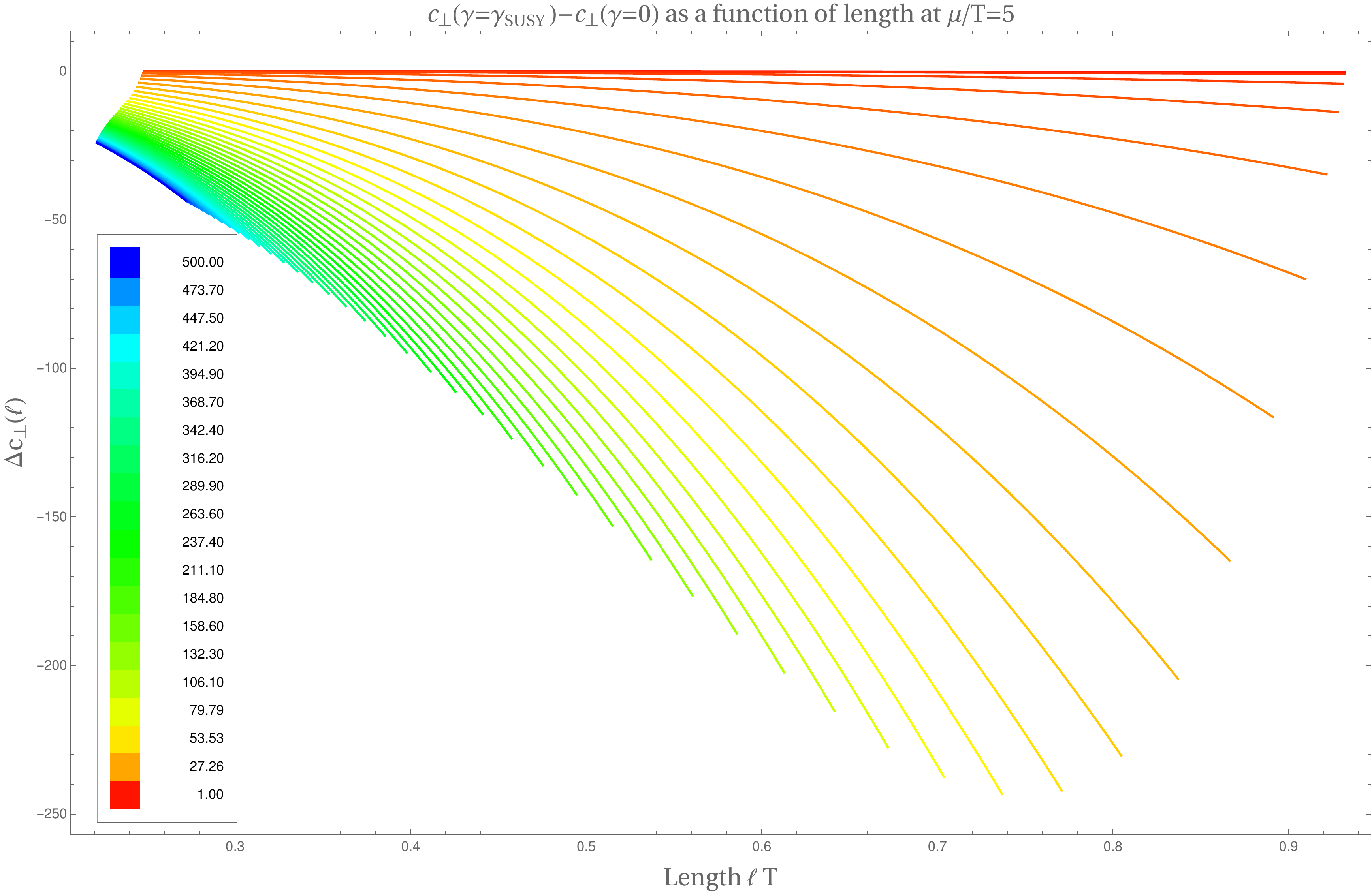}
\end{center}
\end{subfigure}
 \begin{subfigure}[b]{.5\linewidth}
  \begin{center}   
\includegraphics[width=2.9 in]{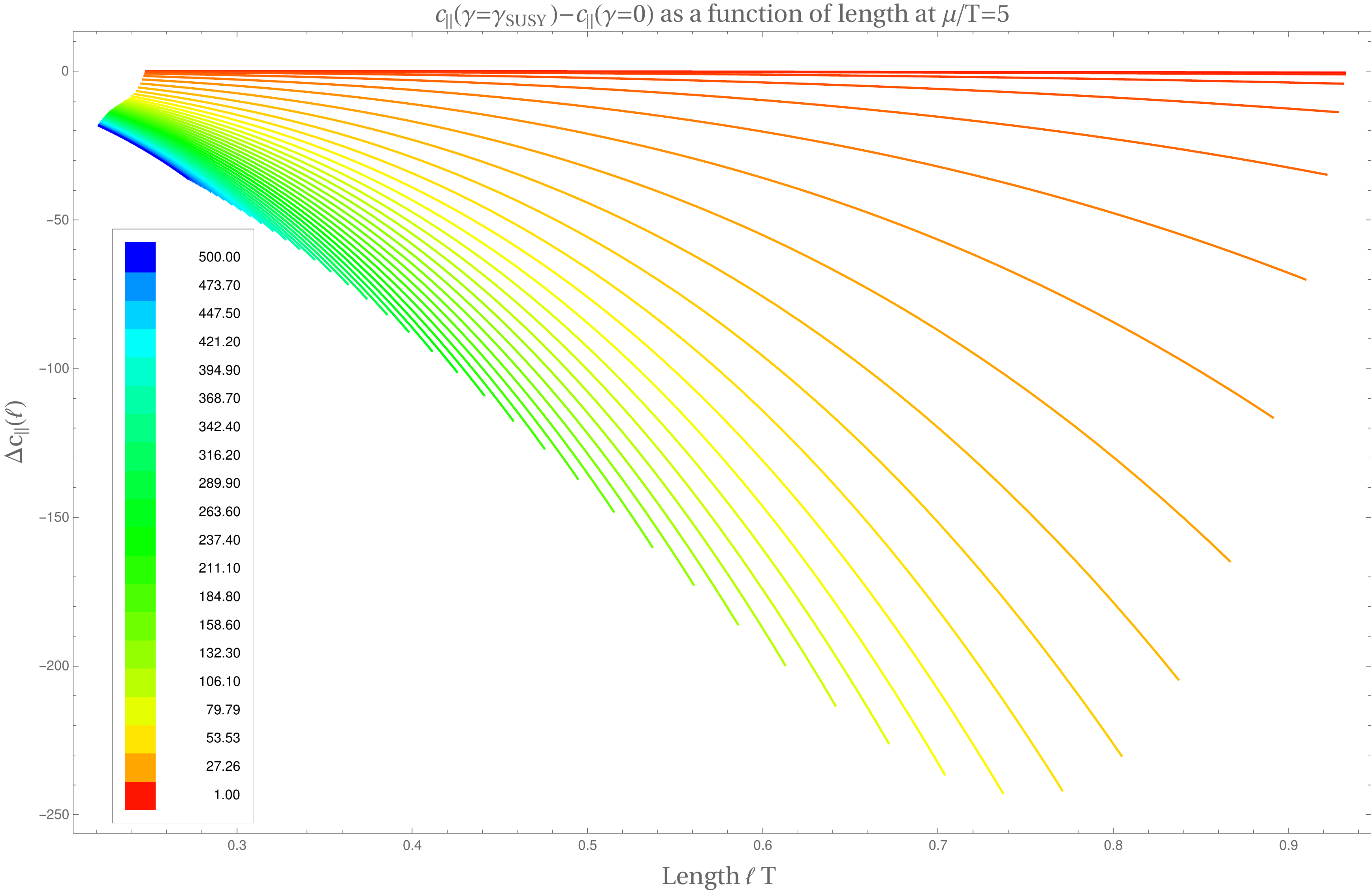}
  \end{center}
\end{subfigure}
   \caption{We display the difference between the c-functions, $\Delta c$ as computed with and without Chern-Simons coupling. \textit{Left:} $\Delta c_{\perp}$ \textit{Right:} $\Delta c_{||}$. In both images these are displayed as a function of dimensionless length $lB^{1/2}$ at fixed $\mu/T=5$ for $B/T^2$ varying from $\color{red}{B/T^2=1.00}$ to $\color{Blue}{B/T^2=500.00}$.
    \label{fig:delta_c_gamma}}
\end{figure}

\begin{figure}[htb]
  \begin{subfigure}[b]{.5\linewidth}
  \begin{center}
\includegraphics[width=2.9 in]{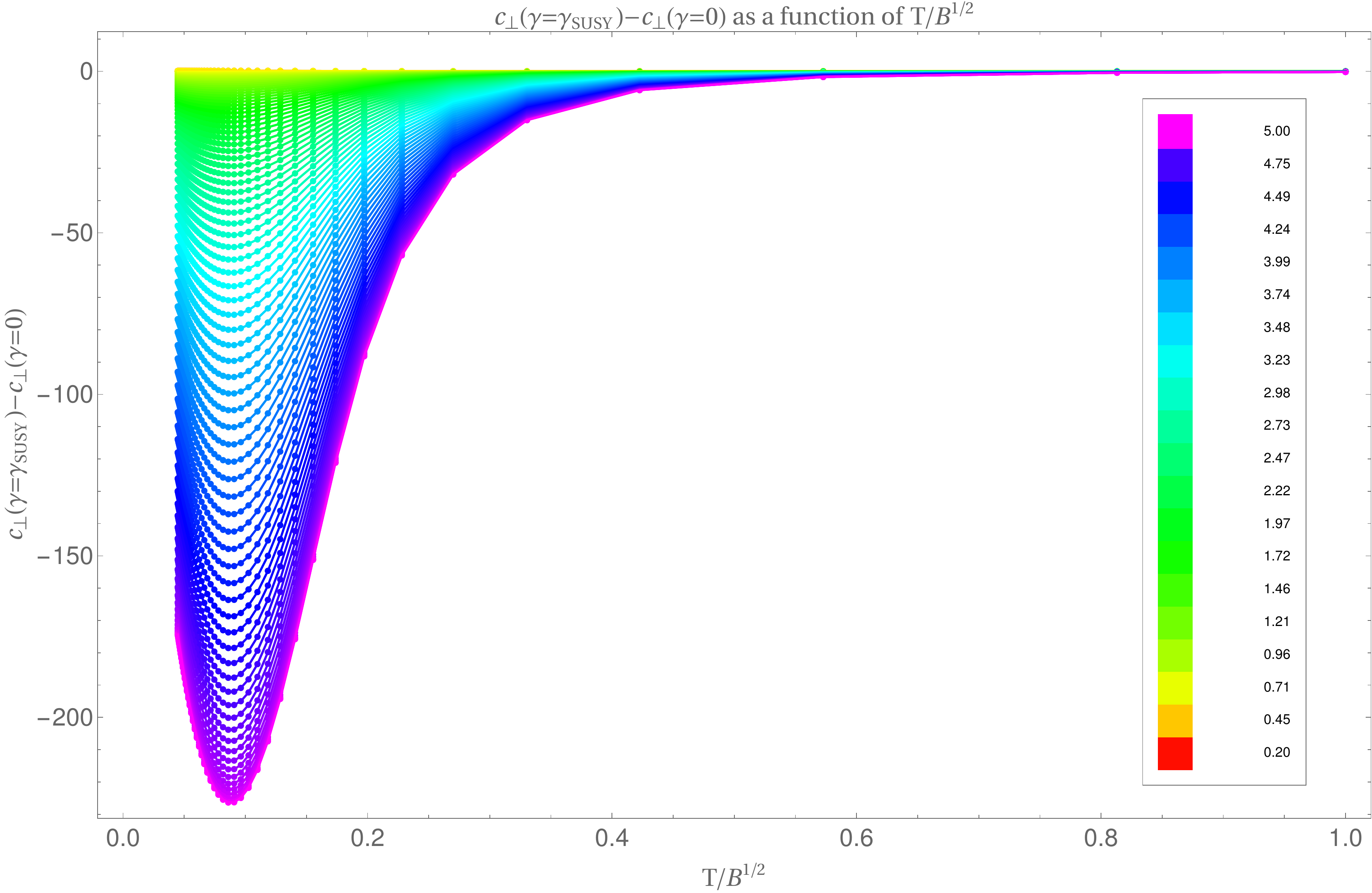}
\end{center}
\end{subfigure}
 \begin{subfigure}[b]{.5\linewidth}
  \begin{center}   
\includegraphics[width=2.9 in]{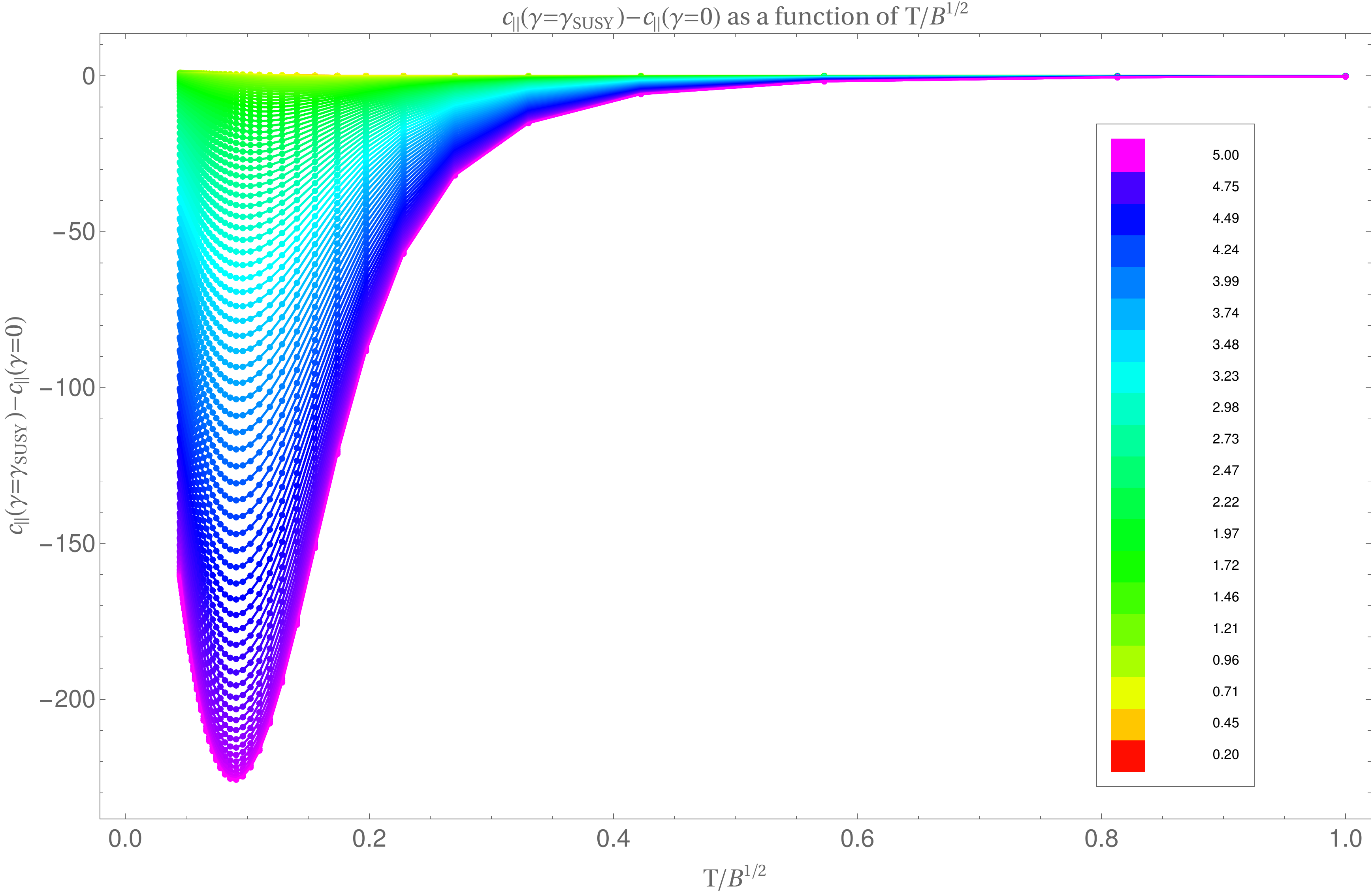}
  \end{center}
\end{subfigure}
   \caption{We display the difference between the c-functions, $\Delta c$ as computed with and without Chern-Simons coupling. \textit{Left:} $\Delta c_{\perp}$.  \textit{Right:} $\Delta c_{||}$. In both images these are displayed as a function of dimensionless Temperature $T/B^{1/2}$ at fixed $\mu/T=5$ for $\ell$ varying from $\color{red}{\ell=2/10}$ to $\color{Blue}{\ell=5}$. 
    \label{fig:delta_c_gammaTB}}
\end{figure}
\paragraph{c-function in $\mathcal{N}=4$ SYM theory in an anisotropic thermal state: }We begin by realizing that the equations for the entanglement entropies in the charged magnetic EMCS black brane can be cast in the form displayed in~\cite{Myers:2012ed,Chu:2019uoh}. We will start with the Lagrangian given in eq.~(\ref{eq:affine_parallel}) and follow closely~\cite{Myers:2012ed}. Let us focus on $L_{||}$ first in the coordinate parameterization
\begin{equation}
  L_{||} =\frac{1}{z(x)^3}\sqrt{v(z(x))^4 \left(\frac{z'(x)^2}{U(z(x))}+w(z(x))^2\right)} \, ,
\end{equation}
where we notice that the Lagrangian does not explicitly depend on the coordinate $x$, which implies there is a conserved quantity of motion. 
\begin{equation}
    H=p_z z'(x)-L_{||},\quad p_z=\frac{\partial L_{||}}{\partial z'(x)}.\label{eq:Conserved} 
\end{equation}
Following~\cite{Myers:2012ed} we choose to identify the constant of motion as $\kappa=1/H$ which is given by,
\begin{equation}
    \kappa=\frac{L z(x)^6}{v(z(x))^4 w(z(x))^2} \label{eq:kappa} \, .
\end{equation}
The c-function is then as given in eq.~(\ref{eq:c_function}),
\begin{equation}
    c_{||}=\frac{2\pi \ell^3}{l_P^3}\frac{\delta S}{\delta \ell}\, ,
\end{equation}
where we use $\delta$ rather then $\partial$ to emphasize the variational nature of this derivative. Given the action functional,
\begin{equation}
    S=\int_{0}^{\frac{\ell-\epsilon}{2}}dx L_{||}(z(x),z'(x);x) \, .
\end{equation}
It is a short variational exercise to show that,
\begin{equation}
    \frac{\delta S}{\delta \ell}=\frac{\partial L_{||}}{\partial z'}\frac{\partial z}{\partial \ell}+L_{||}\left(z(x),z'(x);\frac{\ell-\epsilon}{2}\right)\frac{\partial}{\partial l} \left(\frac{\ell-\epsilon}{2}\right) \, .
\end{equation}
Applying this equation to the case at hand we can find that,
\begin{equation}
   c_{||}= -\frac{2\pi \ell^3}{l_P^3}\frac{\partial_l z(x,\ell)}{\kappa  \partial_x z(x,\ell)} \label{eq:Pre_C} \, ,
\end{equation}
where we have used both the behavior of the metric near the conformal boundary and the fact that during the variational calculation we wish to keep the value of the embedding coordinate $z=z(x,\ell)$ fixed at the cutoff surface $z_c=\epsilon>0$, that is,
\begin{equation}
  \left.  \frac{\exd z_c}{\exd \ell}\right|_{x=\frac{\ell-\epsilon}{2}}=0,\quad \rightarrow\quad \frac{1-\partial_\ell\epsilon}{2}=-\frac{\partial_\ell z}{\partial_x z} \, .
\end{equation}
Finally we can evaluate the ratio of derivatives given in eq.~(\ref{eq:Pre_C}) by considering the constant of motion near the conformal boundary where to leading order $w=v=U=1$,
\begin{equation}
    \frac{\exd z}{\exd x}=-\frac{\sqrt{-z^6+\kappa^2}}{z^3} \, ,
\end{equation}
which can be directly integrated to a hypergeometric function,
\begin{equation}
    x-\ell/2=-\frac{z^4}{4\kappa^2} \,_2F_1\left(\frac{1}{2},\frac{2}{3},\frac{5}{3};\frac{z^6}{\kappa}\right).
\end{equation}
The solution can then be differentiated with respect to both $x$ and $l$ to construct the ratio of derivatives in the limit of $z\rightarrow 0$ in eq.~(\ref{eq:Pre_C}) for which we find,
\begin{equation}
    \frac{\partial_\ell z(x,\ell)}{ \partial_x z(x,\ell)} =-\frac{1}{2} \, .
\end{equation}
Finally this leaves us with, 
\begin{align}
    c_{||}&= \frac{\pi \ell^3\beta_4}{l_P^3}\frac{1}{\kappa(\zs )} \, , \label{eq:C} \\
      c_{\perp}&= \frac{\pi \ell^3\beta_4}{l_P^3}\frac{1}{\kappa_{\perp}(\zs )} \, ,\label{eq:Cperp}
\end{align}
where we have already noted that this same derivation goes through in the transverse case. We can now see that our result matches precisely the expression found in~\cite{Myers:2012ed,Chu:2019uoh}. 

\paragraph{Entropic nature of the c-function:} 
It is worth pointing out the similarity of this result as compared with the area associated with the event horizon,
\begin{align}
    A&= \int \exd^3 x \sqrt{-g_{\text{spatial}}}=\int\exd x_1\exd x_2\exd x_3 \frac{v(z_h)^2w(z_h)}{z_h^3} \, , \\
    A&=V \frac{v(z_h)^2w(z_h)}{z_h^3}, \quad V=\int\exd x_1\exd x_2\exd x_3\, .
\end{align}
To obtain an expression for $\kappa$ we can evaluate at any location $z$, the most useful location to evaluate this expression is at the turning point of surface, $\zs$, defined by $z(x_s)=\zs$ such that, $z'(x_s)=0$, 
\begin{equation}
    \frac{1}{\kappa}=\frac{v(\zs)^2w(\zs)}{\zs^3} \, .
\end{equation}
From this equation we can see the truly ``entropic'' nature of the c-function as defined via the entanglement entropy. This c-function essentially evaluates the area of a spatial hypersurface along the radial $AdS$ direction. 

\paragraph{Effect of the chiral anomaly on the c-functions:} 
The results for the c-functions computed in the charged magnetic black brane background are given in figures~\ref{fig:purenumreics_c}, \ref{fig:purenumreics_cSUSY}, \ref{fig:delta_c_gamma}, and \ref{fig:delta_c_gammaTB}. There, we display the c-functions computed in two ways, firstly via direct numerical calculation utilizing  eq.~(\ref{eq:c_function_para}) and eq.~(\ref{eq:c_function_perp}) (lines), and secondly via the conserved charge as given in eq.~(\ref{eq:C}) and eq.~(\ref{eq:Cperp}) (dots), and find excellent agreement up to expected numerical errors, see appendix~\ref{sec:appendix_convergence}. 

As expected, all c-functions asymptote to the central charge $a_4=2\pi^2$ in the UV-limit at small $\ell T$. The impact of the chiral anomaly is highlighted in figures~\ref{fig:delta_c_gamma} and \ref{fig:delta_c_gammaTB}. In both figures the difference between the c-function evaluated at supersymmetric value for the chiral anomaly ($\gamma=\gamma_{SUSY}$) and vanishing chiral anomaly ($\gamma=0$) is shown. Figure~\ref{fig:delta_c_gamma} shows that the impact of the anomaly on the c-function is much stronger in the IR, i.e.~at large $\ell T$. Interestingly, the effect of the anomaly is strongest at an intermediate magnetic field value, around $T/B^{1/2}\approx 0.087$ (at fixed $\mu/T$), as figure~\ref{fig:delta_c_gammaTB} shows. At smaller magnetic fields the effect dies off quickly and it also decreases towards larger magnetic fields. 
The maximal effect of the anomaly, i.e.~the extremum in figure~\ref{fig:delta_c_gammaTB} is shifted to smaller $T/B^{1/2}$ for larger $\mu/T$. For larger $\mu/T$ that peak is also narrowing, which means that an increasingly narrow range of 
temperature values (normalized to the magnetic field value) is affected by the anomaly. Specifically, for $\mu/T=1/5$, the extremum is located at $T/B^{1/2}=0.109$, while in figure~\ref{fig:delta_c_gammaTB} for $\mu/T=5$ it is located around $T/B^{1/2}\approx 0.087$. This may be the onset of quantum critical behavior, as this system possesses a quantum critical point~\cite{DHoker:2009mmn,DHoker:2009ixq,DHoker:2010zpp,Ammon:2016szz}. In a different system with a quantum critical point, the entanglement entropy was shown to get peaked near the critical point~\cite{Baggioli:2020}. 

In summary, the chiral anomaly affects the system differently at different energy scales. In the case we study here, all the energy scale dependence comes from the thermal state because the theory is conformal and has trivial RG-flow (its couplings do not run). This highlights the importance of the state for the c-function defined in eq.~\eqref{eq:c_function} (and~\eqref{eq:c_function_para}, \eqref{eq:c_function_perp}) which it inherits from the entanglement entropy.

\subsection{Monotonicity}
\begin{figure}[H]
    \begin{subfigure}[b]{0.5\textwidth}
    \includegraphics[width=7cm]{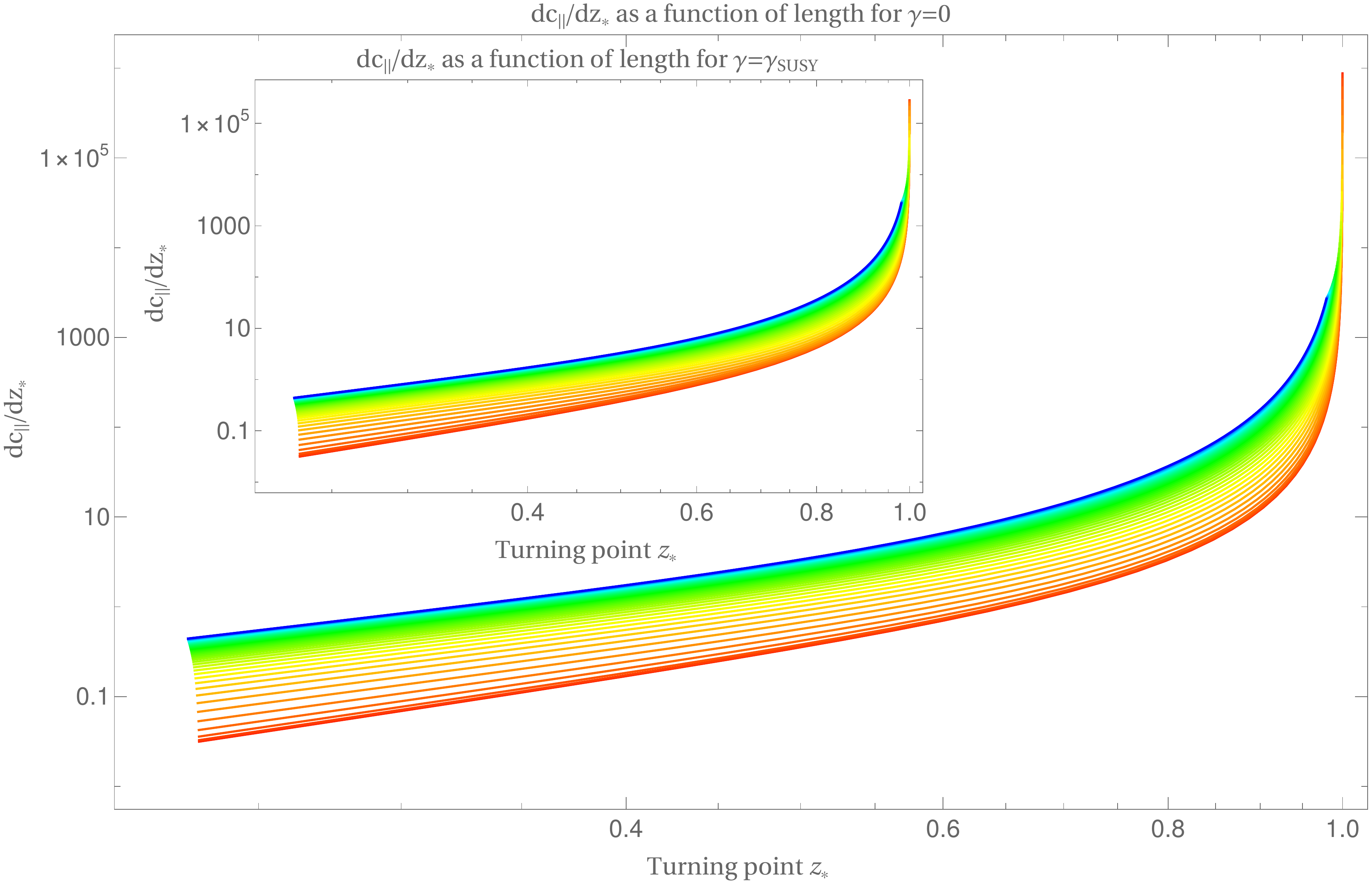}
\end{subfigure}
\begin{subfigure}[b]{0.5\textwidth}
    \includegraphics[width=7cm]{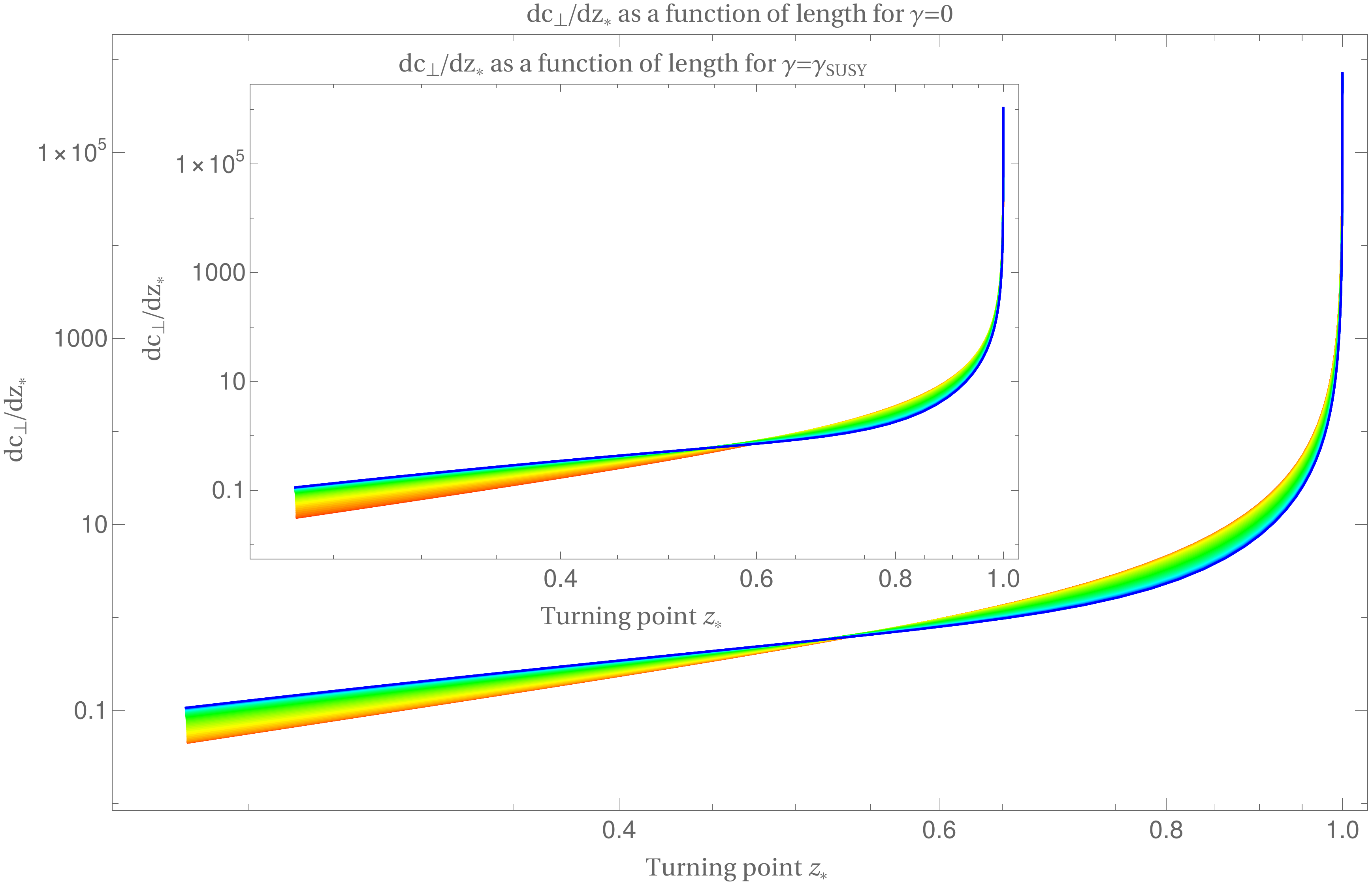}
    \end{subfigure}
    \caption{The slope of the c-function with respect to the bulk turning point $\zs$, which encodes the energy scale of the system. \textit{Left:} $dc_{||}/d\zs$. \textit{Right:} $dc{\perp}/d\zs$. In both images the inset graphic displays the same result at $\gamma=\gamma_{SUSY}$. In addition both images are displayed at $\mu/T=1/5$ for $B/T^2$ varying from $\color{red}{B/T^2=1.00}$ to $\color{Blue}{B/T^2=500.00}$. In both cases we see that $dc/d\zs\geq 0$ and hence $c_{||}$ and $c_\perp$ are  monotonic functions of $\ell$ or equivalently of $\zs$, whose monotonous relation is given in figure~\ref{fig:conjugate}. 
    \label{fig:Mono}} 
\end{figure}

To consider the monotonicity of our proposed c-function it is useful to consider the change of variables given in eq.~(\ref{eq:RG_Coordinates}) under which the line element becomes,
\begin{equation}
   \exd s^2 =-\exd t^2 e^{2 A(\beta )} A'(\beta )^2+e^{2 A_1(\beta )} \left(\exd x_1^2+\exd x_2^2\right)+e^{2 A_2(\beta )} (\exd t c(\beta )+\exd x_3)^2+\exd \beta^2\, , \label{eq:RG_Variables}
\end{equation}
where the similarity to~\cite{Freedman:1999gp,Myers:2010tj,Myers:2012ed} and in particular to the anisotropic case~\cite{Chu:2019uoh} is obvious. Furthermore under this change of variables $\kappa(\zs)$ becomes
\begin{equation}
    \kappa=e^{k(\beta_s)}, \quad k(\beta)=2 A_1(\beta)+A_2(\beta) \, ,
\end{equation}
where we use the shorthand $k_s=k(\beta_s)$. With this identity the c-function can be written as,
\begin{equation}
  c_{||,\perp}= \frac{\pi}{l_P^3}  e^{k(\beta_s)}\ell^{d_{||,\perp}}.
\end{equation}
Note that the energy-scale $\zs$ after this transformation is given by $\beta_s$. In order to determine if this c-function behaves monotonically with that energy scale, we now consider the derivative of this quantity with respect to the turning point $\beta_s$,
\begin{equation}
    \frac{\exd c_{||,\perp}}{\exd \beta_s}= e^{k(\beta_s)}\ell^{d_{||,\perp}-1}\left( \ell(\beta_s)\frac{\exd k}{d\beta_s}+d_{||,\perp} \frac{\exd \ell}{d\beta_s}\right) \, .
\end{equation}
Consider the length $\ell$ of the strip in terms of the bulk depth $\zs$,
\begin{equation}
    \frac{\ell}{2}=\int_0^{\frac{\ell}{2}}\exd x=\int_{\zs}^0\frac{\exd z}{z'(x)}=  \int_{\zs}^0\exd z   \frac{z(x)^3}{\sqrt{U(z(x))} w(z(x)) \sqrt{\kappa ^2 v(z(x))^4 w(z(x))^2-z(x)^6}}\label{eq:length_in_z} \, ,
\end{equation}
where in the final identity we have inverted the expression for $\kappa$ given in eq.~(\ref{eq:kappa}). While eq.~(\ref{eq:length_in_z}) does not look amenable to further analysis the change of variables made in eq.~(\ref{eq:RG_Variables}) significantly reduces these complications,
\begin{equation}
      \frac{\ell}{2}=\int_{\zs}^0\exd z   \frac{z(x)^3}{\sqrt{U(z(x))} w(z(x)) \sqrt{\kappa ^2 v(z(x))^4 w(z(x))^2-z(x)^6}}=\int_{\beta_s}^{\infty} \frac{\exd \beta e^{-A_2(\beta )+k_s}}{\sqrt{e^{2 k(\beta )}- e^{2 k_s}}}\label{eq:length_in_beta} \, ,
\end{equation}
with $k_s=k(\beta_s)$. Upon inspection we see we have arrived at an expression already derived in~\cite{Chu:2019uoh}. The transverse direction goes through identically to this with $A_2$ in the expression in eq.~(\ref{eq:length_in_beta}) replaced with $A_1$. Given that the two resulting expression are identical to those given in~\cite{Chu:2019uoh}, we will not repeat their derivation but simply quote their final result,
\begin{align}
   \frac{ \partial c_{\perp}}{\partial \beta} &= \frac{\beta_4}{4G_5} e^{k_s}\ell^{d_{\perp}-1}d_{\perp}\left(k'_m\int_0^\ell\exd x_1 \frac{1}{k'(\beta)}\left(\frac{k'(\beta)}{d_{\perp}}-A_1'(\beta)-\frac{k''(\beta)}{k'(\beta)}\right)\right)\label{eq:mono_perp}\, ,\\
   \frac{ \partial c_{||}}{\partial \beta} &=\frac{\beta_4}{4G_5} e^{k_s}\ell^{d_{||}-1}d_{||}\left(k'_m\int_0^\ell\exd x_1 \frac{1}{k'(\beta)}\left(\frac{k'(\beta)}{d_{||}}-A_2'(\beta)-\frac{k''(\beta)}{k'(\beta)}\right)\right)\label{eq:mono_para} \, .
\end{align}
What remains to be shown is that,
\begin{equation}
     \frac{ \partial c_{\perp}}{\partial \beta}\leq 0, \quad  \frac{ \partial c_{||}}{\partial \beta} \leq 0\, .
\end{equation}
While this is a simple calculation in the case where $\mu=B=\gamma=0$ (see appendix~\ref{sec:mono_schwarzschild}), we have to resort to numerical computations in order to verify our claim of monotonicity (when $\mu,B,\gamma>0$ are chosen for definiteness). 
The results of this analysis are displayed in figure~\ref{fig:Mono}. 

\section{Discussion}\label{sec:discussion}
%
In this work, we holographically computed entanglement entropies on strips. 
We first computed the effect of a chiral anomaly, charge, and magnetic field on entanglement entropies 
in $\mathcal{N}=4$ Super-Yang-Mills theory at strong coupling, see figures~\ref{fig:embedding} and~\ref{fig:EE_bt2}. 
From those entanglement entropies we computed c-functions defined by equations~\eqref{eq:c_function}, see figures~\ref{fig:purenumreics_cSUSY}, \ref{fig:purenumreics_c}, \ref{fig:delta_c_gamma} and~\ref{fig:delta_c_gammaTB}. Here we have extended the definition of anisotropic c-functions~\cite{Chu:2019uoh} to off-diagonal metrics and to theories including gauge fields, see~\eqref{eq:c_function_para} as well as \eqref{eq:c_function_perp}. 
Notably, these c-functions are evaluated in thermal states as opposed to vacuum states. We demonstrate that for the conformally invariant $\mathcal{N}=4$ SYM theory the c-function defined by~\eqref{eq:c_function},  \eqref{eq:c_function_para}, and  \eqref{eq:c_function_perp}, is monotonically increasing along the energy scale, parametrized by the width $\ell$ of the entanglement strip (or equivalently by the turning point $\zs$ of the geodesic, named $\beta_s$ in another coordinate system). Since the RG-flow of the theory is trivial, we may think of the total RG-flow as the RG-flow of only the thermal states that we consider. In the IR-limit (large $\ell$) the c-functions are proportional to the thermal entropy, see figure~\ref{fig:Thermal_c_subtraction}. 
We expect that these are generic features of conformal field theories in thermal states. 

More specifically, within Einstein-Maxwell-Chern-Simons theory we have shown analytically for thermal states corresponding to the Schwarzschild $AdS$ geometry that the c-functions are monotonically increasing from the ultraviolet towards the infrared regime, see appendix~\ref{sec:mono_schwarzschild}. 
This monotonic increase we also demonstrate numerically in the charged magnetic black brane case, see figure~\ref{fig:Mono}. 
Although not displayed here, we also confirmed that the pure Reissner-Nordstr\"om $AdS_5$ black brane leads to a monotonically increasing c-function. 
%
It is demonstrated analytically how the chiral anomaly, in form of the Chern-Simons coupling non-trivially affects the entanglement entropies in equations~\eqref{eq:Cpara2} and~\eqref{eq:Cperp2}, as well as the c-functions~\eqref{eq:c_function_para} and~\eqref{eq:c_function_perp}. These entanglement entropies and c-functions are thus sensitive to the gauge degrees of freedom. Compared to the study of entanglement entropies which did not depend on the chiral anomaly in~\cite{Nishioka:2015uka} we consider a different state that leads to the dependence on the chiral anomaly.  

While our analytic proof of principle was conducted in the UV limit (expansion in small $\ell$), the full numerical computation reveals that the full effect of the chiral anomaly on the entanglement entropies and c-functions is strongest at a particular intermediate value of the external magnetic field, $T/B^{1/2}\approx 0.1$ according to figure~\ref{fig:delta_c_gammaTB}. At lower temperatures (larger chemical potentials) the c-function becomes narrowly peaked.  
We tested  this statement numerically by computing figure~\ref{fig:delta_c_gammaTB} at lower temperature and observed a narrower peak. 
This may be the onset of quantum critical behavior present in this system, which  
requires a minimal non-zero Chern-Simons coupling~\cite{DHoker:2010zpp,DHoker:2009ixq}  and thus the chiral anomaly needs to be present. 
A peaking behavior of c-functions was observed in a different system~\cite{Baggioli:2020}, 
where they have been proposed as an indicator for quantum critical points in anisotropic systems~\cite{Baggioli:2020}. 
Interestingly, in our case the contribution of the chiral anomaly to the thermal entropy is not peaked, figure~\ref{fig:Spacetime_Comp_entropy}, while the c-function derived from the entanglement entropy is peaked near the critical point, figure~\ref{fig:delta_c_gammaTB}. 
Answering the question if our thermal c-functions may serve as order parameters for quantum phase transitions requires an 
exploration of the region around the quantum critical point. This is a numerically challenging project, as seen in~\cite{Ammon:2016szz}, which we leave for future work.  
We find that the chiral anomaly affects the IR more than the UV regime as shown in figure~\ref{fig:delta_c_gamma}.  
This indicates that a c-function, if taken as an order parameter for a quantum phase transition, may show a stronger effect when evaluated in the IR of the system, i.e.~by probing the system in low energy states, and close to the IR fixed point of the underlying QFT. This more pronounced impact in the IR is in agreement with~\cite{Baggioli:2020}. An exciting application would be the calculation of c-functions in the topologically non-trivial thermal states constructed within Einstein-Yang-Mills theory in~\cite{Cartwright:2020yoc}. 

%
For the c-functions~\eqref{eq:c_function}, \eqref{eq:c_function_para}, and~\eqref{eq:c_function_perp} (defined through the entanglement entropies) one is tempted to conjecture a c-theorem for conformal field theories in thermal states (in presence of a chiral anomaly, charge, as well as a magnetic field in anisotropic states). All thermal states we considered lead to a monotonous increase.  
This is in agreement with the expectation that in thermal equilibrium at fixed temperature, $T$, the states with energies, $E$, lower than the temperature, $E<T$ are occupied with higher probability than those with higher energy, as described by the Boltzmann (thermal) distribution function, $e^{-E/T}$.\footnote{Note that there is a multitude of micro-canonical states with the energy $E$, which is measured by the entropy, i.e.~the probability is multiplied by a multiplicity factor counting the number of states with that energy, $e^{S}$, the full partition function is given by $e^{S-E/T}$, and $S-E/T$ is minimized in thermal equilibrium. We thank M. Stephanov for discussion on this point.}   
Let us extend the interpretation of these c-functions from counting the degrees of freedom of the system, to counting the number of occupied states. Interpreting the c-function as a measure for the occupation of states, it is reasonable that the c-function is large in the IR, i.e.~at low energies, $E\ll T$, where the thermal Boltzmann probability for a state to be occupied is high. 

\begin{figure}[ht]
\centering
    \includegraphics[width=14cm]{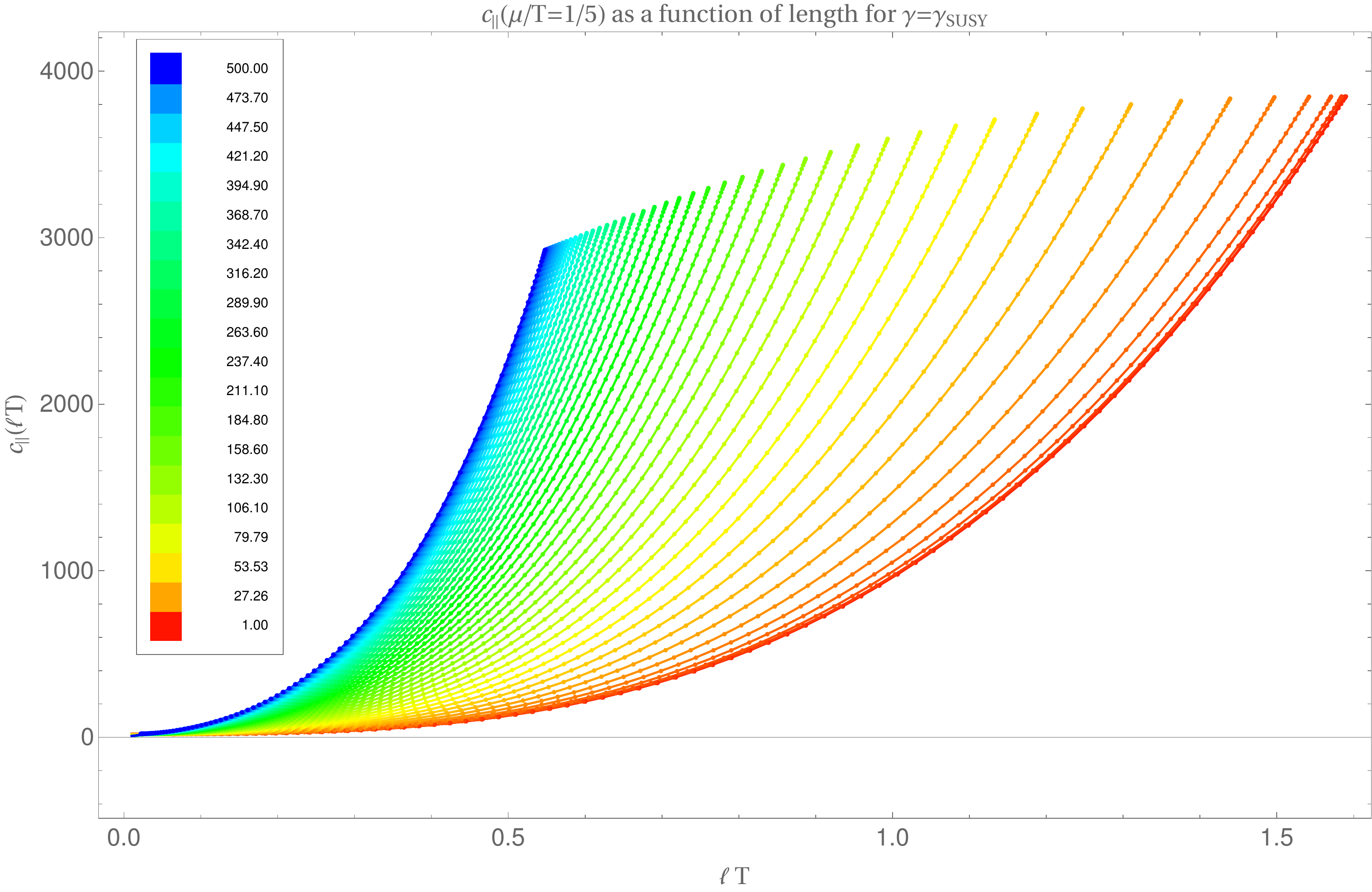}
\caption{
We display the $c_{||}$-function at $\gamma=\gamma_{SUSY}$ as dots along with lines indicating a cubic fit of the numerical data $c_{||}=\sum_{i=0}^3b_i (\ell T)^i$. The c-function is displayed as a function of dimensionless length scale $\ell T$ at fixed $\mu/T=1/5$ for $B/T^2$ varying from $\color{red}{B/T^2=1.00}$ to $\color{Blue}{B/T^2=500.00}$. The horizontal axis, $\ell \, T$, can be understood as RG - or energy scale of the system. 
\label{fig:fit_c_para_g}}
\end{figure}

We find the c-functions to be proportional to $(T\,\ell)^3$, as shown in 
figure~\ref{fig:fit_c_para_g}. 
This growth is in agreement with the volume-scaling expected for the entanglement entropy in thermal states~\cite{Rangamani:2016dms}, which predicts the entanglement entropy of a region ``$a$'' to be proportional to the volume of that region: $S_{||,\perp} \propto T^3 \, \text{Vol}(a) \propto T^3 \ell\, A_2$, where $A_2$ is the infinite area spanned by the other two spatial directions of the strip. 
This leads to the cubic scaling of the c-functions in the IR: 
$c_{||,\perp}\propto \ell^3 dS_{||,\perp}/d\ell \propto (\ell T)^3$. 

We also discussed the way in which the RG-flow of a quantum field theory and a system's state is encoded holographically in the corresponding geometry. Our data displayed in figure~\ref{fig:embedding} gives numerical evidence for  figure~\ref{fig:rgPicture} which displays schematically how the energy scale of the boundary system is encoded in the bulk geometry. The magnetic field scale $B/T^2$ (at fixed $\mu/T$) determines the charged magnetic black brane geometry. One may think of the magnetic field as setting the RG-scale of the theory and the energy scale of the corresponding state. Large $B/T^2$ pushes the transition from IR-like geometry (deformed $AdS_3\times R^2$) to UV-like geometry ($AdS_5$) towards the $AdS$-boundary in most of the metric components (with the exception discussed above). 
Now we introduce minimal surfaces as probes, indicated by thick grey lines in figure~\ref{fig:rgPicture}. They are dual to the entanglement entropy of a strip of width $\ell$ in the boundary theory. They probe either the UV or the IR or intermediate scales of the boundary theory, because they dive down into the bulk to a depth $z=z_*$ which depends on the value of $\ell$. The larger the entanglement width $\ell$, the deeper the minimal surfaces probe into the IR depths of the bulk. 
So, we can ask the question ``what do we probe at a fixed $z_*$?'' If we choose a large value of $B/T^2$ we probe IR-like behavior of the system, while at small $B/T^2$ and at the same $z_*$ we probe more UV-like behavior of the system. In this sense, we may think of $B/T^2$ as ``setting the RG-scale of the system''. Here we stress again that the {\it system} is comprised of the theory and a particular state. The conformal theory has trivial RG-flow, while the state has non-trivial energy scale dependence encoded in the metric components as functions of $z$. 
The CFT and the anomaly are the same at all energy scales, however, the states are influenced differently at different energy scales, and the anomaly affects the states differently at different energy scales.

Our c-functions of a CFT in thermal states monotonically increase towards the IR. The same c-functions had been shown to decrease along the RG-flow in vacuum states~\cite{Myers:2012ed} when the CFT is deformed away from its UV fixed point. Evaluating such an RG-flow away from a CFT in a thermal state, we expect a competition between the increasing and decreasing behavior in the c-functions. 
Computing the c-functions in those systems will likely show that~\eqref{eq:c_function}, \eqref{eq:c_function_para}, and \eqref{eq:c_function_perp} are {\it not} monotonic. Instead, we expect them to have maxima and minima along the energy scale of the theory, similar to the quantities discussed in~\cite{Gursoy:2018umf}. 
Venturing out into the theory space of non-conformal deformations of CFTs would likely entail a metric such as those presented in~\cite{Gursoy:2018umf}, which simultaneously features a domain-wall and a blackening factor. With such a setup one could truly study the effect of the chiral anomaly on the behavior of the c-function  as it flows from one fixed point to another as indicated in figure~\ref{fig:rgFlow}.  
We stress that these c-functions are still highly restricted and thus useful measures, although monotonicity may not be expected to hold in general. This is because entropic c-functions inherit all properties from the entanglement entropies they are derived from.

From a statistical physics point of view it is rather natural to consider the occupation number of thermal states along a certain energy scale, and to consider an RG-flow. 
Consider a conformal field theory (CFT) which describes a given system at a thermal phase transition. At the phase transition the system is in a thermal state in which the occupation of states at all energy scales are equally probable. 
Its c-function in this state is a nonzero constant with the value of the central charge of that CFT. 
If we now consider the same system slightly away from the phase transition by tuning the order parameter, we add operators to that conformal field theory and in this sense an RG-flow is induced in this thermal system.  We propose that the c-function defined from entanglement  entropy is a good measure for the occupation of states of particular energies in a system at fixed temperature. 
As such, it counts degrees of freedom of the underlying QFT as well as the occupation number at a particular energy for a given state. 

Quantum systems exposed to magnetic fields display physics succinctly described by Landau level physics. In a small region of the spacetime the magnetic field lines single out a particular direction, the $x_3$-direction in our model, and leave an $SO(2)$ invariance about this axis. This effectively reduces the dimensions of the theory to (2+1)-dimensional quantum theory. Charged particles are forced onto circular orbits around the magnetic field, filling Landau levels. In recent work~\cite{Ammon:2020rvg} it has been shown that the ratio $\eta_\perp/s$ has same $B$-dependence as $\eta_{||}/s$. That is to say entropy produced in shear processes per degree of freedom is identical in both directions. These results demonstrate the impact, or lack there of, of Landau physics on shear transport phenomena of a quantum system in a strong magnetic field. It is of course then interesting to understand what Landau physics implies about the entanglement spectrum as this is clearly a property of the state itself. It would be natural to expect that the entanglement entropy within large entangling regions behaves, as a function of the magnetic field, in the same way as the thermal entropy. It is in the small regions where we may expect a significant impact on the behavior of the entanglement entropy. Here it is likely to be simpler to work with discs as our entangling region. 
Therefore it is unclear how a strip-like partition as we use in this work, would effect the resulting entanglement entropy. However it is natural to consider an extension of our work in which we take a disc as our entangling region as we consider a quantum Hall effect, a response of the system in an external magnetic field to electric field or charge gradients. 
The seminal work of~\cite{Kitaev:2005dm} then demonstrates that the entanglement entropy should be given as $S=\alpha L-\log(\mathcal{D})+\mathcal{O}(L^{-1})$ where $L$ is the perimeter of the disc, $\mathcal{D}$ is the total quantum dimension and $\alpha$ is a non-universal constant, see also~\cite{Srednicki:1993im,Levin:2006}. In the case of a Laughlin state~\cite{Laughlin:1983fy}, with filling fraction $\nu=1/q$, the quantum dimension is $\mathcal{D}=\sqrt{q}$. Hence for an integer quantum Hall system $\mathcal{D}=1$ and $S=\alpha L+\mathcal{O}(L^{-1})$ and the universal term, $\gamma=\log(\mathcal{D})$, known as the topological entanglement entropy, vanishes. It is not clear if the same formulation of the c-theorem holds, that is $c\approx L^d \partial_L S$, or what effect the temperature has, or how exactly to think of interacting systems such as fractional quantum Hall states~\cite{Hasque:2007,Fujita:2009kw}. We leave these points to be investigated in the future. 

Certain changes to a system can be viewed either as a deformation of the theory by adding an operator to the Lagrangian, or as a modification of the boundary conditions~\cite{Landsteiner:2011tf}. For example a $U(1)$ charge may be introduced in either of those two ways. It is then interesting to ask the question how this affects the c-functions and how this ambiguity is reflected holographically in the geometry. 

There has been increasing experimental interest in the entanglement entropy across a quantum critical point or quantum phase transition, discussing its use as a sort of order parameter, see for example~\cite{Knaute:2017lll,Vidal:2002rm,Rougemont:2017tlu,Liu:2019npm}; and in the case of the QCD deconfinement  transition~\cite{Nishioka:2006gr,Klebanov:2007ws}.  
Experimentally, the material $Sr_3Ru_2O_7$ has a quantum phase transition, its thermal entropy peaks near the quantum critical point~\cite{Rost1360}, and there are proposals for the direct measurement of entanglement entropy~\cite{Abanin:2012dem,Islam2015}. 
Combining these concepts, an interesting future avenue may lead to the direct measurement of the thermal c-functions which we have discussed here. 

Surveying the literature one finds there is no unique definition of a holographic c-function, see for examples~\cite{Myers:2010tj,Freedman:1999gp,Sahakian:1999bd,Banerjee:2015coc,Paulos:2011zu}. 
While we have chosen here to study the entanglement entropy as a suitable definition, this begs the question of whether we have chosen an appropriate definition of the c-function in this system. An even more fundamental question would be under what assumptions or conditions do these definitions agree or disagree? Specifically, is the central charge as calculated via the entanglement entropy in the UV equal to the geometric measure as given in~\cite{Freedman:1999gp} or as calculated via the expansion of a congruence of null geodesics~\cite{Banerjee:2015coc}? And do all of these measures correctly asymptote to the central charge $a_4$ from eq.~\eqref{eq:traceT} near fixed points? 
Perhaps we have not considered computing the entanglement entropy on the adequate entanglement region in the dual theory. Recent work points to spherical entangling regions being the adequate shape for the entanglement region with which we can unify the $c$-, $a$- and $F$-theorems~\cite{Casini:2017vbe}. It would be an interesting exercise to consider, instead of a strip-like region, a spherical region in the dual theory. While the bulk spacetime is anisotropic, the boundary metric is fixed to the homogenous, isotropic, with Minkowski metric. The spherical boundary region would be deformed as one moves deeper into the bulk leading to a set of partial differential equations for the embedding. This procedure is a numerically challenging exercise which we leave open as a future task.

Entanglement entropy and c-functions are going to be very useful for tracking the (re)arrangement of degrees of freedom in systems out of equilibrium. Both quantities are well-defined out of equilibrium in contrast to thermal entropy which is well-defined only in thermal equilibrium. For systems out of equilibrium such as the quark-gluon-plasma (QGP) the rate of entropy production and the relevant mechanisms of entropy production are currently under investigation. In holographic models of Bjorken-expanding QGP one may utilize the apparent horizon area as a crude measure for entropy in the dual field theory~\cite{Chesler:2008hg,Fuini:2015hba}. However, since entropy is not well defined out of thermal equilibrium, entanglement entropy and a c-function may be a better measure for the time-evolution of degrees of freedom. In particular, the entanglement entropy can be used to probe degrees of freedom at any energy scale of the theory, from the UV into the IR, or equivalently any length scale, $\ell$. 
This way the entanglement entropies or c-functions may allow conclusions on the thermalization mechanism in a given system being bottom-up, i.e.~if the IR scales thermalize before the UV scales~\cite{Baier:2000sb}, or not. 
In hindsight, this provides a strong motivation for our study of the entanglement entropy and c-functions as functions of the energy scale of the system in this work.

\acknowledgments
We thank C.~Ecker and J.~Ingram for very helpful discussions during the early stages of this project. 
We are very grateful to H.~Casini, K.~Landsteiner, T.~Nishioka, and T.~Takayanagi for very helpful comments on the manuscript. 
This work was supported, in part, by the U. S. Department of Energy grant DE-SC-0012447.

\appendix
\section{Details of the holographic model}
\subsection{Equations of motion}
\label{sec:eom}
The equations of motion are as follows:
\begin{align}
0&=c'(z) \left(\frac{2 v'(z)}{v(z)}+\frac{3 w'(z)}{w(z)}-\frac{3}{z}\right)+c''(z)-\frac{z^2 p'(z) \left(c(z) p'(z)+E'(z)\right)}{w(z)^2} \, , \\
0&=\frac{c(z) \left(w(z) \left(c(z) p'(z)+E'(z)\right) \left(v(z)^2 w(z) c'(z)+\gamma  z B\right)-v(z)^2 p'(z) U'(z)\right)}{U(z)v(z)^2} \nonumber\\
&+ \left(c'(z) p'(z)+\frac{w'(z) \left(2 c(z) p'(z)+E'(z)\right)}{w(z)}-\frac{E'(z)}{z}+E''(z)\right)\nonumber\\
&+\frac{2 v(z) E'(z) v'(z)-\frac{\gamma  z B p'(z)}{w(z)}}{v(z)^2} \, ,\\
0&=-\frac{w(z) \left(c(z) p'(z)+E'(z)\right) \left(v(z)^2 w(z) c'(z)+\gamma  z B\right)}{U(z) v(z)^2}+\frac{p'(z) U'(z)}{U(z)}\nonumber\\
&+p'(z) \left(\frac{2 v'(z)}{v(z)}-\frac{w'(z)}{w(z)}-\frac{1}{z}\right)+p''(z) \, , \\
0&=-\frac{v(z)^4 \left(w(z)^2 \left(z \left(2 z^3 \left(c(z) p'(z)+E'(z)\right)^2+15 U'(z)-3 z U''(z)\right)-24 U(z)+24\right)\right)}{3 z^2 v(z)^4w(z)^2}\nonumber\\
&+\frac{z^4 U(z) p'(z)^2-3 z w(z) \left(z U'(z)-2 U(z)\right) w'(z)}{3 z^2 v(z)^4w(z)^2}- w(z)^2 c'(z)^2\nonumber\\
&+\frac{-6 z v(z)^3 \left(z U'(z)-2 U(z)\right) v'(z)+z^4 B^2}{3 z^2 v(z)^4} \, ,\\
0&=\frac{v(z)^4 \left(z^4 \left(c(z) p'(z)+E'(z)\right)^2-6 z U'(z)-24\right)+6 z^2 v(z)^3 U'(z) v'(z)+2 z^4 B^2}{6 z^2U(z) v(z)^3}\nonumber\\
&+\frac{v(z) \left(24-\frac{z^4 p'(z)^2+6 z w(z) w'(z)}{w(z)^2}\right)+6 z \left(v'(z) \left(\frac{z w'(z)}{w(z)}-5\right)+z v''(z)\right)+\frac{6 z^2 v'(z)^2}{v(z)}}{6 z^2} \, , \\
0&=\frac{v(z)^4 w(z) \left(3 z^2 w(z)^3 c'(z)^2+w(z) \left(z^4 \left(c(z) p'(z)+E'(z)\right)^2-6 z U'(z)-24\right)\right)}{6 z^2 U(z) v(z)^4 w(z)}\nonumber \\
&+\frac{U'(z) w'(z)}{ U(z) v(z)^4 w(z)}+\frac{ v(z) \left(z^4 p'(z)^2+3 w(z) \left(z \left(z w''(z)-4 w'(z)\right)+4 w(z)\right)\right)}{3 z^2 w(z)} \nonumber\\
&+\frac{2 v'(z) \left(z w'(z)-w(z)\right)}{ z  v(z)}+\frac{-z^4 B^2 w(z)}{6 z^2 U(z) v(z)^4 } \, .
\end{align}

\subsection{More on boundary and horizon geometries}
\label{sec:geometries}
Charged magnetic black branes, uncharged magnetic black branes, non-magnetic charged black branes, and non-magnetic uncharged black branes all asymptote to specific geometries near the horizon and near the boundary. 

\noindent \textbf{Near horizon} $\mathbf{AdS_2\times \mathbb{R}^3}$ \textbf{- vanishing charge and magnetic field:} This simple solution is the well known $AdS$-Schwarzschild solution. In our choice of parameterization this is given by
\begin{equation}
    U=-m z^4+1,\quad v=w=1 \, , 
\end{equation}
with all other components of the metric and gauge field vanishing.  \\

\noindent \textbf{Near horizon} $\mathbf{AdS_2\times \mathbb{R}^3}$ \textbf{- vanishing magnetic field:} The simple solution is the well known $AdS$-Reissner-Nordstr\"om solution. In our choice of parameterization this is given by
\begin{equation}
    U=-m z^4+\frac{q^2 z^6}{12}+1,\quad v=w=1, \quad E(z)= \mu+\frac{z^2 q}{2} \, ,
\end{equation}
with all other components of the metric and gauge field vanishing.

\paragraph{\textbf{Near horizon} $\mathbf{AdS_3\times \mathbb{R}^2}$ \textbf{-vanishing charge:}} There exists an analytic solution to the system as obtained in~\cite{DHoker:2009ixq}. In our choice of parameterization these can be represented as
\begin{equation}
    U=3+z_h z^2,\quad w=1, \quad v=\frac{z \sqrt{B}}{\sqrt{2} 3^{1/4}} \, ,
\end{equation}
with all other components of the gauge field and metric vanishing.  

\paragraph{\textbf{Warped} $\mathbf{AdS_3}$\textbf{:}} This special solution only occurs for a particular value of the Chern-Simons coupling $\gamma < \gamma_{susy}$ at $\gamma=\pm 1$. In our choice of parameterization this is given by
\begin{equation}
    U=\frac{\left(q^2+2 \left(B^2+6\right)\right) (z-z_{-}) (z-z_{+})}{3 z_{-} z_{+}},\quad v=w=z, \quad E(z)= \mu -\frac{q}{z} \quad c=c_2+\frac{B}{z} \, ,
\end{equation}
where $z_-,z_+$ are the inner and outer horizon locations and all other components of the metric and gauge field are vanishing. In addition this particular solution only exists provided that the magnetic field and the charge density satisfy,
\begin{equation}
    q^2+2 B^2-24=0 \, .
\end{equation}

\begin{figure}[H]
    \centering
    \includegraphics[width=14cm]{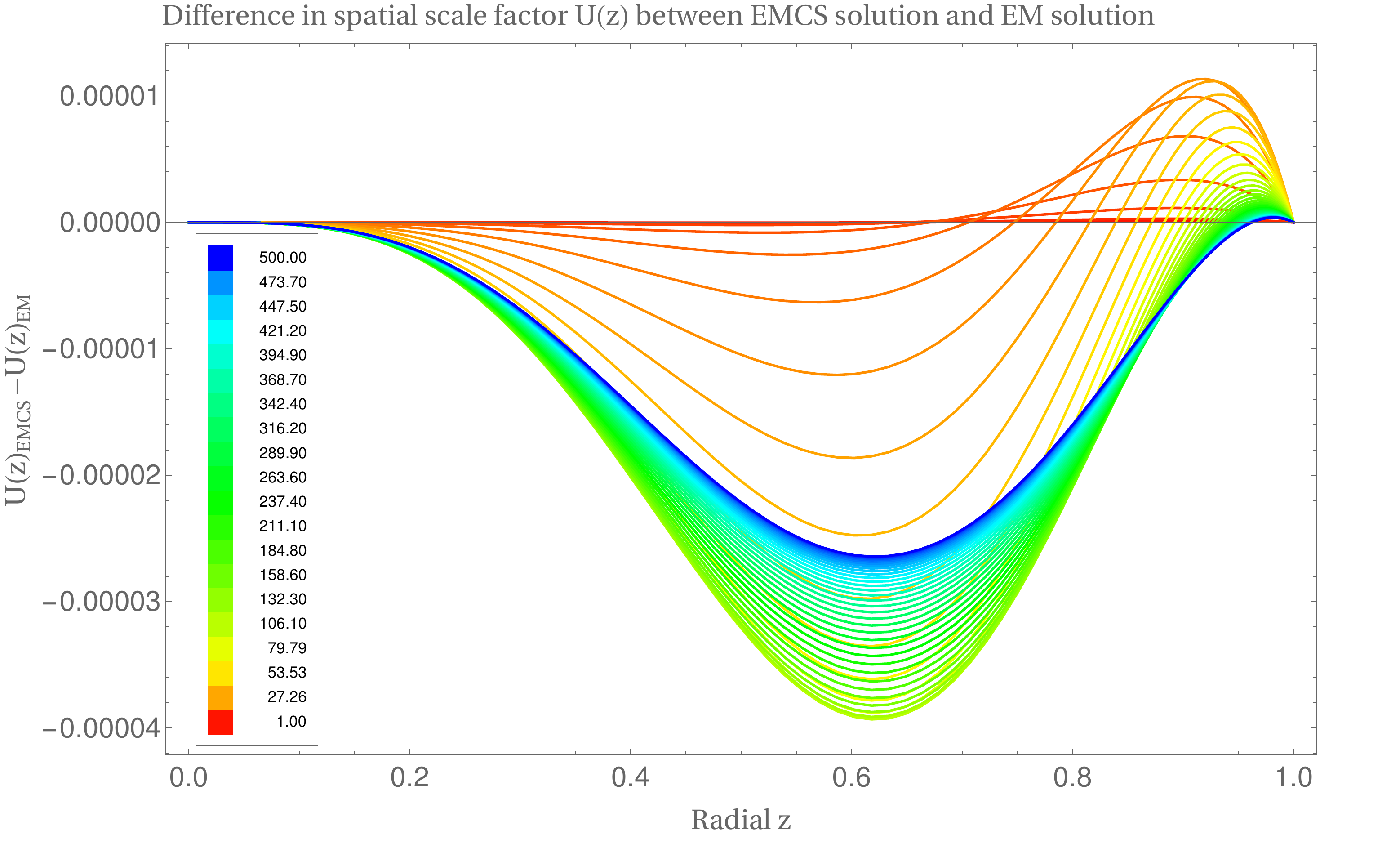}
    \caption{We display the difference between the blackening factor for full solutions to Einstein-Maxwell-Chern-Simons theory and solutions to the Einstein-Maxwell theory. The inset legend displays the coloring for different values of $B/T^2$. The difference between these spacetimes can be seen to be very large deep in the bulk as we move towards $B/T^2\gg 1$. 
    \label{fig:Spacetime_CompEM}}
\end{figure}

\begin{figure}[H]
    \begin{subfigure}[b]{0.5\textwidth}
    \includegraphics[width=75mm,scale=0.5]{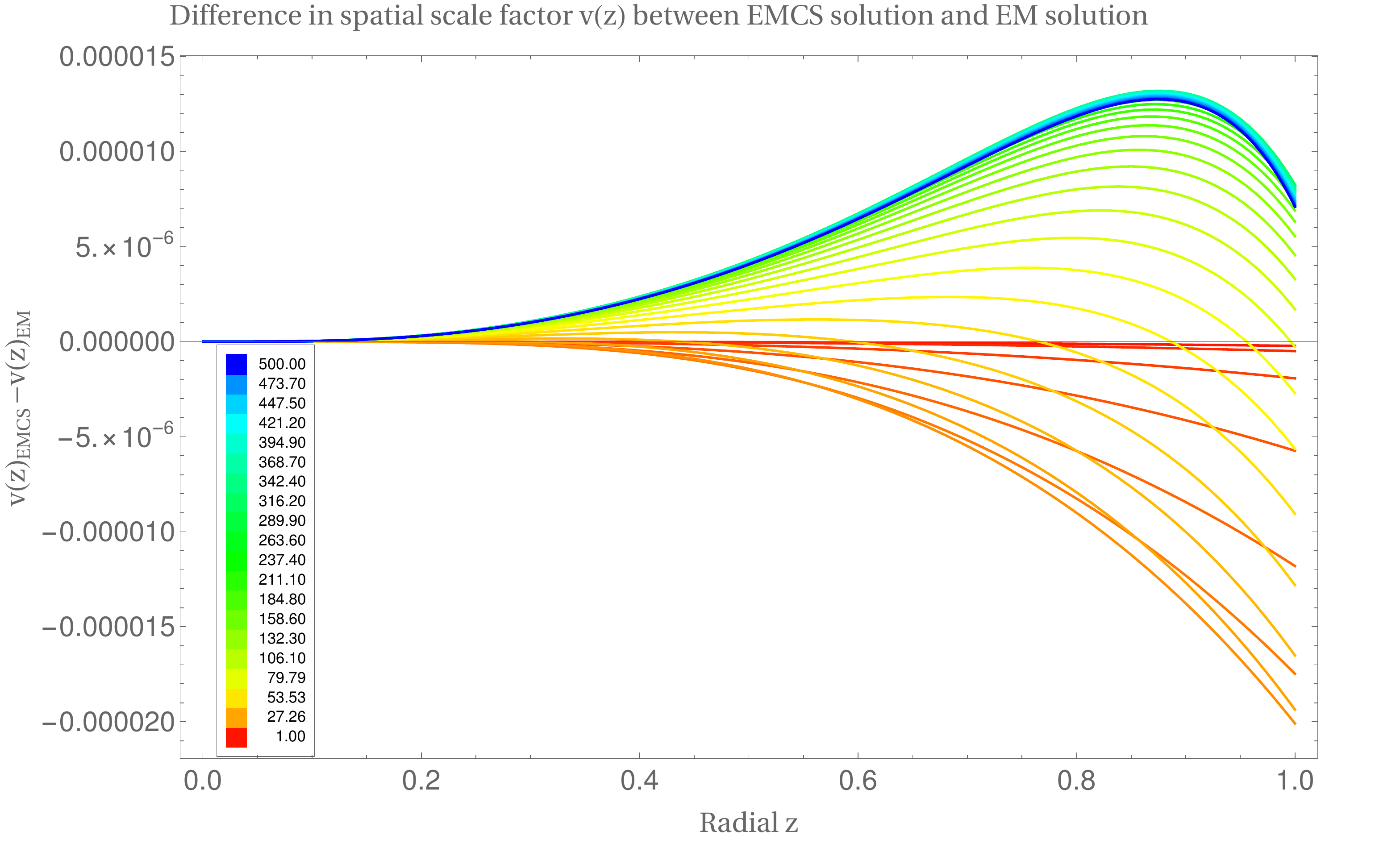}
 \end{subfigure}
 \begin{subfigure}[b]{0.5\textwidth}
    \includegraphics[width=75mm,scale=0.5]{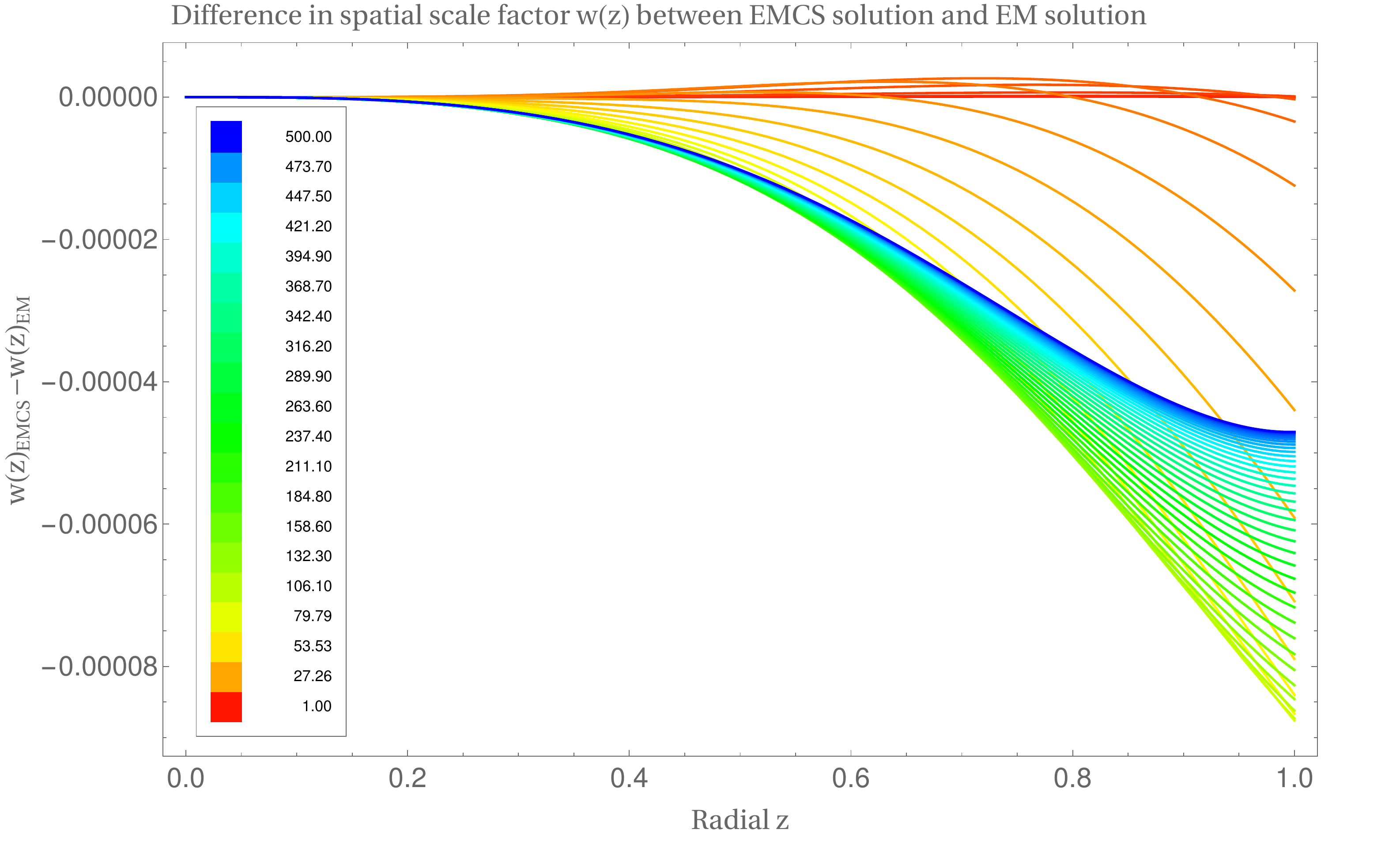}
    \end{subfigure}
    \caption{We display the difference between the spatial scale factors $v(z)$ and $w(z)$ for full solutions to Einstein-Maxwell-Chern-Simons theory and those for Einstein-Maxwell theory. The inset legend displays the coloring for different values of $B/T^2$. The difference between these spacetimes can be seen to be very large deep in the bulk as we move towards $B/T^2\gg 1$. 
    \label{fig:scalingEM}}
    \end{figure}

\begin{figure}[H]
    \centering
    \includegraphics[width=14cm]{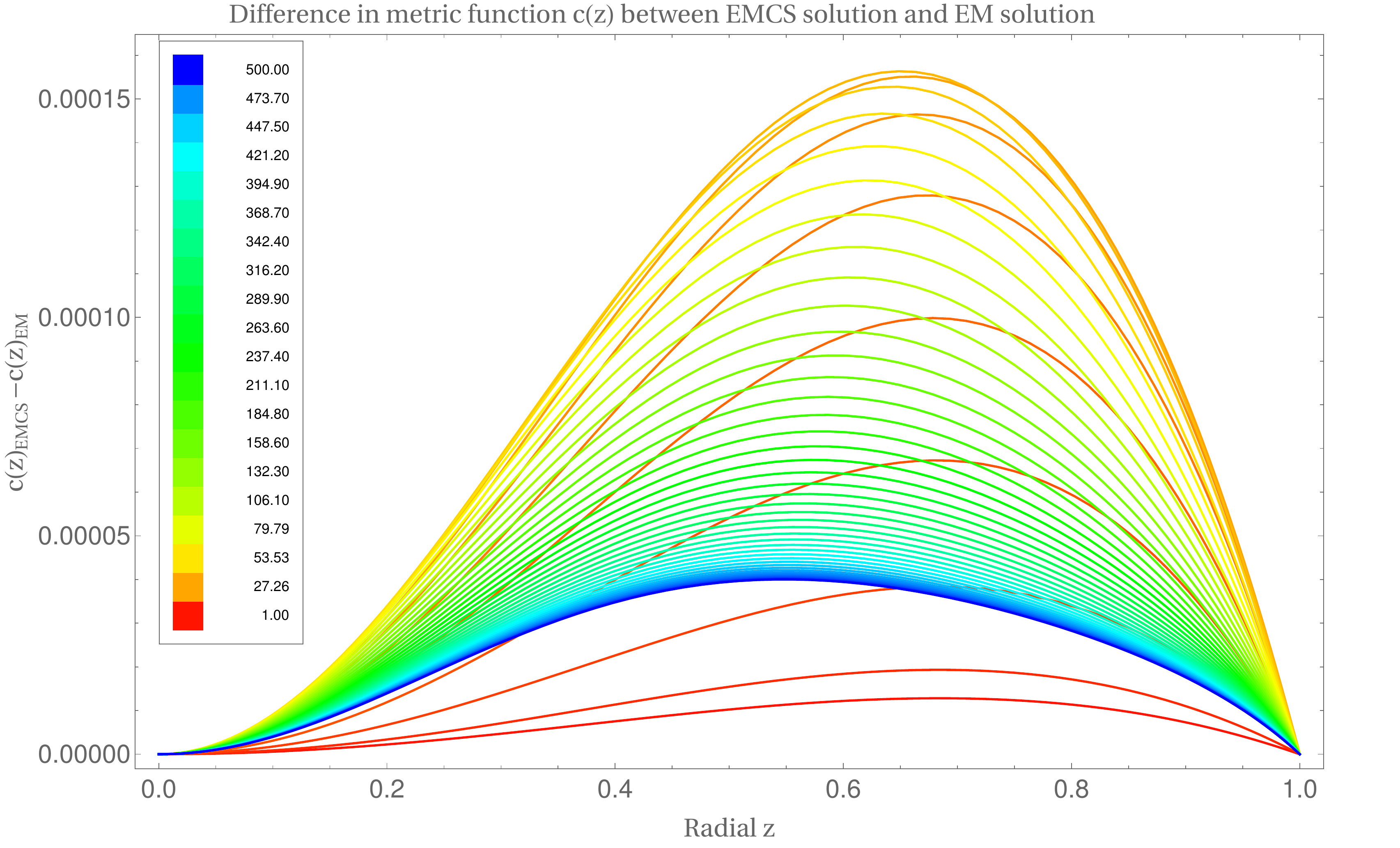}
    \caption{We display the difference between the metric function  $c(z)$ for full solutions to Einstein-Maxwell-Chern-Simons theory and the Einstein-Maxwell theory.  The inset legend displays the coloring for different values of $B/T^2$.
    \label{fig:Spacetime_Comp_C}}
\end{figure}

\paragraph{Comparison of EMCS ($\gamma=2/\sqrt{3}$) to EM ($\gamma=0$)}  
The effect of the Chern-Simons (CS) term can be seen clearly when comparing charged magnetic solutions to the Einstein-Maxwell (EM) system to the charged magnetic solutions of the EMCS system. See figures~\ref{fig:Spacetime_CompEM} and~\ref{fig:scalingEM}. 
In the EMCS solution there is an additional metric function, $c(z)$, which is not present in the EM- or RN-solution. We display this function in figure~\ref{fig:Spacetime_Comp_C} where we have taken $c_{EM}(z)=0$. 

\subsection{Coordinate parameterization}\label{sec:appendix_Coordinate}
\begin{figure}[htb]
    \begin{subfigure}[b]{0.5\textwidth}
    \includegraphics[width=75mm,scale=0.5]{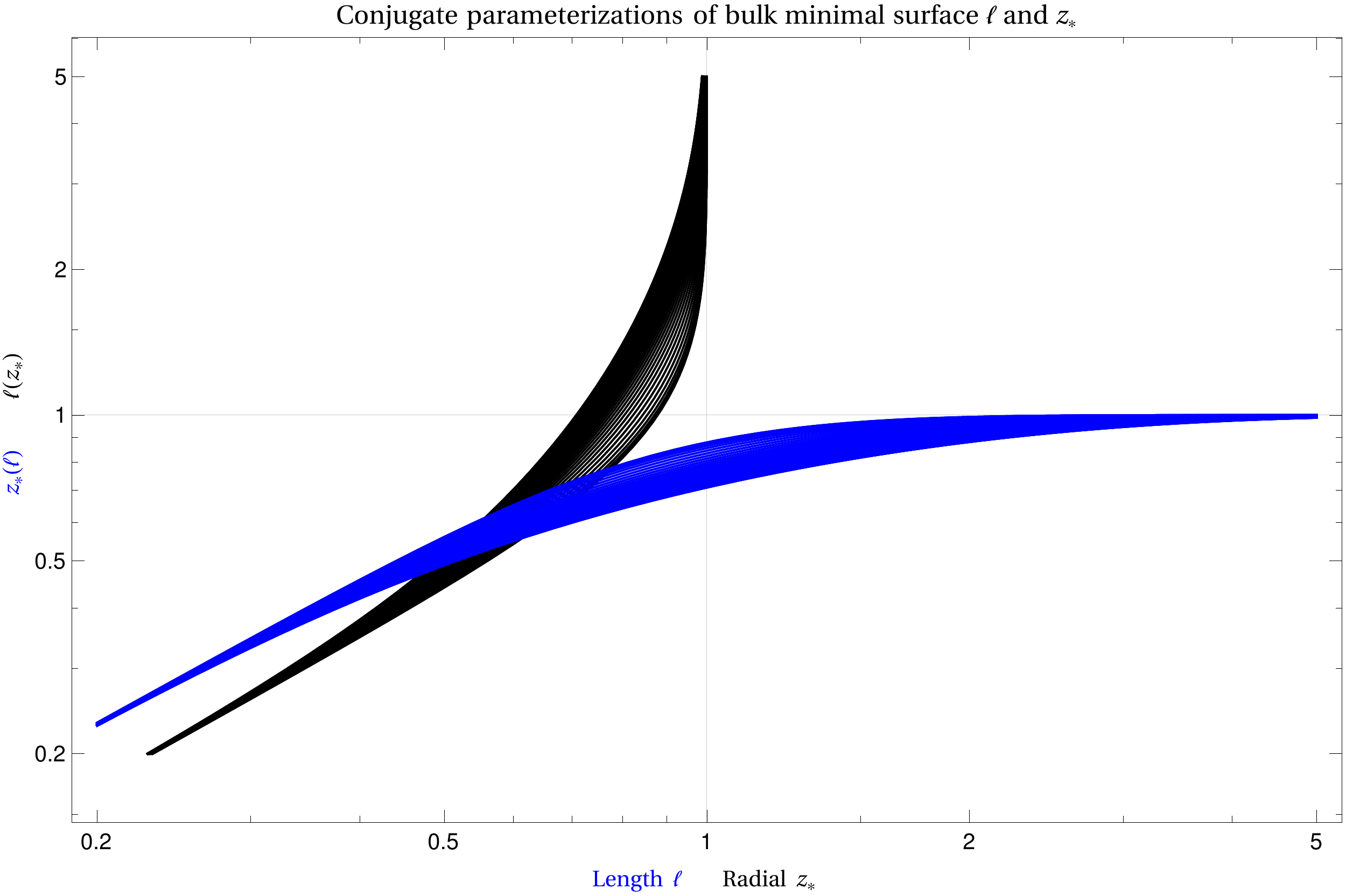}
 \end{subfigure}
 \begin{subfigure}[b]{0.5\textwidth}
    \includegraphics[width=75mm,scale=0.5]{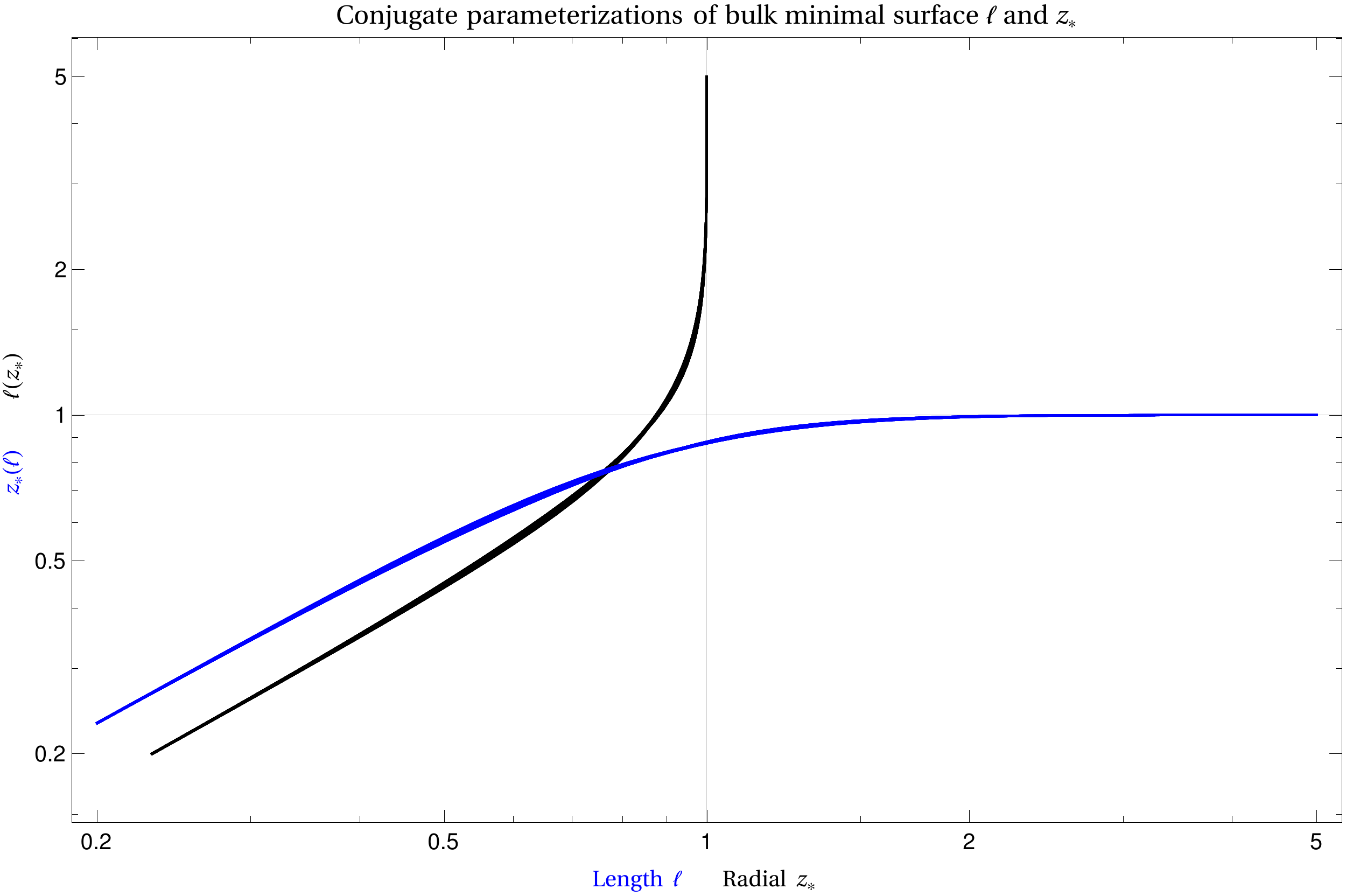}
    \end{subfigure}
    \caption{The conjugate variables $\ell$ and $\zs$ are displayed for bulk minimal surfaces with $\gamma=0$ for a range of magnetic field values.  \textit{Left:} Parallel strips with embedding coordinates  $(z(\sigma),t(\sigma),x_3(\sigma))$. \textit{Right:} Transverse strips with embedding coordinates $(z(\sigma),t(\sigma),x_1(\sigma))$. Despite displaying the figure on a log-log scale the deviations due to change in magnetic field values is vary small for strips in the transverse directions. While noticeable variation occurs for strips oriented in the parallel direction. 
    \label{fig:conjugate}}
\end{figure}

In this section we collect a separate method of calculating the entanglement entropy which is also popular in the literature. Here we directly compute the entanglement entropy in terms of the bulk radial coordinate $z$, or the field theory coordinates $x_1,x_3$. Removing the affine parameter we can express the integral in eq.~(\ref{eq:affine_parallel}) as
\begin{align}\label{eq:Coor_EE}
    S_{\perp}&=\frac{1}{2G_5} V_{\perp}\int_{\epsilon}^{\zs} \exd z\sqrt{\frac{v(z)^2 w(z)^2}{z^6}\left(\frac{1}{U(z)}+v(z)^2 x_1'(z )^2\right)}  \nonumber\\
    &=\frac{1}{4G_5}V_{\perp}\int_{-(\ell-\epsilon)/2}^{(\ell-\epsilon)/2} \exd x_1\sqrt{\frac{v(z)^2 w(z)^2}{z^6}\left(\frac{z'(x_1)}{U(z)}+v(z)^2\right)}\, , \end{align}
and the integral in eq.~(\ref{eq:affine_transverse}) as,
\begin{align}
      S_{||}&=\frac{1}{2G_5} V_{||}\int_{\epsilon}^{\zs} \exd z\sqrt{\frac{v(z)^4}{z^6}\left(\frac{1}{U(z)}+w(z)^2 x_3'(z )^2\right)} \nonumber\\
     & =\frac{1}{4G_5}V_{||}\int_{-(\ell-\epsilon)/2}^{(\ell-\epsilon)/2} \exd x_3\sqrt{\frac{v(z)^4}{z^6}\left(\frac{z'(x_3)}{U(z)}+w(z)^2 \right)}  \, ,
\end{align}
where we have introduced a finite cutoff $\epsilon\ll 1$ and the turning point $\zs$, the point at which the surface reaches its maximum depth in the bulk $AdS$ spacetime. The depth into the bulk which the surface probes is not independent of the width of the strip $\ell$. We can relate the width $\ell$ to the bulk depth $\zs$ by considering
\begin{equation}
    \frac{\ell}{2}=\int_0^{\ell/2}\exd x=\int_{\zs}^{\epsilon}\frac{\exd x}{\exd z}\exd z \, .
\end{equation}
The values of the strip width $\ell$ and the conjugate bulk depth $\zs$ for the surfaces calculated in this work are displayed in figure~\ref{fig:conjugate}.

\subsection{Alternate embedding}\label{sec:appendix_ALT}
\begin{figure}[ht]
    \begin{subfigure}[b]{0.5\textwidth}
    \includegraphics[width=75mm,scale=0.5]{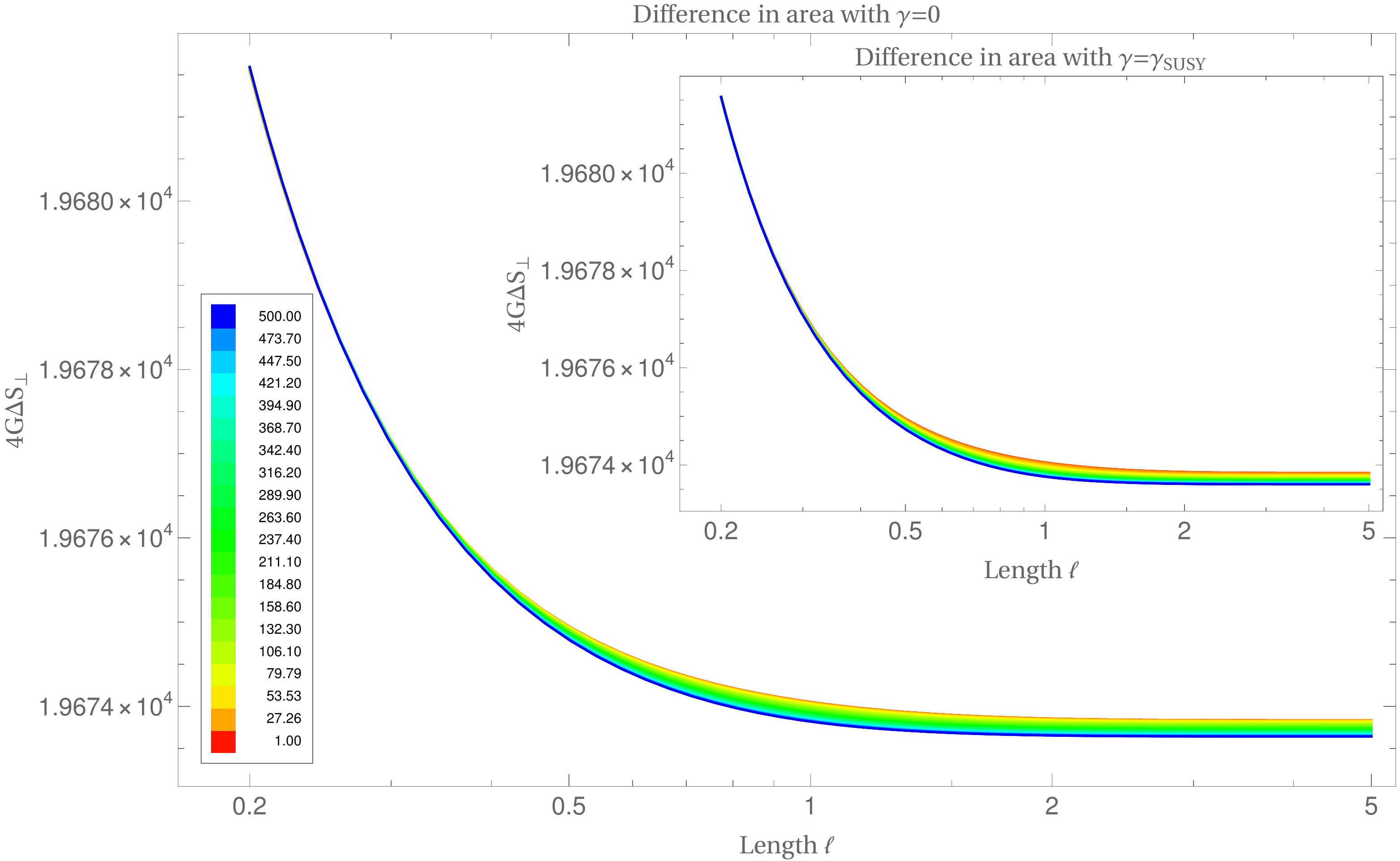}
 \end{subfigure}
 \begin{subfigure}[b]{0.5\textwidth}
    \includegraphics[width=75mm,scale=0.5]{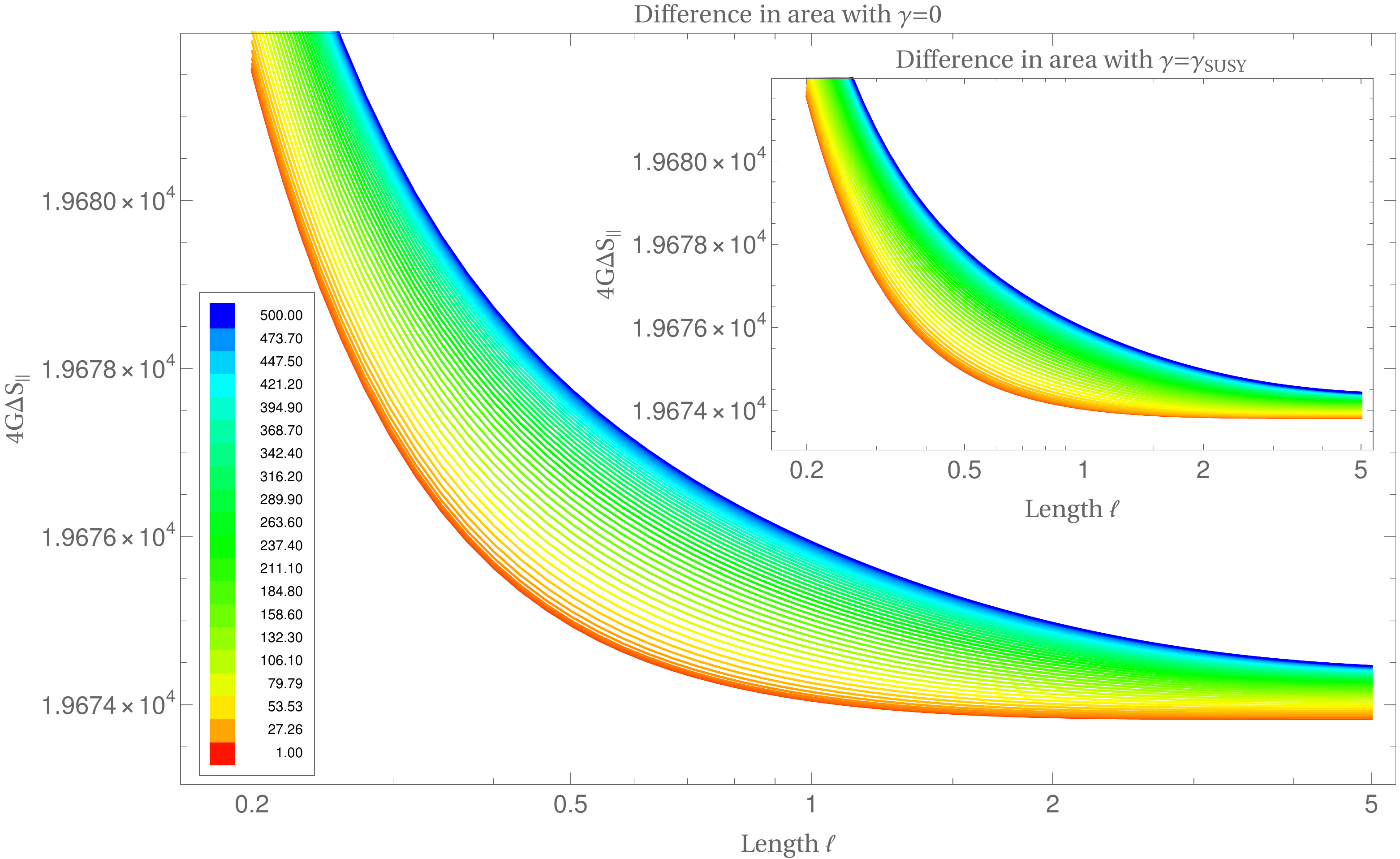}
    \end{subfigure}
    \caption{Difference in the area between two possible embeddings $\Delta S$  \textit{Left:} Transverse strips $\Delta S_{\perp}$.  \textit{Right:} Parallel strips $\Delta S_{||}$. In both images the inset graphic shows the same information with $\gamma=\gamma_{SUSY}$. The differences are displayed as a function of length scale $\ell$ at fixed $\mu/T=5$ for $B/T^2$ varying from $\color{red}{B/T^2=1.00}$ to $\color{Blue}{B/T^2=500.00}$.
    \label{fig:ALT}}
\end{figure}

In our system there exists an additional possible solution which is given by three segments,
\begin{equation}
    x_{i}=\pm \ell/2,\, z, \qquad x_{i},\, z=z_h\, .
\end{equation}
Here the subscript, $i= 1, \, 3$, indicates either parallel ($x_3$) or perpendicular ($x_1$) strips~\cite{Zhang:2019zbf}. Constructing the EE via eq.~(\ref{eq:surface}) we find
\begin{align}
    \hat{S}_{\perp}&=\frac{1}{4G_5} V_{\perp}\left(2\int_{z_h}^{0}\exd z \frac{v(z)w(z)}{z^3 U(z)^{1/2}} + \ell \frac{v(z_h)^2 w(z_h)}{z_h^3}\right)\, , \\
     \hat{S}_{||}&=\frac{1}{4G_5} V_{||}\left(2\int_{z_h}^{0}\exd z \frac{v(z)^2}{z^3 U(z)^{1/2}} + \ell \frac{v(z_h)^2 w(z_h)}{z_h^3}\right)\, .
\end{align}
To test for a transition in the EE due to a change in embedding we construct,
\begin{equation}
    \Delta S_{\perp}=\hat{S}_{\perp}-S_{\perp}, \quad \Delta S_{||}=\hat{S}_{||}-S_{||} \, .
\end{equation}
The result of this construction is displayed in figure~\ref{fig:ALT}. The figure shows that there is no transition to a more favorable embedding within the regime we probe in this work.

\subsection{Monotonicity of Schwarzschild $AdS$}
\label{sec:mono_schwarzschild}
In this section we will display in $d>2$ asymptotically anti-de-Sitter Schwarzschild $AdS_{d+1}$ spacetime the c-function as defined in eq.~(\ref{eq:c_function}) is monotonically increasing. The line element in eq.(\ref{eq:Metric_sch}) can be rewritten as,
\begin{equation}
    ds^2=-\frac{r^2}{L^2} f(r) \exd t^2+\frac{\exd r^2}{\frac{r^2}{L^2}f(r)}+\frac{r^2}{L^2}\exd x_i\exd x^i \, .
\end{equation}
We perform the transformation shown in eq.~(\ref{eq:RG_Coordinates}) to obtain,
\begin{equation}
    ds^2=e^{2A}A'^2\exd t^2+\exd \beta^2+e^{2A}\exd x_i\exd x^i \, ,
\end{equation}
where we have put the $AdS$ radius $L=1$. If we consider the entanglement of a strip directed along the $x_1$ direction we use eq.~(\ref{eq:mono_perp}) and we find in general that we must show
\begin{equation}
    \frac{ \partial c}{\partial \beta} = \frac{\beta_4}{4G_5} e^{k_s}\ell^{d_{\perp}-1}d_{\perp}\left(k'_m\int_0^\ell\exd x_1 \frac{1}{k'(\beta)}\left(\frac{k'(\beta)}{d_{\perp}}-A_1'(\beta)-\frac{k''(\beta)}{k'(\beta)}\right)\right) \leq 0 \, , \label{eq:mono_sch_int}
\end{equation}
where we have dropped the perpendicular subscript as the value will be the same for any choice of orientation. In this case $A_1=A_2=A$, $e^{2B}=e^{2A}A'^2$ and $k=(d-1)A$. Furthermore, $d_1+d_2=d-1$ and $d_\perp=d_{||}=d-1$. The quantity in the brackets becomes
\begin{equation}
   \frac{1}{k'(\beta)} \left(\frac{k'(\beta)}{d_{\perp}}-A_1'(\beta)-\frac{k''(\beta)}{k'(\beta)}\right)=-\frac{1}{d-1}\frac{A''}{A'^2}\, . 
   \label{eq:mono_sch_suff}
\end{equation}
To understand how this quantity behaves we recall that $f(r)=A'^2$. With this in mind we find that
\begin{equation}
    f'(r)=2 e^{-A}A'', \quad \text{or} \quad A''(\beta)=\frac{r f'(r)}{2}\, .
\end{equation}
Inserting our result into eq.~(\ref{eq:mono_sch_suff}) we find
\begin{equation}
    -\frac{1}{d-1}\frac{A''}{A'^2}=-\frac{r f'(r)}{2(d-1)f(r)} \, .
\end{equation}
The blackening factor is given by $f(r)=1-(r_0/r)^d\geq 0$ and $f'(r)=dr_0^d/r^{d+1}\geq 0$, hence we have shown that
\begin{equation}
    -\frac{r f'(r)}{2(d-1)f(r)} \leq 0 \, .
    \label{eq:mono_sch_result}
\end{equation}
Since $A'_m>0$, which can be seen from the defining relation, we find that all of the terms multiplying the integral are greater then our equal to zero. Hence, eq.~(\ref{eq:mono_sch_result}) proves that $ \frac{ \partial c_{\perp}}{\partial \beta}\leq 0$, i.\ e.\ our c-function is monotonically decreasing along $\beta$. That is to say, as we flow from the UV to the IR our c-function monotonically increases. This is exactly the opposite of what can be found in the RG geometries of \cite{Freedman:1999gp,Myers:2010tj}. The increasing flow of the c-function confirms the results of~\cite{Paulos:2011zu} who showed that differently defined type of c-functions also increase in thermal states.  
In terms of the temperature of the black brane $T=dr_0/(4\pi)$ we can recast our final result in the form,
\begin{equation}
     \frac{ \partial c}{\partial \beta} = \frac{\beta_4}{8G_5} r(x_m)^{d-1}\ell^{d-2}(d-1)\sqrt{f(r(x_m))}\left[\int_0^\ell\exd x_1 \left(-\frac{d (4\pi T)^{d}}{(d r(x_1))^d-(4\pi T)^d}\right)\right] \leq 0 \, , 
\end{equation}
where we recall that $x_m$ is the location of the turning point of the minimal surface in the $AdS$ bulk. 
Notice that taking the zero mass limit $r_0\rightarrow 0$ reduces the blackening factor to the standard form of empty $AdS$ i.\ e.\ $f(r)\rightarrow 1$. In this limit the temperature in our formula goes as $T\rightarrow 0$ and hence, 
\begin{equation}
    \lim_{r_0\rightarrow 0}\frac{\partial c}{\partial \beta}=0
\end{equation}
where we see this formula correctly reduces to the result displayed in figure~\ref{fig:empty}. The c-function is a constant in the vacuum state of the dual field theory when there are no operators added to the spectrum to drive an RG flow. 

\subsection{Null energy condition}
\label{sec:NEC}
In general the NEC requires that the energy-momentum tensor satisfy,
\begin{equation}
    n_{\mu}n_{\nu}T^{\mu\nu}\geq 0,\quad \text{with} \quad  n_{\mu}n_{\nu}g^{\mu\nu}=0 \, ,
\end{equation}
for all null vectors $n^{\mu}$. In our work the energy momentum tensor can be written as
\begin{equation}
    T_{\mu\nu}\exd x^{\mu}\exd x^{\nu}=  T_{zz}\exd z\exd z+ T_{tt}\exd t\exd t+ T_{11}\exd x_1\exd x_1 + T_{22}\exd x_2\exd x_2+ T_{33}\exd x_3\exd x_3+ 2 T_{t3}\exd t\exd x_3 \, .
\end{equation}
Using the null condition to eliminate $n_0$ and using the fact that the remaining components of $n$ are independent, the NEC reduces to the following conditions in general,
\begin{align}
0 & \leq g_{33} g_{tt} (\tensor{T}{^r_r}-\tensor{T}{^t_t})+g_{t3}^2 (\tensor{T}{^3_3}-\tensor{T}{^t_t})-g_{t3} g_{tt} \tensor{T}{^t_3}  \, ,\\
0 & \leq g_{33} g_{tt} (\tensor{T}{^1_1}-\tensor{T}{^t_t})+g_{t3} (g_{t3} (\tensor{T}{^3_3}-\tensor{T}{^t_t})-g_{tt} \tensor{T}{^t_3}) \, , \\
 0 & \leq g_{33} g_{tt} (\tensor{T}{^2_2}-\tensor{T}{^t_t})+g_{t3} (g_{t3} (\tensor{T}{^3_3}-\tensor{T}{^t_t})-g_{tt} \tensor{T}{^t_3}) \, , \\
0 & \leq 2 \sqrt{g_{33} g_{tt}+g_{t3}^2} (g_{t3} (\tensor{T}{^t_t}-\tensor{T}{^3_3})+g_{tt} \tensor{T}{^t_3})+g_{33} g_{tt} (\tensor{T}{^3_3}-\tensor{T}{^t_t}) \nonumber  \\
&+2 g_{t3}^2 (\tensor{T}{^3_3}-\tensor{T}{^t_t})-2 g_{t3} g_{tt} \tensor{T}{^t_3} \, .
\end{align} 
If we set $g_{t3}=T_{t3}=0$, we recover the result of~\cite{Chu:2019uoh}, 
\begin{align}
  0 & \leq \tensor{T}{^r_r}-\tensor{T}{^t_t}\, , \\
  0 & \leq \tensor{T}{^1_1}-\tensor{T}{^t_t}\, , \\
   0 & \leq \tensor{T}{^2_2}-\tensor{T}{^t_t}\, , \\
   0 & \leq \tensor{T}{^3_3}-\tensor{T}{^t_t}\, ,
\end{align}
for the NECs when working in an anisotropic $AdS_{5}$ geometry with three spatial directions. Contracting the Einstein equations with the null vectors $\xi$ the authors write the NEC in terms of the Ricci curvature
\begin{equation}
 (\tensor{R}{^i_i}-\tensor{R}{^0_0})\geq 0\, ,\quad  (\tensor{R}{^j_j}-\tensor{R}{^0_0}) \geq 0\, ,\quad   (\tensor{R}{^r_r}-\tensor{R}{^0_0}) \geq 0 \, .
\end{equation}

For the case we will discuss in this work, the energy-momentum tensor is due to a bulk $U(1)$ gauge field and takes the standard form 
\begin{equation}
  T_{\mu\nu}=  \frac{1}{2}\left(F_{\mu\alpha}\tensor{F}{_{\nu}^{\alpha}}-\frac{1}{4}g_{\mu\nu}F_{\alpha\beta}F^{\alpha\beta}\right)\label{eq:Energy_Momentum}.
\end{equation}
Given a null vector $n^{\mu}$ we find 
\begin{equation}
     n^{\mu}n^{\nu}T_{\mu\nu}=\frac{1}{2}\psi^{\alpha}\psi_{\alpha}, \qquad \psi^\alpha=n^{\mu}\tensor{F}{_\mu^\alpha}\, .
\end{equation}
Notice that the contraction of $\psi$ with the null vector $n$ vanishes
\begin{equation}
    n^{\mu}\psi_{\mu}=n^{\mu}n^{\nu}\tensor{F}{_\nu_\mu}=0,
\end{equation}
since it is the contraction of a symmetric and anti-symmetric tensor. Since $n^{\mu}$ is null, this implies that $\psi^{\alpha}$ is either null or spacelike and hence\footnote{This can be seen quickly by constructing a null vector $n$ and choosing an arbitrary $\psi$ such that $n^{\mu}\psi_{\mu}=0$. Due to the signature of the metric and condition $n^{\mu}\psi_{\mu}=0$, one finds that $\psi^{\mu}\psi_{\mu}\geq 0$ no matter what choice is made for the remaining components of $\psi^{\mu}$ or $n^{\mu}$. } 
\begin{equation}
     n^{\mu}n^{\nu}T_{\mu\nu}=\frac{1}{2}\psi^{\alpha}\psi_{\alpha}\geq 0\, .
\end{equation}
Therefore, the Maxwell-Chern-Simons matter which we couple to the Einstein Gravity action in our theory obeys the NEC. Here we stress again however that the matter which we couple to our gravitational theory is not used to trigger an RG flow, rather, it is included to prepare a specific state of the dual CFT. 

\section{Solution strategies}
In this appendix we collect a brief description of the various numerical methods and techniques we have used in this work. 

\subsection{Numerical methods}
\label{sec:appendix_numerics}
\textbf{Einstein-Maxwell-Chern-Simons spacetime solutions:}
We construct solutions to the equations displayed in appendix~\ref{sec:eom} via a pseudo-spectral method~\cite{boyd}. We employ a truncated Chebyshev representation of the metric components on a Gauss-Lobatto grid of $N$ grid points~\cite{boyd}. In order to deal with the nonlinear nature of the equations we utilize a Newton-Raphson technique. In essence we represent the equations of motion as a matrix
\begin{equation}
    L \phi = S \, ,
\end{equation}
where $L$ is a linear operator, $S$ is a source function and we represent the functions as a vector in a function space as $\phi=(U,v,w,c,E,P)$. We extract the linear operator by replacing $\phi(z)$ by $\phi(z)+\epsilon\Delta \phi(z)$ where $\epsilon$ is a small number and $\Delta\phi$ represents a correction to an initial given $\phi$. Representing the equations of motion by the vector $e^i$, and replacing $\phi$ we linearize the equations around an initial field configuration
\begin{equation}
    \frac{\exd e^i}{\exd \epsilon}\left|_{\epsilon=0}\right.=-e^i \, .
\end{equation}
The resulting equations that we solve are 
\begin{equation}
    L^i_j \Delta\phi_j=-e^i, \quad \Delta\phi=(\Delta U,\Delta v,\Delta w,\Delta c,\Delta E,\Delta p) \, .
\end{equation}
By providing an initial field configuration we can then iteratively improve the initial guess by solving the linear equation. On each step of the algorithm we add a correction back to the total solution since by definition
\begin{equation}
    \Delta\phi_j=\phi^{(n+1)}_j-\phi^{(n)}_j \, ,
\end{equation}
where the index $(n)$ on the fields represent the $n$th iteration of the solution process. 

We typically start our scheme at low values of the magnetic field for which a Schwarzschild black brane is good initial guess. However we may suspect that our guess can be far away from the true solution since we have generated our linear operator only to first order in $\epsilon$. Hence we do not add the full correction to the solution initially. Instead we use,
\begin{equation}
   \phi^{(n+1)}_j = \alpha_1\Delta\phi_j+\phi^{(n)}_j,\quad \alpha_1=\begin{cases}
   \frac{1}{N n}\sum_{i=1}^{n}\sum_{k=1}^{N}  e^i_k >1/20, & 1/20 \\
    \frac{1}{N n}\sum_{i=1}^{n}\sum_{k=1}^{N}  e^i_k <1/20, & 1
   \end{cases} \, .
\end{equation}
Here, $N$ is the number of points in the Chebshyev grid and $n$ is the number of equations of motion. 
We continue to iteratively solve the system of equations until $e^i$ is as small as we need it to be; we typically continue until
\begin{equation}
    \frac{1}{N n}\sum_{i=1}^{n}\sum_{k=1}^{N} e^i_k <10^{-15} \quad\text{\&} \quad\xi<10^{-15} \, ,
    \label{eq:convergence_Background}
\end{equation}
where we have introduced $\xi$ as a measure of the convergence of the method constructed from the left-over (thus far unused) Einstein equation as discussed in the next section. 

\paragraph{Solutions of fixed dimensionless quantities:} In order to work at a fixed $\tilde{\mu}=\mu/T$ and $\tilde{B}=B/T^2$ we again employ a Newton-Raphson-like scheme only in $\mu/T(\mu,B)$ and $B/T(\mu,B)^2$. We repeatedly solve the EMCS equations of motion until we find a solution with the desired values. However in order to fully utilize a Newton method we need to know the Jacobian generated from the equations of motion. Unfortunately, we \textit{a priori} do not know $T(\mu,B)$ as it is known only numerically. To circumvent this issue we use a 2D secant method. This method is identical in methodology to the Newton method only we take the addition step to write derivatives in the Jacobian as finite differences. Explicitly, for $h=(\mu/T(\mu,B),B/T(\mu,B)^2)$ and $y=(\mu,B)$,
\begin{equation}
    J^i_j=\frac{\partial h^i}{\partial y^j} \, .
\end{equation}
With the Jacobian in hand we linearize by replacing 
\begin{subequations}
 \begin{align}
 T(\mu,B)&\rightarrow 
\frac{1}{2}(T\left(\mu_{k-1},B_{k-1}\right)+T\left(\mu_{k-2},B_{k-2}\right))\, , \\
 \partial_\mu T&\rightarrow \frac{T_{k-1}-T_{k-2}}{\mu_{k-1}-\mu_{k-2}},\quad \partial_B T\rightarrow \frac{T_{k-1}-T_{k-2}}{B_{k-1}-B_{k-2}} \, , \\
  B&\rightarrow \frac{1}{2}(B_{k-1}+B_{k-2}),\quad
\mu\rightarrow \frac{1}{2}(\mu_{k-1}+\mu_{k-2}) \, ,
\end{align}\label{eq:linearize}
\end{subequations}
in order to construct a linear finite difference operator, $L^i_j$, representing the Jacobian of the system. While this method does not require an explicit form of the derivative it does require information about the previous two steps for each iteration. To start our scheme we simply compute two arbitrary solutions. To advance forward to the next iteration we compute
\begin{equation}
    y^i_{k}=y^i_{k-1}-\alpha_2 (L^{-1})^i_jh_{k-1}^j,\quad \alpha_2=\begin{cases}
   \frac{1}{2}| h^i-(\tilde{\mu},\tilde{B})|^2 >1/20, & 1/20 \\
   \frac{1}{2}| h^i-(\tilde{\mu},\tilde{B})|^2 <1/20, & 1
   \end{cases} \, .
\end{equation}
We continue to iterate until our solutions satisfy both $\frac{1}{2}| h^i-(\tilde{\mu},\tilde{B})|^2<10^{-15}$ and the conditions given in eq.~(\ref{eq:convergence_Background}). To construct explicit solutions for the minimal surfaces dual to the entanglement entropy we follow two different approaches.

\paragraph{Affine (non-affine) parameterization:} In the first approach we follow~\cite{Cartwright:2020qov,Ecker:2015kna,Cartwright:2019opv}. We first compute geodesic equations associated with the line element\footnote{We repeat briefly our analysis given in~\cite{Cartwright:2019opv} for completeness.},
\begin{equation}
    \frac{\exd^2 x^{\mu}}{\exd \sigma^2}+\tensor{\Gamma}{^{\mu}_{\alpha}_{\beta}} \frac{\exd x^{\alpha}}{\exd \sigma} \frac{\exd x^{\beta}}{\exd \sigma} =J \frac{\exd x^{\mu}}{\exd \sigma} \, .
\end{equation}
This leaves us with two equations for $(z(\sigma),x_1(\sigma))$ or $(z(\sigma),x_3(\sigma))$. We then discretize by replacing
\begin{subequations}
 \begin{align}
 f(z(\sigma))&\rightarrow 
\frac{1}{2}(f\left(z_{k-1}\right)+f\left(z_{k}\right)) \, , \\
 f'(\sigma)&\rightarrow \frac{f_{k}-f_{k-1}}{\sigma_{k}-\sigma_{k-1}} \, , \\
  f(\sigma)&\rightarrow \frac{1}{2}(f_{k}-f_{k-1}) \, , \\
  J f(\sigma)&\rightarrow \frac{1}{2}(J_{k-1}f_{k-1}+J_{k}f_k),\hspace{1cm} 
J=\frac{\exd^2 \lambda}{\exd \sigma^2}/\frac{\exd \lambda}{\exd \sigma} \, .
\end{align}\label{eq:discrete}
\end{subequations}
We can then construct a matrix representation of the equations by working with purely first order equations at the grid points and minimizing the correction needed to each grid point. We work with 
\begin{equation}
 E^{\mu}\equiv\begin{cases} \frac{\exd x^{\mu}}{\exd\sigma} - p^{\mu} \, , \quad 
\hspace{.2cm} \text{if} \hspace{.2cm}\mu=1,2 \, , \\
 \frac{\exd p^{\mu}}{\exd 
\sigma}+\tensor{\Gamma}{^{\mu}_{\alpha\beta}}p^{\alpha}p^{\beta}
 -J p^{\mu}\, , \quad \hspace{.2cm}\text{if} \hspace{.2cm}\mu=3,4  \, .
\end{cases}
\end{equation}
We then
expand $E^{\mu}_n$ in a Taylor series up to first order and require this Taylor 
series to vanish as displayed in~\cite{numericalrecipes}. As with our background spacetime solutions we require an initial guess. For our first step we take an empty $AdS$ spacetime solution as the guess~\cite{Ecker:2015kna,Cartwright:2019opv} and then use our solution to seed our next relaxation. We use the same metric of success as with our background solutions by investigating
\begin{equation}
    \frac{1}{N_r n_r}\sum_{\mu=1}^{n_r}\sum_{k=1}^{N_r}  E^{\mu}_k <10^{-15}  \, ,
\end{equation}
where $N_r$ is the number of grid points in our relaxation grid and $n_r$ is the number of equations we solve.  

\paragraph{Coordinate parameterization:} In the coordinate parameterization we directly compute the relevant integrals for the Entanglement entropy via Mathematica's NIntegrate command. We begin by computing the background solutions as described above and then directly compute lengths of the strips in the dual theory by assigning a collection of values for $\zs$ via
\begin{equation} 
    \frac{\ell}{2}=\int_{\zs}^{\epsilon}\frac{\exd x}{\exd z}\exd z \, .
\end{equation}
We can invert the relation to obtain $\zs(\ell)$. We then solve eq.\ (\ref{eq:Conserved}) with the condition $z'(\zs)=0$ for $H_0$. Inserting the result back into eq.\ (\ref{eq:Conserved}) gives a relation for $z'(x)$ which can be inserted into eq.\ (\ref{eq:Coor_EE}) resulting in an integral which gives the value of the entanglement entropy for a fixed value of $\zs$. We can numerically integrate via Mathematica's NIntegrate command~\cite{Mathematica},
\begin{equation}
    S=\text{NIntegrate[}L,\{z,\zs,\epsilon\}\text{]} \, .
\end{equation}
We have used this method as a further independent verification of the EE we compute in this work.

\subsection{Numerical convergence}
\label{sec:appendix_convergence}
We have checked the numerical convergence of both the solutions to the background equations of motion as well the solutions to the geodesic equations. The background equations of motion contain a left over constraint equation $k(U,v,w,c;z)=0$. We use this left over first order differential equation as a separate means of checking the solutions of the equations of motion. We examine,
\begin{equation}
 \xi=  \log\left(\left|Max_{z\in (0,1)}\left(\frac{k(z)}{|k(z)|}\right)\right|\right).
\end{equation}
Where by $|k(z)|$ we mean that we take the equation $k(z)$ and expand all of the terms in the equation and sum the absolute value of each term. By doing so we obtain a relative measure for the residual of the leftover Einstein equation. During the solution of the equations of motion we monitor $\xi$ as well as the mean of the residual of the equations of motion at all grid points with typical residuals on the order of $10^{-20}$. 

\begin{figure}[H]
    \centering
    \includegraphics[width=13cm]{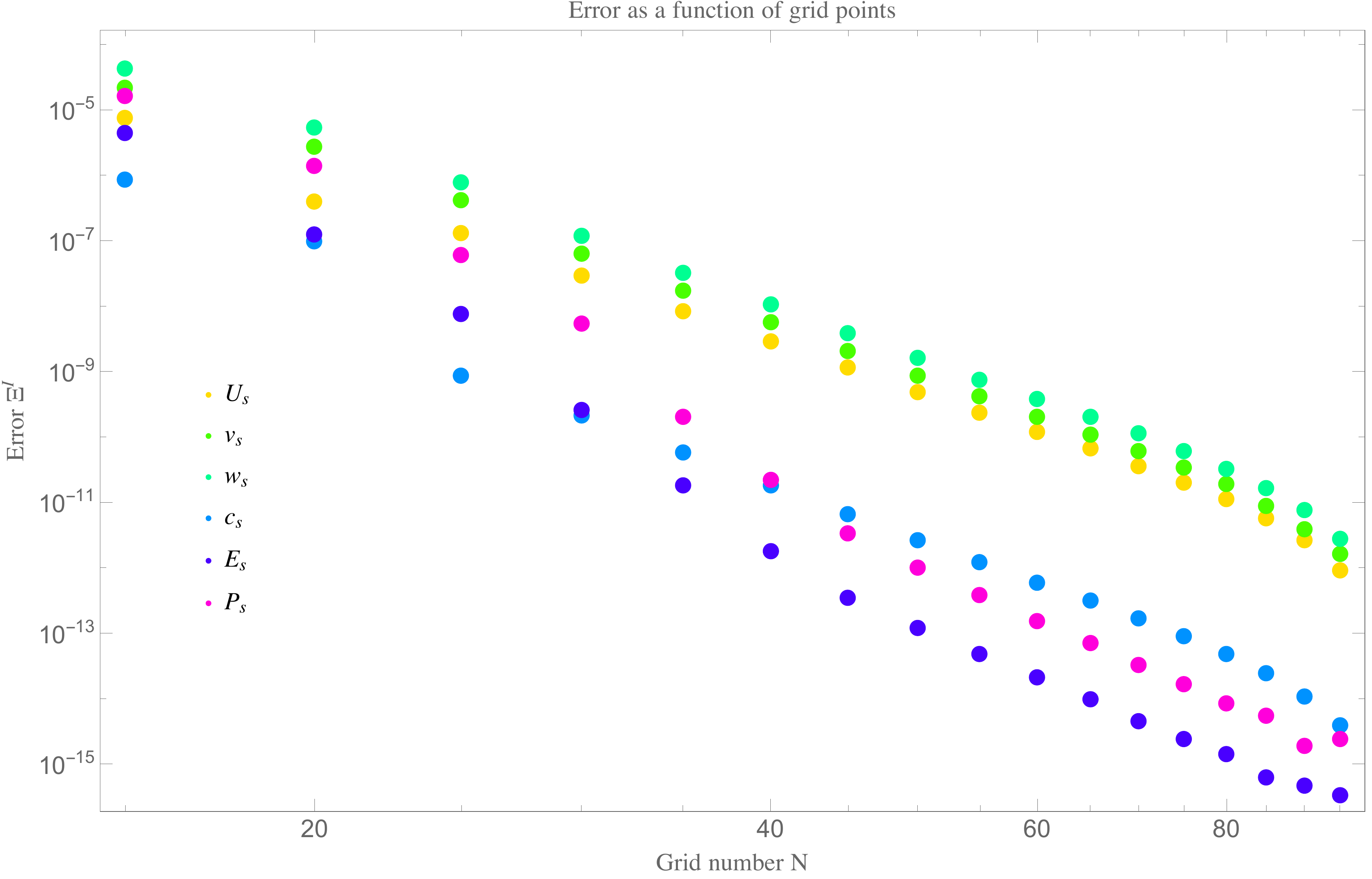}
    \caption{We display an example of the grid point convergence of the solutions for the spacetime metric functions, $\Xi^{I}$, as defined in eq.~(\ref{eq:convergence}). The solutions we display are computed at example values of $\mu/T=0.32772$ and $B/T^2=
10.740155$ with $\gamma=2/\sqrt{3}$. 
    \label{fig:grid_convergence}}
\end{figure}

In addition to monitoring the residual of the differential equations we also check the accuracy of our method by checking the convergence of our solution as a function of the number of grid points. We compute
\begin{equation}
    \Xi^{I}=\max_{z\in[0,1]}\left| \phi^I(z;N) -\phi^{I}(z,N_{max})\right| \, ,
    \label{eq:convergence}
\end{equation}
where $\phi^{I}(z,N_{max})$ is a reference solution at the grid size we used to compute all figures in this work given by $N_{max}=100$. An example of the convergence of the solutions is displayed in figure~\ref{fig:grid_convergence}.

During the solution of geodesic equations of motion we monitor the mean of the residual over all grid points. In addition, our method to compute the entanglement entropy relies on an ultra-violet cutoff. We computed the entanglement entropy for a variety of cutoff values $z_{UV}$ and display the  results in table~\ref{tab:cut_dep_trans_highT}. We have also checked the dependence of our results on the number of grid points used in the discretization.

To ensure convergence of our solutions we also employ an analytic mesh refinement of the radial Chebyshev grid given by~\cite{Ammon:2016szz,Macedo:2014bfa}
\begin{equation}
    z=1-\frac{\sinh{\lambda(1-\zeta)}}{\sinh\lambda} \, .
\end{equation}
Explicitly using this transform leads to very large equations of motion. The resulting integration of the field equations takes a large amount of time. To circumvent this we do not explicitly change variables in the equations of motion. Rather we only change variables when computing the differentiation matrices. This is most easily seen in an example,
\begin{equation}
    A(z)f''(z)+B(z)f'(z)+C(z)f(z)=S(z) \, .
\end{equation}
The differential operator associated with this linear problem is,
\begin{equation}
  L=  A(z)\frac{d^2}{dz^2}+ B(z)\frac{d}{dz}+ C(z)\, ,
\end{equation}
which we render discrete by replacement of the derivatives by their associated Chebyshev matrix representation,
\begin{equation}
   L_{discrete}=\bar{A}(z)\cdot D_Z^2+ \bar{B}(z)\cdot D_z+ \bar{C}(z)\cdot I \, ,
\end{equation}
where $I$ is an $N\times N$ identity matrix. The bar notation indicates constructing a matrix with each entry representing the value of the function at each grid point along the diagonal of an $N\times N$ matrix. The speed up of our routine is accomplished by replacing the derivative operators $D$ with transformed derivative operators,
\begin{equation}
    D_z= \bar{\frac{d \zeta}{d z}}\cdot D_\zeta, \quad  D_z^2= \left(\bar{\frac{d \zeta}{d z}}\right)^2\cdot D_\zeta^2 + \bar{\frac{d^2 \zeta}{d z^2}}\cdot D_\zeta \, .
\end{equation}
The bar notation again indicates evaluating the quantity at each grid point and placing it along the diagonal of an $N\times N$ matrix. The operators $D_\zeta$ are again Chebyshev differentiation matrices created with a Gauss-Lobatto grid for the variable $\zeta$. The transformed derivative operator then reads,
\begin{align}
L_{discrete}&=\bar{A}\left(1-\frac{\sinh{\lambda(1-\zeta)}}{\sinh\lambda}\right)\cdot \left[ \left(\bar{\frac{d \zeta}{d z}}\right)^2\cdot D_\zeta^2+ \bar{\frac{d^2 \zeta}{d z^2}}\cdot D_\zeta \right] \nonumber \\
&+ \bar{B}\left(1-\frac{\sinh{\lambda(1-\zeta)}}{\sinh\lambda}\right)\cdot \left[\bar{\frac{d \zeta}{d z}}\cdot D_\zeta\right]+ \bar{C}\left(1-\frac{\sinh{\lambda(1-\zeta)}}{\sinh\lambda}\right)\cdot I \, . \label{eq:Discrete_Transformations}
\end{align}
where we now evaluate the functions $A,B$ and $C$ locations of the transformed grid. While at first glance this appears more complicated there is a dramatic increase in the speed with which one can find solutions. The transformed derivative operators in the square brackets need only be computed once and the equations of motion do not need to be transformed, only evaluated as described in eq.~(\ref{eq:Discrete_Transformations}). 
\begin{table}[ht]
\centering\begin{tabular}{c|c}
 $z_{uv}$ & $S-Svac$\\
 \hline
 0.00001 & 0.401279 \\
 0.00005 & 0.105756 \\
 0.0001 & 0.103175 \\
 0.0005 & 0.102675 \\
 0.001 & 0.102671 \\
 0.005 & 0.102656 \\
 0.01 & 0.102613 \\
 0.05 & 0.101552 \\
\end{tabular}
\caption{The regulated entanglement entropy at various cutoff surfaces for $l=0.4$ in the transverse direction. We can see variation on the order of $10^{-4}$.  
    \label{tab:cut_dep_trans_highT}}
\end{table}

\bibliographystyle{JHEP}
\bibliography{ref}

\end{document}